\documentclass[11pt,letterpaper]{article}

\usepackage[T1]{fontenc}
\usepackage[english]{babel}
\usepackage{times}

\usepackage[letterpaper,margin=1in]{geometry}

\usepackage{amsmath,amstext,amsbsy,amsopn,amsthm,amsfonts,amssymb}
\usepackage{mathtools}
\usepackage{url}
\usepackage{tabularx,booktabs}
\newcolumntype{C}{>{\centering\arraybackslash}X}
\usepackage{booktabs,array,colortbl}
\usepackage[]{graphicx}
\usepackage{rotate}
\usepackage{wrapfig}
\usepackage{caption}
\usepackage[caption=false]{subfig}
\usepackage{lastpage}

\usepackage[small]{titlesec}
\usepackage{paralist}

\usepackage{verbatim} 
\usepackage[usenames,dvipsnames,svgnames]{xcolor}
\usepackage{multirow}
\usepackage{algorithmic}
\usepackage[linesnumbered,ruled,vlined]{algorithm2e}

\usepackage{fancyhdr}
\renewenvironment{thebibliography}[1]{%
	\begin{oldthebibliography}{#1}%
		\setlength{\parskip}{0.0cm}%
		\setlength{\itemsep}{0.0cm}%
	}%
	{%
	\end{oldthebibliography}%
}

\begin{document}


\title{\Huge{\textbf{PUFchain}: Hardware-Assisted Blockchain for Sustainable Simultaneous Device and Data Security in the Internet of Everything (IoE)}}

\maketitle

\author{
\begin{center}
\begin{tabular}{cc}
Saraju P. Mohanty  &			Venkata P. Yanambaka \\
Department of Computer Science and Engineering &	 School of Science and Engineering \\
University of North Texas, USA.	&		Central Michigan University, USA.  \\
\texttt{saraju.mohanty@unt.edu} &  \texttt{yanam1v@cmich.edu} \\\\
\end{tabular}
\begin{tabular}{cc}
Elias Kougianos & Deepak Puthal \\
Department of Electrical Engineering & School of Computing\\
University of North Texas, TX 76203. &  Newcastle University, UK \\
Email: \texttt{elias.kougianos@unt.edu} &  Email: \texttt{Deepak.Puthal@newcastle.ac.uk} \\
\\\\
\end{tabular}
\end{center}
}


\cfoot{Page -- \thepage-of-\pageref{LastPage}}

\begin{abstract}
\textit{This article presents the first-ever blockchain which can simultaneously handle device and data security, which is important for the emerging Internet-of-Everything (IoE)}.
This article presents a unique concept of blockchain that integrates hardware security primitives called Physical Unclonable Functions (PUFs) to solve scalability, latency, and energy requirement challenges and is called PUFchain.
Data management and security (and privacy) of data, devices, and individuals, are some of the issues in the IoE architectures that need to be resolved. Integrating the blockchain into the IoE environment can help solve these issues and helps in the aspects of data storage and security. 
This article introduces a new blockchain architecture called PUFchain and introduces a new consensus algorithm called ``Proof of PUF-Enabled Authentication'' (PoP) for deployment in PUFchain. The proposed PoP is the PUF integration into our previously proposed Proof-of-Authentication (PoAh) consensus algorithm and can be called ``Hardware-Assisted Proof-of-Authentication (HA-PoAh)''. However, PUF integration is possible in the existing and new consensus algorithms.
PoP utilizes PUFs which are responsible for generating a unique key that cannot be cloned and hence provide the highest level of security. A PUF uses the nanoelectronic manufacturing variations that are introduced during the fabrication of an integrated circuit to generate the keys. Hence, once generated from a PUF module, the keys cannot be cloned or generated from any other module. PUFchain uses a PUF and Hashing module which performs the necessary cryptographic functions. Hence the mining process is offloaded to the hardware module which reduces the processing times.  PoP is approximately 1,000$\times$ faster than the well-established Proof-of-Work (PoW) and 5$\times$ faster than Proof-of-Authentication (PoAh). The transaction time reduction compared to PoAh is 79.15\%. An optimized ultra low power design of the PUF and hashing module also provides a significant decrease of power consumption. 

%
\end{abstract}


\section{Introduction}
\label{SEC:Introduction}

The introduction of the Internet  aided the development of E-commerce and helped with the financial transactions among different entities \cite{Puthal_IEEECEM_2018,puthal2018blockchain}. These electronic financial transactions helped fuel trade growth. A central entity takes responsibility to perform secure communication between two entities and execute a secure financial transaction between them. This central entity is also responsible for, and can be questioned in the case of a failure or fraud. In some cases of centralized systems, there is also a chance that a single point of failure can cause a system failure at a catastrophic level. A central entity also raises trust, privacy and security issues. There is also a delay added by the central entity during the transaction process unlike a peer-to-peer (P2P) communication \cite{Puthal_IEEECEM_2018,puthal2018blockchain}.

\begin{figure}[htbp]
	\centering
	\includegraphics[width=0.80\textwidth]{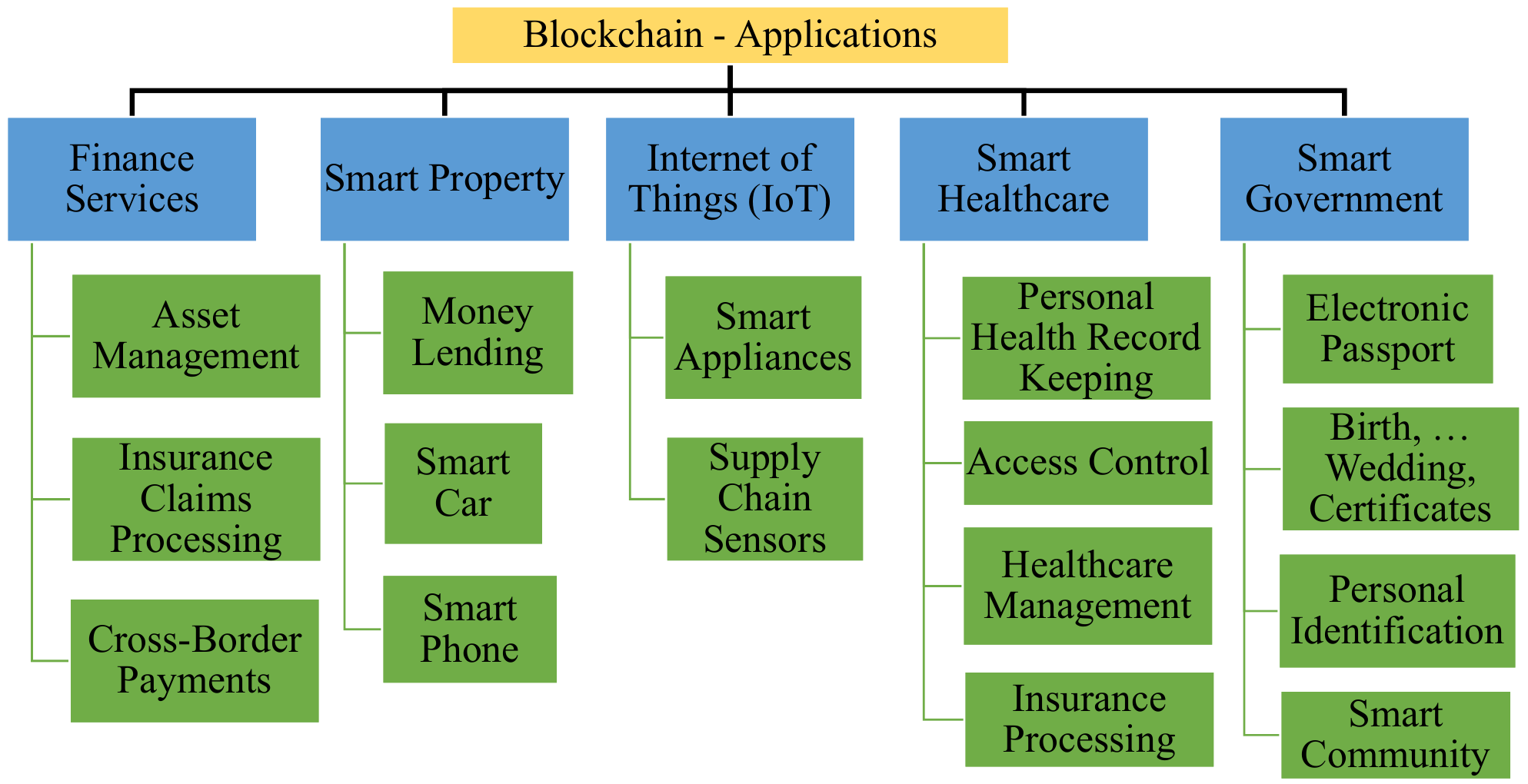}
	\caption{Possible Applications of Blockchain Technology \cite{Puthal_IEEECEM_2018,puthal2018blockchain}.}
	\label{FIG:Blockchain_Technology_Applications}
\end{figure}

Blockchain technology was introduced as an answer to all the questions above. The blockchain uses a decentralized ledger, where all the participants in the network maintain a synchronized copy of the complete or partial ledger. A blockchain-based cryptocurrency, Bitcoin, was proposed  
in 2008 \cite{Lee_MCE_2019-Sep,Nakamoto_CML_2009}. Since then multi-frontal research is being undertaken to use blockchain technologies in a variety of applications, as summarized in Fig. \ref{FIG:Blockchain_Technology_Applications} \cite{Puthal_IEEECEM_2018,puthal2018blockchain,Wu_MCE_2018-Jul,Lamberti_MCE_2018-Jul}.
All the transactions performed between different entities are stored in the distributed ledger and after every successful transaction, the ledger is synchronized across every node in the network. This removes the necessity of a central entity, and with it, the many issues discussed above.

The Internet-of-Things (IoT) is the backbone for a variety of smart application domains, including smart cities, smart healthcare, and smart transportation \cite{Mohanty_MCE_2016-July_Smart-Cities, Roy_MCE_2018-Mar}. In essence ``Instrumentation'', ``Interconnections'', and ``Intelligence'', which are referred to the 3Is of a smart city, are due to the IoT \cite{Mohanty_ICCE_2019_Keynote,Mohanty_ISCT_2019_Keynote}. The Internet of Everything (IoE) is a concept that has the IoT integrated as one of the components of its architecture and environment \cite{Fiaidhi_MITP_2019-Jan,Alan_IEEEComSoc_2014}. People, Data and Processes are the other three components in an IoE environment.  Data collection is one of the crucial elements of IoE. Data collection in an IoE environment is carried out in many ways which are discussed in detail in Section \ref{SEC:IoE_Vision}. The amount of data collected in the IoE environment increases daily.  The devices used for such applications are low performance, low power consumption devices which cannot provide high computational power to the architecture \cite{Moreno_IEEETII_2017}. In such scenarios, introducing an ``Edge Layer'' will aid the entire network of devices by reducing the computing load. In some use cases, near real-time processing can also be achieved with the help of an edge datacenter \cite{Shi_IEEEIoT_2016}.


Edge computing helps in developing the applications where data processing needs to be offloaded but huge amounts of data transfers over the network are not a possibility. The devices in an IoE network do not have high computing capabilities, and in most cases are dedicated to data collection. But transferring all the data to a cloud is also not ideal in such scenario because of the network bandwidth limitations. Security and privacy are two other aspects of  an IoT architecture which need paramount attention during the design phase. To address all these issues, researchers around the world are developing various solutions where the computational needs are offloaded to the edge of the network \cite{Stanciu_CSCS_2017}.

Integration of edge datacenters helps break the limitations of resource constrained and low performance devices \cite{Wright_IEEECPSCom_2018}. These architectures are finding their way into many mission-critical application scenarios such as military, healthcare, and industrial IoT \cite{cisco_edge_fog,Li_IEEEN_2018}. Such mission-critical applications also demand the highest reliability, security and privacy. There have been various solutions proposed for the security and privacy aspects of the IoT architectures \cite{Minoli_CCNC_2017,Pan_arXiv_2018}. Cryptographic algorithms were proposed which can be used to strengthen the security of the IoT. But, a central entity is vital if a cryptographic algorithm has to be used as a security measure for the IoT. 

Blockchain technology can be used to remove the  requirement of a central entity from IoT architectures. A decentralized public ledger is used in the blockchain for organizing the data and executing the transactions. Every node connected to the network has a copy of the ledger. This helps in maintaining consistency and security. A blockchain is cryptographically anchored and tamper-proof and is a collection of different transactions that occurred among the participants in the network \cite{Puthal_IEEECEM_2018,Kolokotronis_IEEECEM_2019}. 

The rest of the paper is organized as follows: A detailed discussion on the IoE and how the integration of blockchain can help solve some of the issues, and the challenges in attaining a successful integration, are presented in Section \ref{SEC:IoE_Vision}. Section \ref{SEC:Novel_Contributions} presents the novel contributions of the paper. Section \ref{SEC:Breakthrough_of_Blockchain} reviews existing blockchain consensus algorithms. Section \ref{SEC:Blockchain_Challenges} discuses the challenges in  blockchain technology, which make it difficult to be useful for large, complex IoT systems. Section \ref{SEC:Related_Research} presents the research that is being conducted in these areas. Related prior research on blockchain technologies and PUFs are also presented in the section. Section \ref{SEC:PUF_Chain} presents the architecture of the proposed blockchain, PUFchain, and Section \ref{SEC:PUF_PoP} presents the proposed consensus algorithm, PoP. Properties and figures of merit of PUFs and the PUF design used for implementation of PUFchain are presented in Section \ref{SEC:PUF}. The experimental results are presented in Section \ref{SEC:Experimental_Results}. Conclusions and future directions are discussed in Section \ref{SEC:Conclusion}.

\section{The Internet of Everything (IoE) - The Need for Both Device and Data Security}
\label{SEC:IoE_Vision}

The IoT is a network of devices that are connected to each other, communicate with each other and exchange data with each other for intelligent decision making \cite{Mohanty_MCE_2016-July_Smart-Cities}. But recently, a new idea or architecture has been introduced, called the Internet of Everything (IoE) \cite{Fiaidhi_MITP_2019-Jan,Alan_IEEEComSoc_2014,Evans_Cisco_2012,Miraz_ITA_2015}. This section discusses the concept of IoE, how blockchain can help implementations of the IoE and its challenges.

\begin{figure}[htbp]
	\centering
	\includegraphics[width=0.70\textwidth]{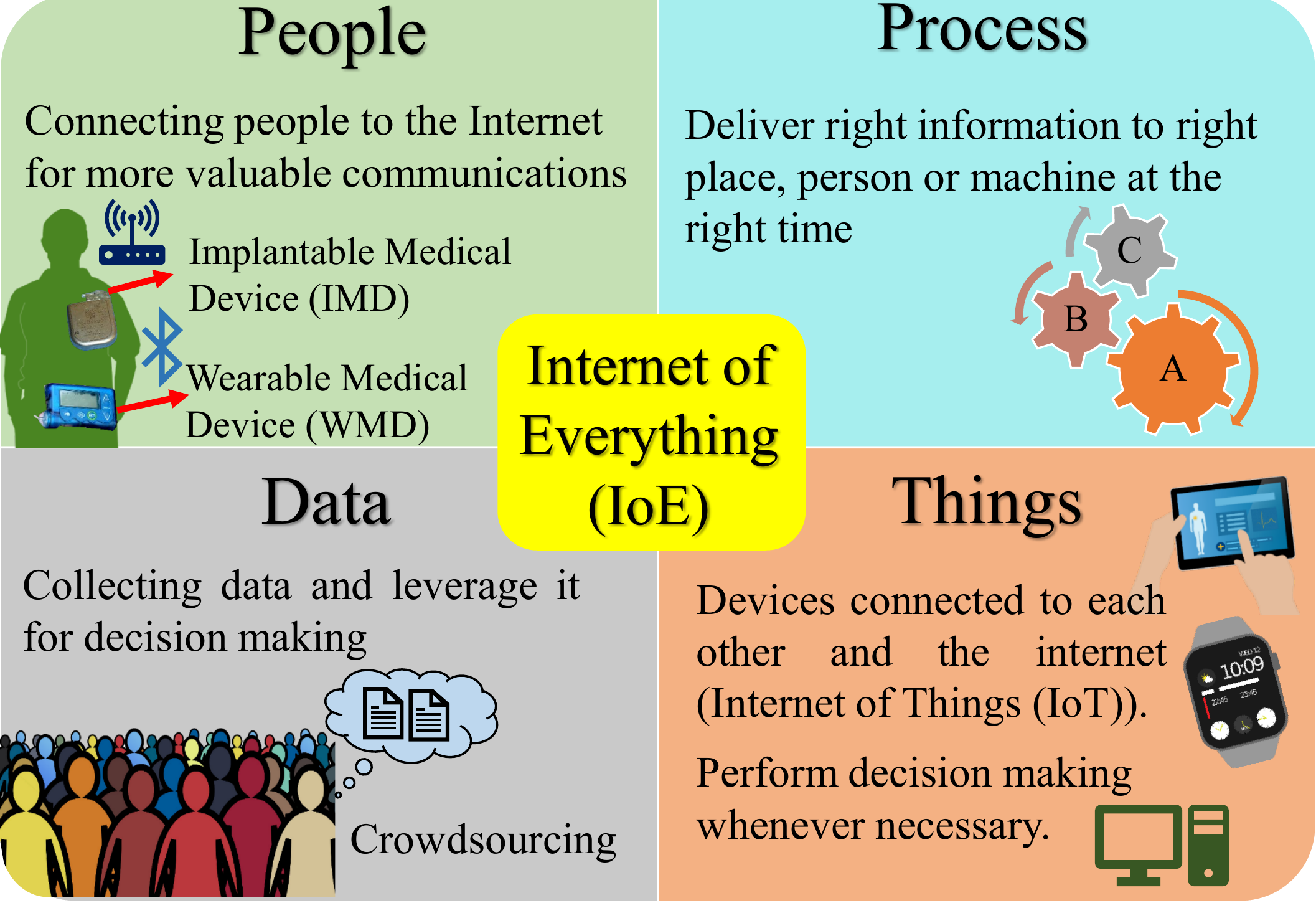}
	\caption{The vision of Internet of Everything (IoE) with Human-in-the-Loop driving serious need for security (including privacy) for device, person, location, and data.}
	\label{FIG:IoE_Vision}
\end{figure}

The IoE is comprised of four different components, \textit{People, Processes, Data,} and \textit{Things}. Fig. \ref{FIG:IoE_Vision} shows a broad perspective of the IoE. The Things in the IoE is the concept of IoT where various devices are connected to the Internet and each other for data exchange and decision making. The other components of the IoE are also crucial parts, where People, Processes, and Data combined bring out a different perspective compared to the IoT alone. 

People in the IoE environment become nodes in the network themselves. Various devices in and on the People are continuously connected to the Internet and participate in establishing more valuable and reliable communications between other People and devices. As an example, Fig. \ref{FIG:IoE_Vision} presents the concept of Smart Healthcare in the IoE, where people have Implantable Medical Devices (IMD) implanted inside them monitoring their health status continuously, such as, for example, a Pacemaker. Indicated are also Wearable Medical Devices (WMD) which a person can wear on their body for various applications such as heart rate monitoring \cite{Prabha_IEEECEM_2018}. These are collectively called Implantable and Wearable Medical Devices (IWMDs) \cite{Zhang_JPROC_2014-Aug}.

The other component in the IoE environment is the Data. Collection of data could be from anywhere. As shown in Fig. \ref{FIG:IoE_Vision}, data collection can be performed using crowd sourcing, which also involves People. The data that is collected is leveraged for making intelligent decisions in various aspects of our day-to-day-life \cite{Evans_Cisco_2012,Minerva_IEEEIoTThings_2015}. The collected data processing and usage also plays a vital role in the case of the IoE. Once the data is collected, routing the ``\textit{right data}'' to the ``\textit{right place}'' or the ``\textit{person}'' or in this case, a possible ``\textit{machine}'', is also important. The data routing also has to be done at the ``\textit{right time}''. This is the ``\textit{Process}'' in the case of the IoE. With all the various components of IoE, data collection, processing and security can face many potential threats. As a solution to all such issues, blockchain integration is being considered and many such architectures are being proposed.

\section{Novel Contributions of this Paper}
\label{SEC:Novel_Contributions}

\subsection{Problem Definition - The Challenge of Blockchain based Security in IoE}


Blockchain technology is one of the exploding technologies around the world which has the potential to be introduced as a solution to various issues in different environments and day-to-day applications, such as the IoT. The blockchain uses cryptographic hashing functions for security aspects, and this also helps in achieving an append-only approach to the data, where once it gets added to the blockchain, it can neither be edited not deleted. 

The blockchain, along with its advantages, also faces some challenges which need solving before it can be integrated into any environment. Some of the challenges are scalability, high computational requirements, high power consumption, and security and privacy. For example, if an IoT environment is considered, almost all these properties, such as, power consumption, computational requirements and security and privacy are bottlenecks for the integration of blockchain. Mining in the blockchain in some consensus algorithms requires very high computational power and dedicated hardware.

%

\subsection{Proposed Novel Solution: Physical-Unclonable-Function-Integrated Blockchain (PUFchain)}

This paper presents a novel IoT architecture where the blockchain is integrated for robust security and data management, called ``\textit{PUFchain}''. A high level of security is attained in this architecture by integrating a hardware assisted security module, a PUF. A hybrid oscillator arbiter PUF presented in \cite{Yanambaka_ALOG_2017-Dec} is used for the implementation of PUFchain in this paper. A PUF and Hashing module is implemented over the IoT device which helps in the computation of cryptographic hashes. Adding an extra module to generate the unique key with a PUF and compute the cryptographic hash reduces the computation requirement of the IoT device. The power consumption can also be reduced significantly by choosing ultra low power designs of PUFs and optimizing the PUF and hashing module. 

\subsection{Proposed Novel Solution: Proof of PUF-Enabled Authentication (PoP)} 

To overcome these issues and integrate the blockchain into IoT applications, this paper also presents a new consensus algorithm. Proof of PUF-Enabled Authentication (PoP) is proposed in this paper to be integrated into an IoT environment. PoP utilizes the unique keys generated by a PUF module and performs the cryptographic hashing functions. The keys generated by a PUF module are unique to the device itself, which makes the entire system robust and resistant to attacks. The keys generated are not transmitted over the network which makes it resilient to various attacks. The PUF and hashing module in PoP also makes it significantly faster compared to the other blockchain architectures by reducing the computing load on the IoT device, as discussed above. This widens the range of devices that can be used for implementation of PUFchain in the IoT environment.

\section{The Breakthrough of Blockchain}
\label{SEC:Breakthrough_of_Blockchain}

\begin{figure}[htbp]
	\centering
	\includegraphics[width=0.90\textwidth]{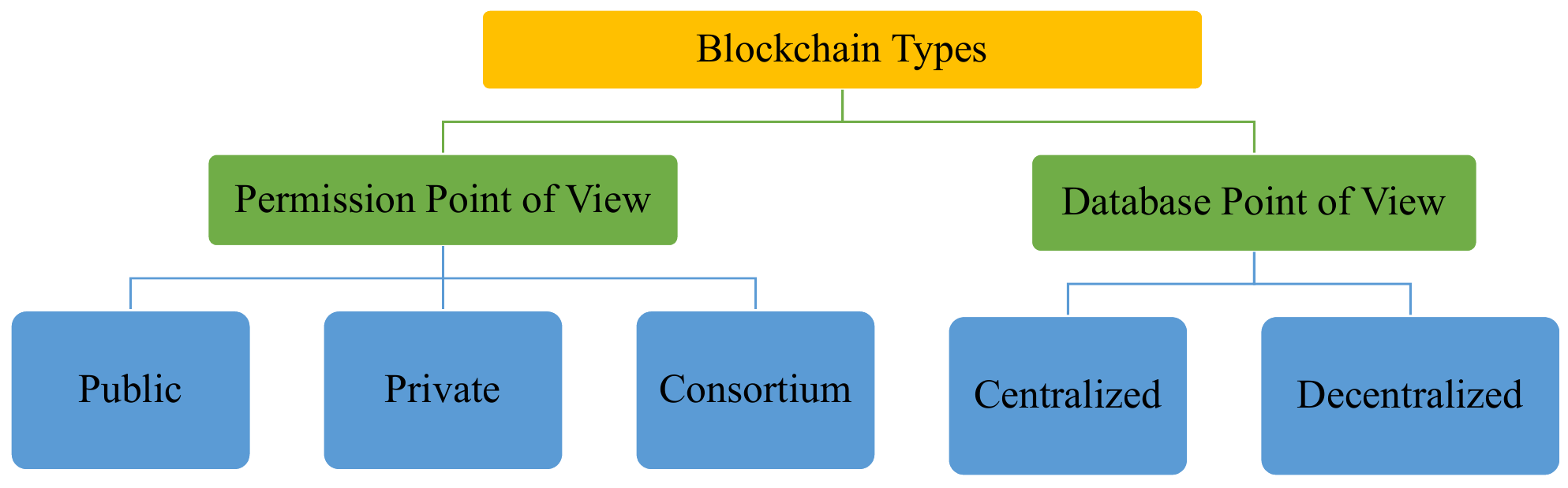}
	\caption{Different Types of Blockchain \cite{Puthal_IEEECEM_2018,puthal2018blockchain}.}
	\label{FIG:Blockchain_Types}
\end{figure}

Blockchain technology can be of various types as shown in Fig. \ref{FIG:Blockchain_Types} \cite{Puthal_IEEECEM_2018,puthal2018blockchain}.
Blockchain uses the concept of distributed ledger where the copy of the entire ledger, or a part of it, will be stored at the local storage of every node in the network. There is no central entity in the case of a blockchain network. The lack of central entity in the blockchain is replaced by a consensus algorithm \cite{Puthal_IEEECEM_2018}. All the participants in the network agree upon a consensus algorithm, a set of rules, required to validate the transactions. For a block of transactions to be validated and added to the blockchain, the ``miners'' in the network should run the consensus algorithm and validate the transactions. The blockchain literature describes several types of consensus algorithms being developed, where individual consensus algorithms have distinct properties, advantages and features. 

\begin{figure}[htbp]
	\centering
	\includegraphics[width=0.90\textwidth]{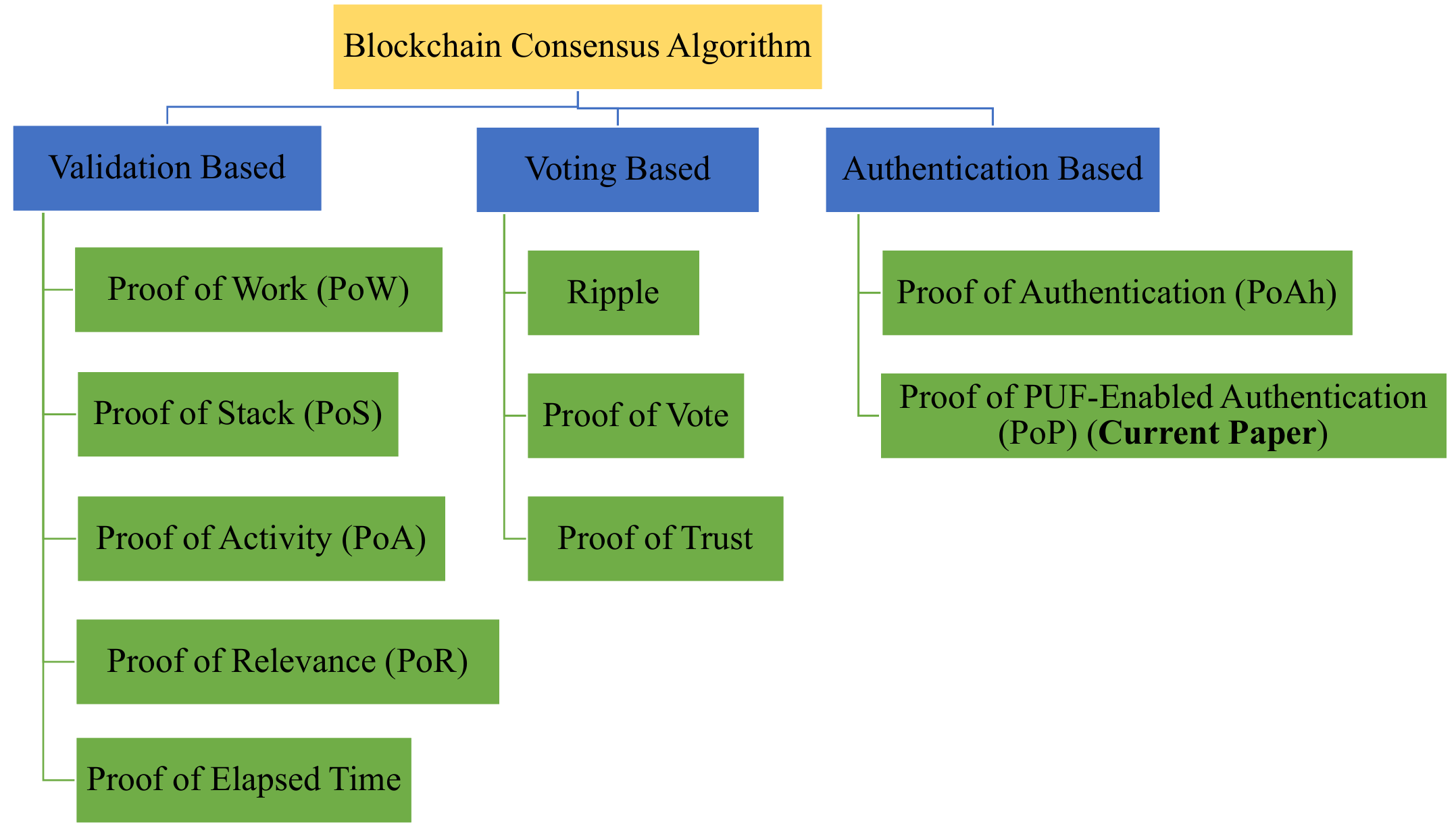}
	\caption{Various Consensus Algorithms used in the Blockchain.}
	\label{FIG:Consensus_Algorithms_Taxonomy}
\end{figure}

Consensus algorithms follow different processes to generate and validate blocks. There are several important consensus algorithms, as identified in Fig, \ref{FIG:Consensus_Algorithms_Taxonomy}. We classify them into 3 groups:
\begin{enumerate}
	\item 
	Validation based,
	\item 
	Voting based, and
	\item
	Authentication based.
\end{enumerate}
Bitcoin uses Proof-of-Work, Etherium uses Proof-of-Stake and Link uses Delegated Proof-of-Stake \cite{Nakamoto_CML_2009,King_decred_URL_PoS,Ozyilmaz_MCE_2019-Mar}. In the blockchain, blocks combine multiple transactions and blocks can be generated by few nodes in the network. All the blocks and transactions should be validated and/or accepted by identified network members as part of the consensus algorithm. After block validation, they are added into the chain and are cryptographically connected with each other. These consensus algorithm adaptabilities depend on several factors, such as resource utilization, transaction speed, and scalability. In some cases, the consensus algorithm is a computationally demanding cryptographic hash, as in the case of Bitcoin \cite{Nakamoto_CML_2009}, and the miners will be rewarded for completing validation of the block. Voting based consensus algorithms include Ripple \cite{Schwartz_White-Paper_2014_Ripple}, Proof-of-Vote \cite{Li_HPCC_2017_PoV}, and Proof-of-Trust \cite{Zou_TSC_2018-May_PoT}. A lightweight consensus algorithm named
Proof of Authentication (PoAh) specifically geared towards the IoT has been introduced \cite{Puthal_ICCE_2019,Puthal_IEEEP_2019}. It may be noted that the blockchain is in the top of the current technology and explored to be used in almost every application. However, there is no such consensus algorithm developed or tested for resource-constrained IoT applications. 
PoW and PoS are heavy-duty consensus algorithms and will not run in light-duty, battery-powered IoT devices. PoAh follows the cryptographic authentication mechanism for mining process.

%


\subsection{Distributed Ledger and the Blockchain}
\label{sec:Distributed_Ledger}

Blockchains are up made of \textit{blocks} which are interconnected in a unidirectional manner, with a very similar structure to the linked-list and called a distributed ledger \cite{Tschorsch_IEEECST_2016}. Every person who wants to utilize Bitcoin will have a virtual wallet which consist of a public and a private key. Each block within a chain is encrypted with the SHA256 and RIPEMD-160 hash functions and contains a cryptographic hash of the previous block. By design, this makes the blockchain very resistant to modification as tampering one block will require the modification of the entire chain. For Bitcoin, each block is 1MB in size \cite{Tschorsch_IEEECST_2016}. 

The main elements required within a blockchain are a \textit{transaction identifier} and all inputs and outputs of the transaction. In conventional currency, each unit of currency can only be used once by a person. This is the same in the blockchain, as input represents currency received by a user and output represents currency used. Each input transaction, representing ownership of the currency, can only have one output transaction. This output transaction is called the \textit{unspent transaction output} (UTOX) \cite{Puthal_IEEECEM_2018}. If there is more than one output transaction, the person is attempting to spend the same coin twice, and the illegal transaction is called a \textit{spent transaction output} (STOX) \cite{Tschorsch_IEEECST_2016}. This ledger is how the decentralized network can differentiate between one unique coin from another  \cite{Puthal_IEEEP_2019,Conti_IEEECST_2018,Gauhar_CS_2019}. Different consensus algorithms for blockchain are described as follows.

\subsection{Proof-of-Work  (PoW) Consensus Algorithm}

In order for transactions to be a secure and consistent, miners within a blockchain network must maintain and record on the same distributed ledger. However, with millions of decentralized nodes and no central server to maintain the network, the question is how this can be done? The blockchain solution to this problem is \textit{Proof of Work (PoW)}. 

Every Bitcoin transaction must be eventually added onto the blockchain. In order to add a new block onto a blockchain, the PoW scheme requires participating nodes to compete against each other to encrypt the new block into the existing chain, and the final encrypted solution must be below a certain target value. Having a lower target value increases the difficulty of the encryption as there are fewer solutions. If a miner successfully encrypts a block, they will transmit their solution to other nodes within the network. Other miners will then verify the new block and add it onto the chain. The miner who finds the correct solution is then rewarded with some Bitcoin \cite{Conti_IEEECST_2018}.

The encryption process is purposely designed to be resource consuming. This design is a security feature as it deters malicious nodes from adding blocks into the chain.  This process makes the verification of new blocks dependent on total computing power, as opposed to total number of nodes. The idea is that it will be more difficult to possess the majority of computing power on a Bitcoin network as compared to a majority of identities \cite{Conti_IEEECST_2018}. The probability of a miner finding the final solution for the newly added block and being awarded some Bitcoin is approximately the ratio of their computing power over the total computing power of the network.

For Bitcoin, the generation of each new block will consume an average of 10 minutes, where the difficultly of the encryption is adjusted to maintain this time \cite{Conti_IEEECST_2018}. \textit{PoW} also resolves the problem of double spending. If a person attempts to spend the same currency twice within a short period of time, miners will find solutions for both transactions. 
When this occurs, the blockchain resolves this by having miners accept the longest known fork, and this then becomes the main chain \cite{Puthal_IEEECEM_2018}. The other fork is orphaned and its transactions are nullified. This solution ensures that in the case of double spending, only one person will receive the currency. Outside of malicious attacks, forks within a chain can also occur if two mining nodes produce a valid solution at the same time \cite{Conti_IEEECST_2018}.

\subsection{Proof-of-Stake (PoS) Consensus Algorithm}

\textit{The Proof of Stake (PoS)} model is used in maintaining crypto-currencies such as Ethereum and Peercoin. Instead of basing the probability of creating a new block on total computational power, as in the PoW scheme, PoS utilizes a concept called \textit{coin age}. Coin age is the amount of time a particular currency has been held by a user. If a user has held that coin for 30 days, the coinage of the coin would be 30, and when a coin is spent, the coin age would be reset to 0 \cite{Tschorsch_IEEECST_2016}.
Similar to the PoW, new blocks are required to be below a target value, however in PoS the target value required of a new block is individually determined. The target value is determined by the coin age. A higher coin age would translate to a higher target value and an easier puzzle for the miner. Every new solution is also accompanied by a \textit{coinstake} block that contains a coin that is sent back to the miner itself. This transaction is used to determine the \textit{coin age}, and thus target value, to be used for the solution. The coin age of the coin contained in the coinstake block is then reset to 0 if the solution is accepted by the blockchain network \cite{Tschorsch_IEEECST_2016}. This system ensures the miner cannot cheat the system, as they are selected based on their provided information. However, a miner always has great chance to win the bitcoin due to it being always selected from a limited number of dedicated miners.  

\subsection{Proof-of-Activity (PoA) Consensus Algorithm}
\label{sec:PoA_Algorihtm}

In the PoS scheme, even if the participant is off-line, a participant's coins continue to grow in coinage \cite{Tschorsch_IEEECST_2016}. This leads to a problem where miners would switch online to generate blocks every few weeks, then go off-line. This significantly reduces the numbers of concurrently online nodes, and coin security is reduced significantly. Another problem with this scheme leading to hoarding of the currency is that it discourages the exchange of the coin as each exchange would reset the coinage \cite{Tschorsch_IEEECST_2016}.

\textit{Proof of Activity (PoA)} is developed to condense the impact of these issues. The major concept of PoW is to reward the fraction of active peers, where peers having high processing capacity increase the probability of `winning' the reward. After solving the PoW puzzle, the peer broadcasts the verified blocks to the networks. Individual peers getting the block are able to derive a number from the solution. If a peer owns the coin corresponding to the ``ticket'', they will sign their signature onto the block and are able to receive a part of the reward. In the case of off-line peers, they are unable to sign the block and this results is not achieving the reward.
With this award system, the PoA scheme incentivizes nodes in the network to persist on the blockchain network, whereby improving the security of the network. This scheme is also less punishing and encourages spending of the currency, as by keeping currency for longer time does not advance the chance of getting a remuneration \cite{Tschorsch_IEEECST_2016}.

\subsection{Proof-of-Authentication (PoAh) Consensus Algorithm}

Proof of Authentication (PoAh) adopts a traditional PoW consensus methodology for lightweight block verification \cite{Puthal_IEEECEM_2018}. The miners in the network could be trusted nodes in the network to authenticate the blocks, followed by authenticated nodes which push the network peers to integrate block into the chain. This algorithm includes two steps:
\begin{enumerate}
	\item 
authenticate the transactions in a block and respective sources, and
\item
upon successful authentication, respective trusted node's trust value is increased by one unit.
\end{enumerate}
The distributed ledger is updated in the network peers based on blocks from miners. Individual transactions in a block are verified by the trusted nodes in the networks. A trusted node turns to normal node in the network after a certain number of false identifications. In each false transaction, the respective trusted node's trust value reduced by a unit.

\section{Challenges of Blockchain Technology}
\label{SEC:Blockchain_Challenges}


\begin{figure}[htbp]
	\centering
	\includegraphics[width=0.95\textwidth]{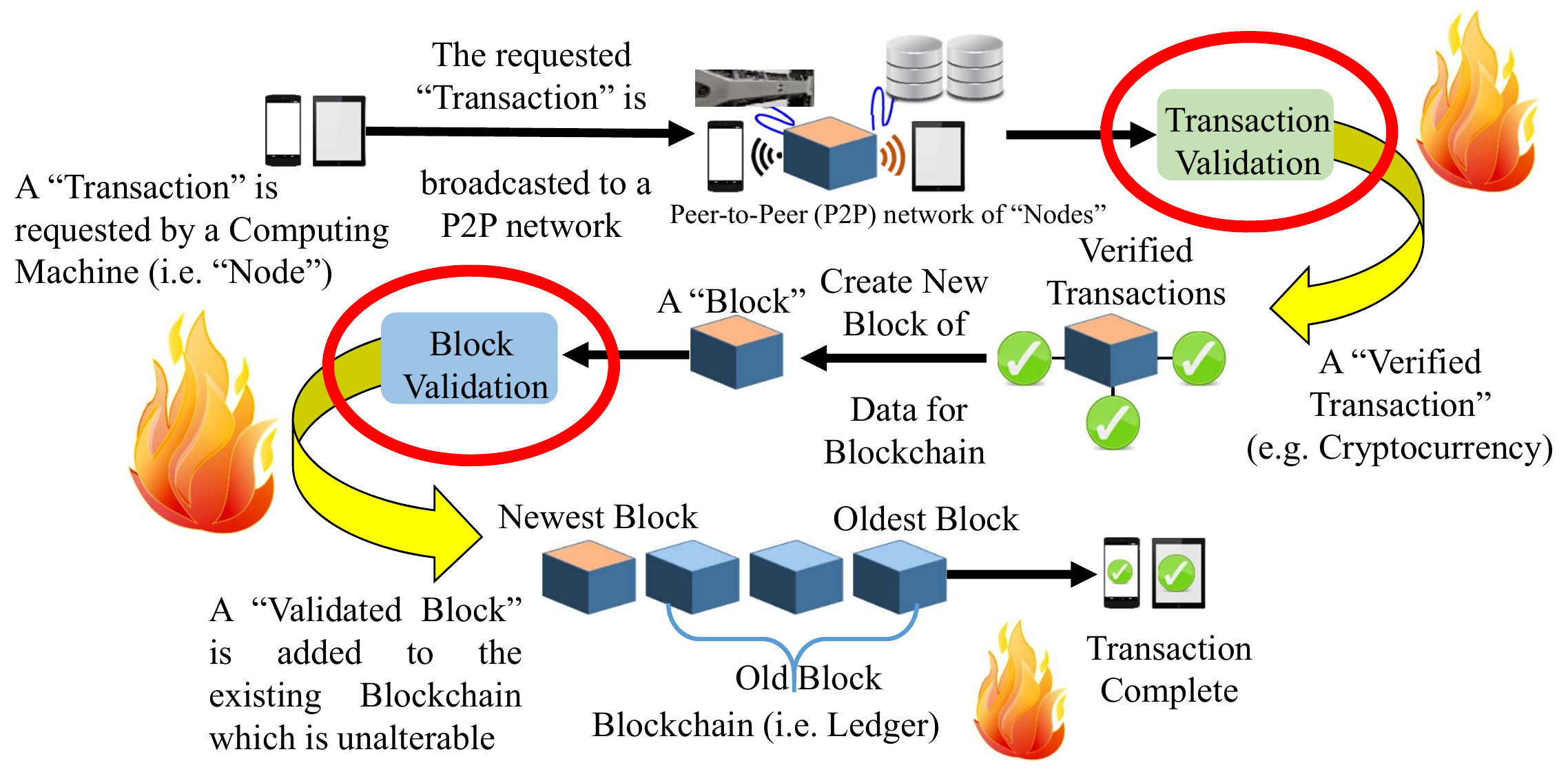}
\caption{Processing Flow of Blockchain showing Energy Overheads \cite{Puthal_IEEECEM_2018,Puthal_IEEEP_2019,Puthal_ICCE_2019,Mohanty_ICCE_2019_Keynote,Mohanty_ZINC_2018_Keynote}.}
	\label{FIG:Blockchain_Processing_Flow-with_Energy-Overhead}
\end{figure}

\begin{figure}[htbp]
	\centering
	\includegraphics[width=0.45\textwidth]{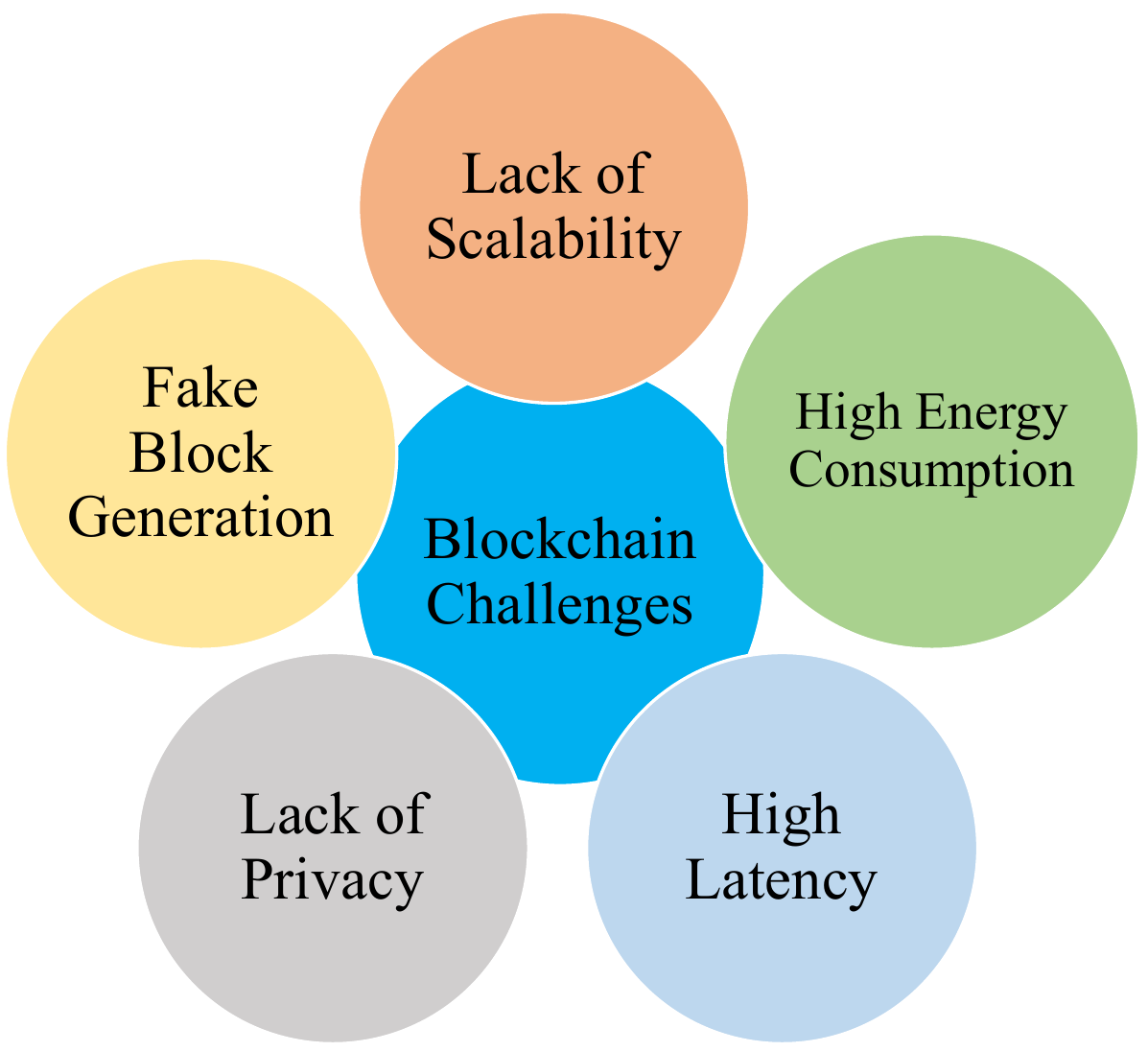}
	\caption{Challenges or Issues of Blockchain Technology.}
	\label{FIG:Blockchain_Challenges}
\end{figure}

The blockchain uses cryptographic hashes to maintain security and consistency, as shown in Fig. \ref{FIG:Blockchain_Processing_Flow-with_Energy-Overhead} \cite{Puthal_IEEECEM_2018}. Once a block has been added to the blockchain, it can neither be edited or deleted. If any data modification is performed on the blocks added to the blockchain, the entire ledger will be broken indicating a discrepancy. There are many use cases for blockchain but there are also many challenges \cite{Zheng_IEEEBDC_2017,Puthal_IEEECEM_2018}. Various challenges of blockchain technology which either prevent its application or make it computationally and energy demanding are shown in Fig. \ref{FIG:Blockchain_Challenges} \cite{Puthal_IEEECEM_2018,Puthal_IEEEP_2019,Puthal_ICCE_2019}. 

The transactions in the blockchain are combined to form the blocks. Once the nodes in the network form blocks with the transactions, the process of mining starts which validates the blocks and the transactions in it. The mining process requires high computational resources and dedicated hardware, which consume a high amount of power. The dedicated hardware requirements also make the scalability difficult \cite{Puthal_IEEEP_2019,Lu_IEEEAccess_2019,Hencry_IEEESP_2018}. With an increased amount of data and nodes in the P2P network, the latency also increases. As the number of transactions increases, the time taken to validate the transactions also increases and this gives rise to more issues. It also becomes more difficult to conceal the identity of the user in the case of a distributed ledger \cite{Hencry_IEEESP_2018}. Observing the transactions, a user can be backtracked to their real-world identity. There are also the issues of attacks on the blockchain where fake blocks can be generated.

\section{The State-of-the-Art in Blockchain based IoT Security}
\label{SEC:Related_Research}

This literature review covers the several implementations of the blockchain technique used to secure crypto-currencies. It must be noted however that not all of the processes reviewed can be feasibly implemented into IoT systems. IoT systems will often consist of many low powered nodes, as opposed to numerous powerful servers used to maintain crypto-currency. Listed are some differences between IoT systems and crypto-currencies that must be accounted for during the implementation of blockchain.


Edge computing and the blockchain are being explored by researchers for applications in IoT environments. The blockchain as a distributed control system and hierarchical system is presented in \cite{Stanciu_CSCS_2017}. A blockchain which is a Hyperledger Fabric is selected for implementation in the paper. The blocks in the blockchain are presented as smart contracts on a supervisor level and the executive level is where the edge node integration is realized. The entire process control is based on the micro-services architecture \cite{Stanciu_CSCS_2017}. Another smart-contract based on Etherium is presented in \cite{Wright_IEEECPSCom_2018}. The paper presents the SmartEdge blockchain architecture, where the nodes are able to utilize the computational resources of the edge devices. Smart contract based blockchains are also integrated into medical devices. Armor Chain, a permissioned, semiprivate decentralized network was proposed for smart home environments and smart healthcare devices in \cite{Paliokas_IEEECEM_2019}. All sensitive data are stored on the medical device itself but not in the blockchain which prevents issues of privacy. Smart contracts are used in Armor Chain for the cyberthreats relating to the identity of the manufacturer, consumer and authorized personnel for making updates or fixing the device. A deep examination of the blockchain as a security solution for various cyberthreats has been examined in \cite{Kolokotronis_IEEECEM_2019}. The paper begins with an explanation on what is blockchain and how a blockchain works but focuses more on how trusted can be a blockchain, how device integrity can be protected using blockchain technology, and the current market situation.

Blockchain is also used as a solution for many security and privacy issues. Although there are some blockchain consensus algorithms which are not capable of completely concealing the identity of the user, there are also blockchain algorithms which make it near to  impossible to trace back to the user. Blockchain uses cryptographic hash functions for maintaining consistency and security in the ledgers \cite{Lu_IEEEAccess_2019}. Various IoT devices are prone to attacks and data safeguarding has become a major issue. The blockchain can be a potential solution with an integration into IoT architectures. Various use cases of the blockchain in safeguarding the security aspects of an IoT environment are presented in \cite{Lu_IEEEAccess_2019}. The paper also presents various malicious attacks that are possible on the blockchain and the potential defenses against them that are already implemented in the consensus algorithms. 

Blockchain research is also extended to various applications and use cases including healthcare. Blockchain integration into the smart home and smart healthcare can help with various aspects, such as device authentication, secure communication and secure storage of collected data. A blockchain framework for healthcare using consumer medical devices was presented in \cite{Paliokas_IEEECEM_2019}. Smart contracts were used for secure communication using the blockchain architecture. Access control is another use case of the blockchain. BlendCAC, a decentralized capability based access control for achieving a high level security in an IoT environment is presented in \cite{Xu_iThings_2018}. Using the proposed blockchain, a large scale IoT environment and devices in the network can be controlled, where the IoT devices themselves are the masters of the resources and no central entity will be monitoring them, which offers a decentralized lightweight and scalable IoT access control solution.

One of the issues with transferring the blockchain technique into IoT systems is the process of adding new blocks onto the chain. The three schemes from the literature review, namely Proof of Work, Proof of Stake and Proof of Activity focus on incentivizing peers to continue to maintain the network \cite{Xin_IEEEIDS_2017}. Especially emphasized in Proof of Work are the large amounts of resources. This is also used to prevent attacks from malicious peers. Within the framework of a private blockchain network, these measures do not seem necessary and, if implemented, may even reduce network security as some nodes may be barred from participating due to real-world hardware constraints.

The proposed design also uses a Physical Unclonable Function (PUF) module which generates unique keys for the devices in the network. The keys generated by a PUF module cannot be replicated using any other module of similar design. All the keys are unique to the PUF module that generated them. Hence, a unique identifier in the blockchain can add an extra layer of security and make it resistant to the attacks. 

There are various architectures of PUF modules that were proposed which can be integrated into different applications \cite{Yanambaka_IEEE-TSM_2018-May,Zhou_EL_2019}. A PUF uses the manufacturing variations that are introduced during the fabrication of an IC to generate the unique keys. A PUF can be designed using different architectures, such as Ring Oscillator \cite{yanambaka_IEEETSM_2017,Rahman_IEEETETC_2016,Zhou_EL_2019}, SRAM \cite{Shifman_IEEESSCL_2018,Liu_IEEEAccess_2018}, and Arbiter PUF \cite{Zalivaka_IEEETIFS_2019,Sahoo_IEEETC_2018}. 

The PUF depends on the variability of the devices at the core of the IC. This means a slight change in the geometry of the nanoelectronic devices can affect the performance and even change the outputs. Aging effects may cause a change in the geometry of the devices and might cause a change in the output. Hence, as a solution, various aging resistant designs of PUF were also designed and proposed. One such aging resistant ring oscillator based PUF was presented in \cite{Rahman_IEEETETC_2016}.

Table \ref{Table:Comparison} shows a comparison of the PUFchain with other blockchain technologies and the consensus algorithms.

\begin{table*}[htbp]
	\caption{Comparison of PUFchain with other Blockchain Technologies.}
	\label{Table:Comparison}
	\centering
	\begin{tabular}{|p{3.5cm}|c|c|p{3.6cm}|p{3.4cm}|}
		\hline
		Consensus Algorithm & Year & Bblockchain Type & Mining & Prone to Attacks \\
		\hline
		\hline
Proof-of-Work (PoW) \cite{Nakamoto_CML_2009} & 2008 & Permission-Less & Based on Computation Power & Bribe attack, Sybil attack, 51\% attack\\

\hline		
Proof-of-Stake (PoS) \cite{King_Online_2019} & 2012 & Permission Less & Validation & DoS, Sybil attacks\\

\hline
		
Ripple \cite{Ripple} & 2014 & Permissioned & Vote-Based Mining & DoS attack, Sybil Attack \\

\hline
Proof-of-Vote \cite{Li_HPCC_2017}& 2017 & Consortium & Vote-Based Mining &  -- \\
		\hline
	
		Proof-of-Trust \cite{Zou_IEEETSC_2018} & 2018 & Permission Based & Probability and Vote-Based Mining& DDoS Attack \\
		\hline

Proof-of-Authentication (PoAh) \cite{Puthal_IEEEP_2019,Puthal_ICCE_2019} & 2019 & Permission Based & Authentication Based & -- \\
		
		\hline
		Proof of PUF-Enabled Authentication (PoP) (\textbf{This Article}) & 2019 & Permission Based & Authentication Based &--\\
		\hline
	\end{tabular}
\end{table*}

\section{The Proposed PUF Integrated Novel Blockchain (PUFchain) Architecture}
\label{SEC:PUF_Chain}

\subsection{Overview of the Proposed PUFchain}

\begin{figure*}[htbp]
	\centering
\subfloat[PUFchain Working Model]{\label{fig:PUFChain_Depiction_Model}\includegraphics[width=0.80\textwidth]{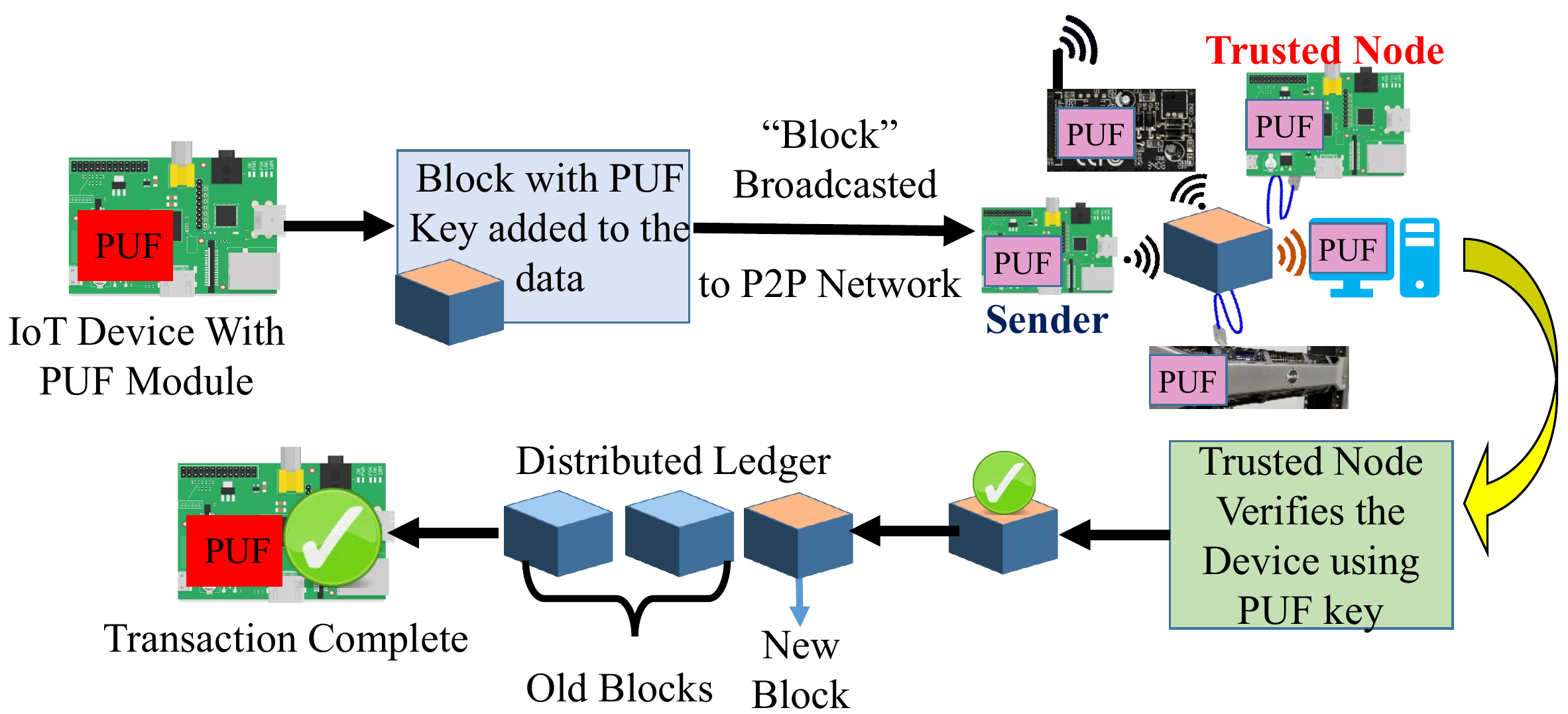}}\\
\hspace{0.2cm}
\subfloat[PUFchain System Model]{\label{fig:PUFChain_System}\includegraphics[width=0.80\textwidth]{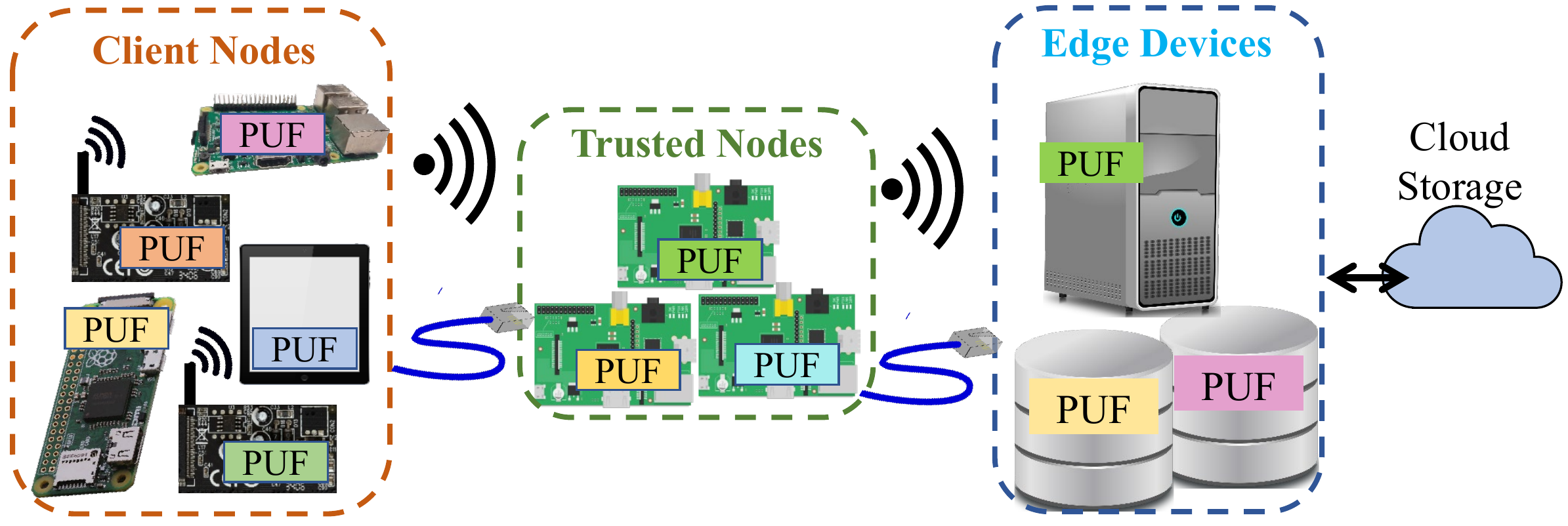}}
\caption{Depiction of the proposed PUF integrated blockchain - PUFchain.}
	\label{FIG:PUFChain_Depiction}
\end{figure*}

This section presents the proposed architecture for PUFchain that we envision as a PUF integrated secure blockchain which can solve energy requirements, scalability, and latency of the existing blockchain.  
A node in the PUFchain network consists of the IoT device and a hardware module which has the PUF generating the key, and also the hashing capabilities. Fig. \ref{FIG:PUFChain_Depiction} shows the architecture of PUFchain blockchain. The IoT device is responsible for gathering the environmental data. The PUF and hashing module is added to the IoT device. This reduces the computing burden on the IoT device itself. The specifications of the IoT device do not affect the security aspects or the performance of the PUFchain. The PUF and hashing module consists of a cryptographic processor and the PUF module. The cryptographic processor gets the data from the IoT device and the PUF module, which supplies the unique key. The cryptographic hashing function is performed in the cryptographic processor. Once the hash is computed, the IoT device transmits the data to the network. 

The current section describes a scenario where an IoT environment was designed and the proposed PUFChain consensus algorithm was implemented. A series of transactions were initiated, authenticated and added to the blockchain which validates the proposed algorithm. A two phase protocol was proposed where during the introduction of a new device into the network, the device has to go through an enrollment phase and when a transaction is broadcast to the network, it undergoes the authentication phase.

\subsection{Why and How PUF Integration in Blockchain?}

\begin{figure}[htbp]
	\centering
	\includegraphics[width=0.45\textwidth]{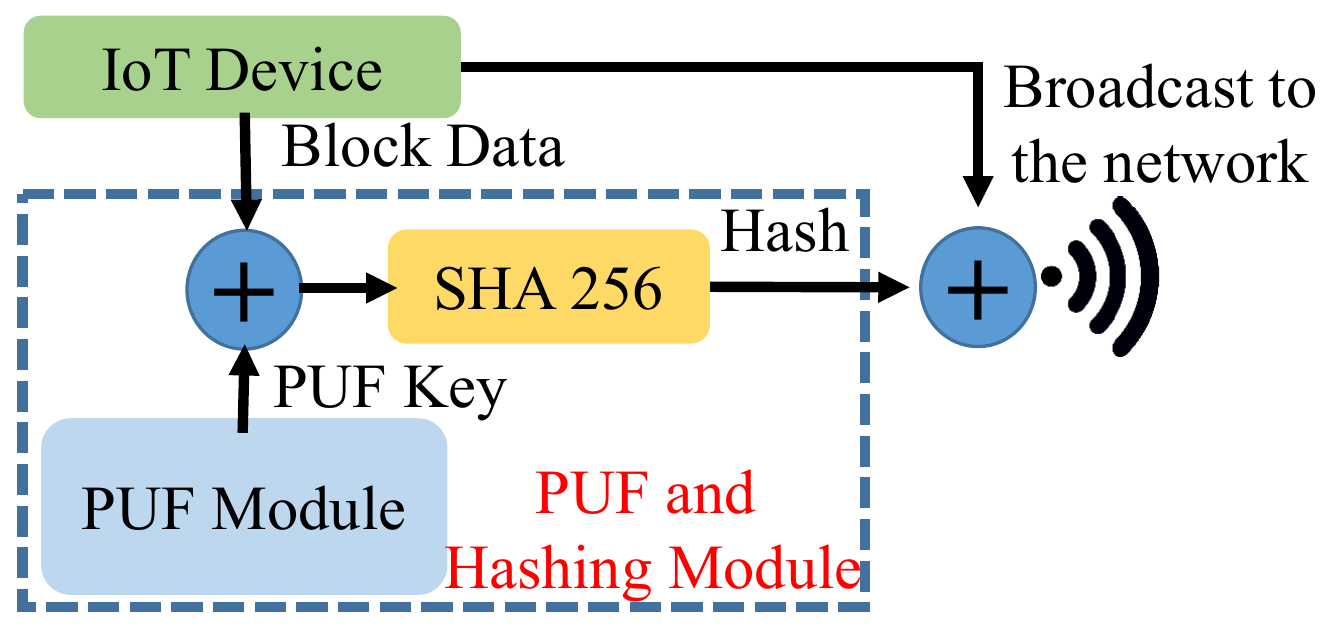}
	\caption{Node in a PUF-Chain Environment.}
	\label{FIG:PUF_Chain_Node}
\end{figure}

Fig. \ref{FIG:PUF_Chain_Node} shows the structure of a node in the network. The main intention of PUFchain is integration of the blockchain consensus algorithm in an IoT network which has low power and low form factor presence. 
The PUFchain network consists of trusted nodes and the client nodes. Client nodes will collect the environmental data and broadcast it to the network. The trusted nodes are responsible for mining and validating the devices that collect the data. Fig. \ref{FIG:PUF_Chain_Node} shows the structure of both the client and trusted node. This ensures that no extra hardware is required for being a trusted node or a client node. If needed, a client node can be granted permissions and escalated to a trusted node position.

A PUF is responsible for generating a unique identity for the IoT device. A PUF can generate a series of unique keys that can only be generated from that PUF module. The output of the PUF key depends on the input and as the challenge input changes, the response from the PUF module also be different. The set of keys generated from a PUF module cannot be cloned or generated form any other module. Hence the name, Physical Unclonable Function. The PUF keys are not stored in the memory of the IoT devices. When the keys are required, they will be generated from the module and the hashing is performed using the module. This makes the IoT device more secure as, depending on the PUF architecture, more than one key can be generated by changing the input. The output of the PUF key can be changed on-the-fly and various security threats can be avoided.

\subsection{Block Structure in PUFchain}

\begin{figure*}[htbp]
	\centering
\subfloat[Proof of Work (PoW) Block Structure]{\label{fig:PoW_Block}\includegraphics[width=0.49\textwidth]{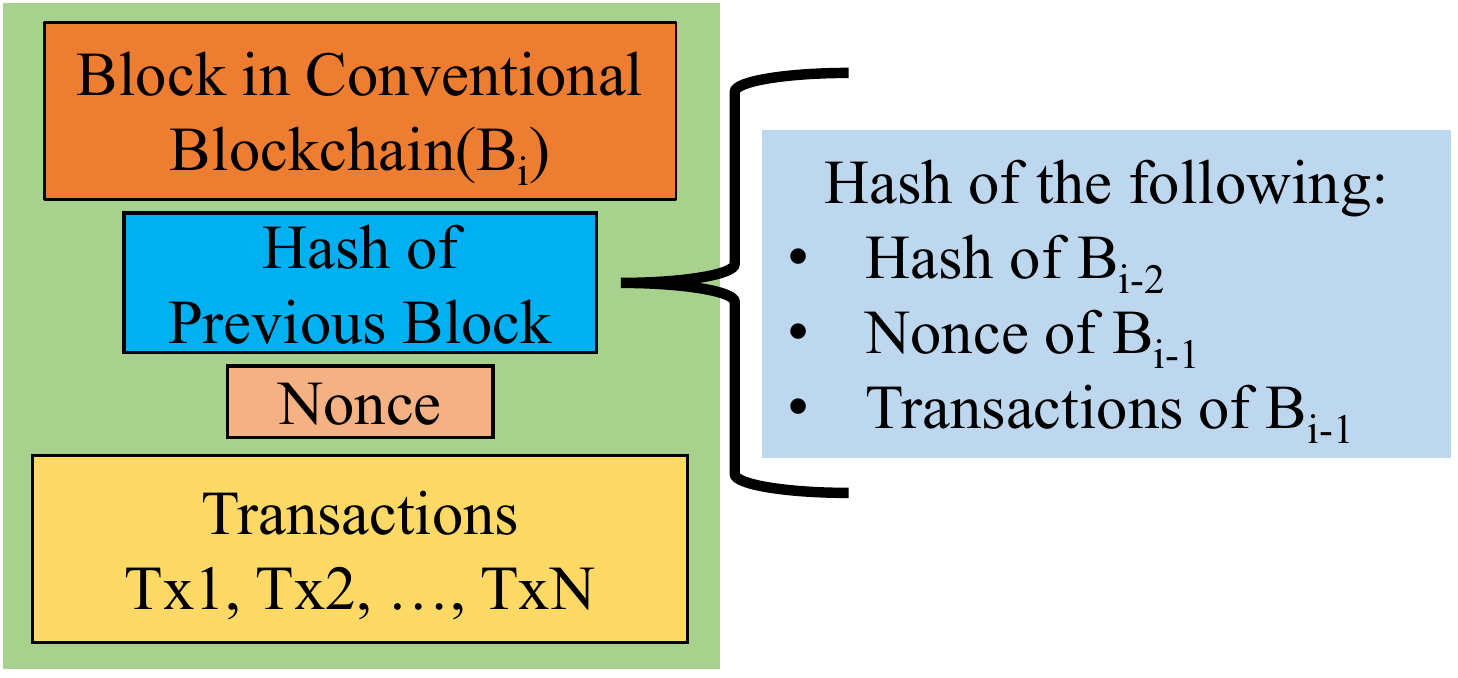}}
	\hspace{0.2cm}
\subfloat[Proof of Authentication (PoAh) Block Structure]{\label{fig:PoAh_Block}\includegraphics[width=0.49\textwidth]{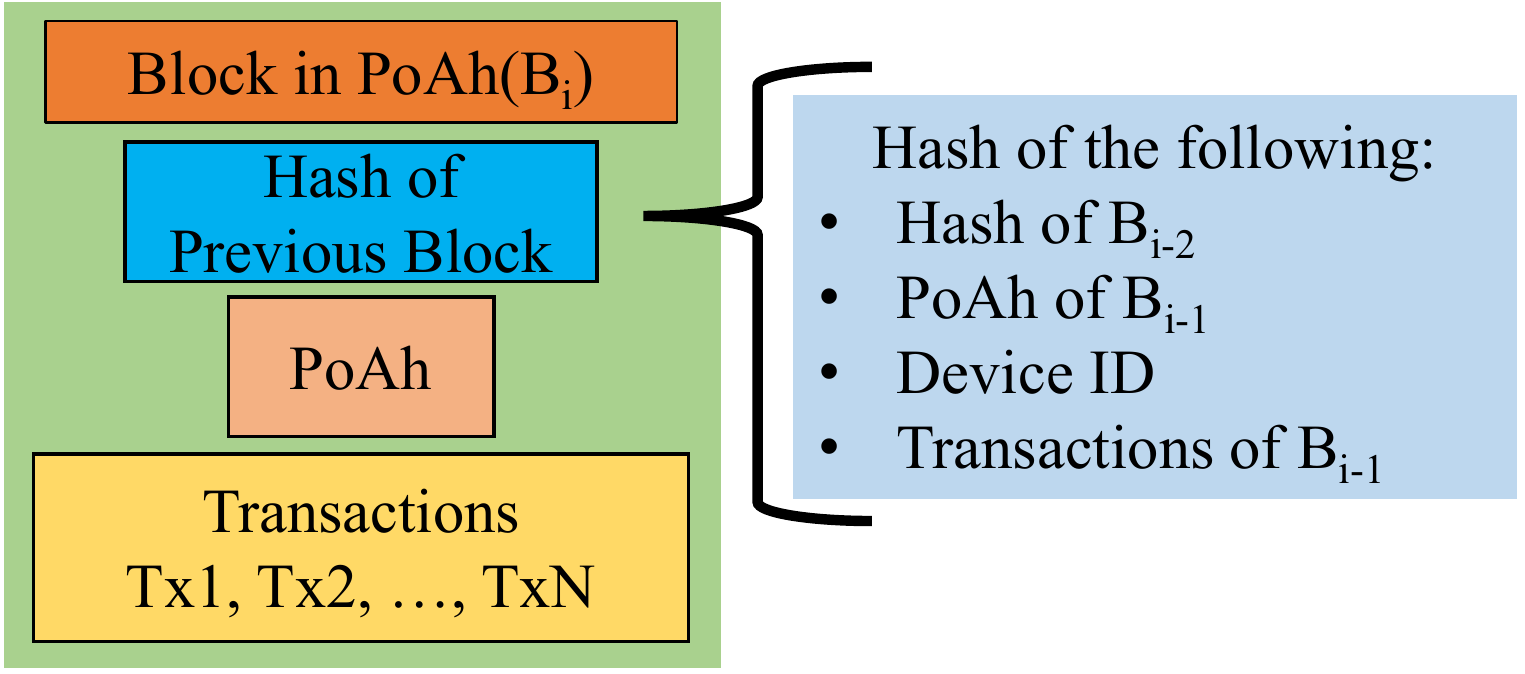}}
	\hspace{0.2cm}
\subfloat[Proof PUF-Enabled Authentication (PoP) Block Structure]{\label{fig:PUFChain_Block}\includegraphics[width=0.50\textwidth]{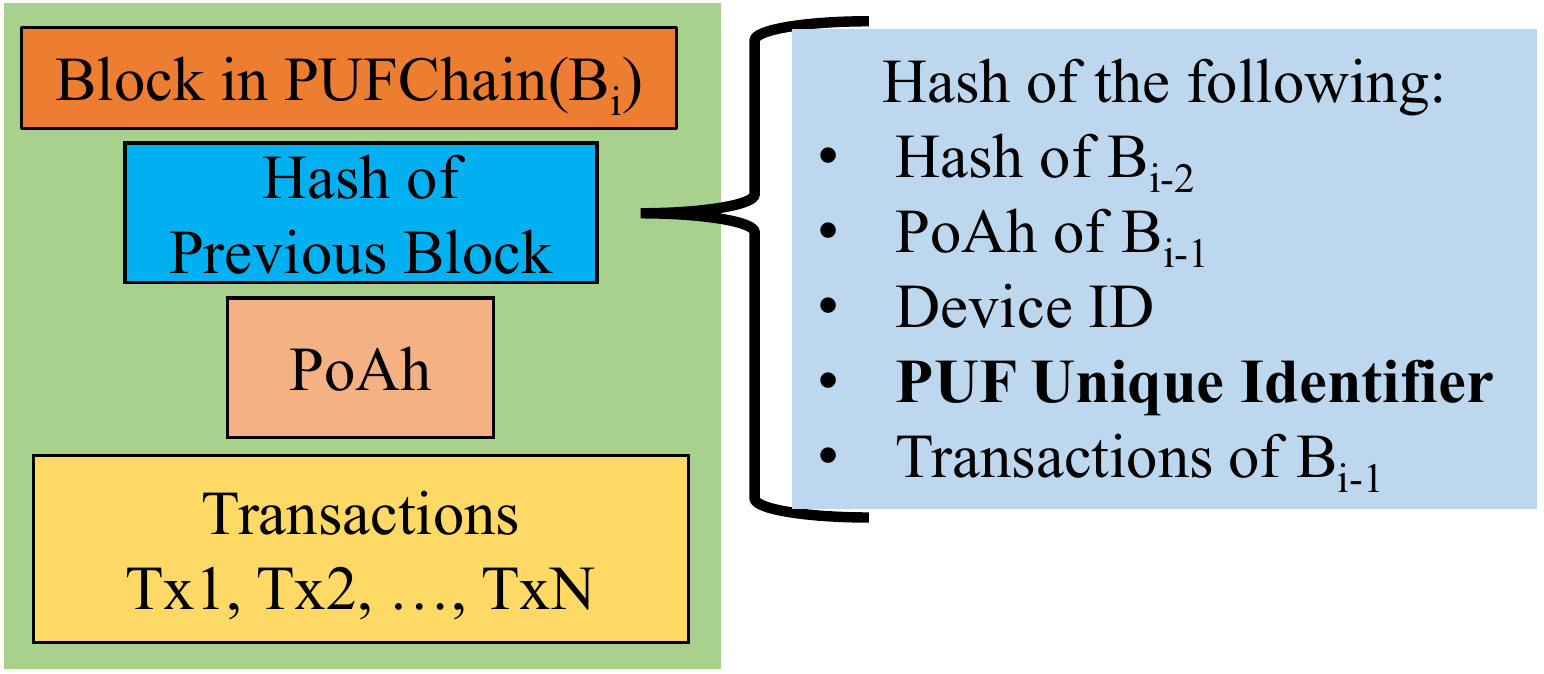}}
	
	\caption{Block Structure in PoW, PoAh, and PoP.}
	\label{FIG:Block_Structures}
\end{figure*}

Fig. \ref{FIG:Block_Structures} shows the comparison of the block structures of PoW, PoAh and PoP. As shown the figure, the PoW block has the hash of the previous block and the ``nonce''. This nonce generation requires high processing power which makes it unsuitable for various IoT architectures. Hence the nonce generation is avoided in the PoAh and also the proposed PoP. The hash computing is similar in all the cases. Hash of the PoW is the cryptographic hash of the previous block, nonce of the previous block and the transactions. But in the case of the PoAh, the hash is the cryptographic hash of the previous block, the PoAh of previous block and the ``device-ID''. This is helpful in the case of device authentication, the PoAh.

\section{The Proposed Novel Proof of PUF-Enabled Authentication (PoP)}
\label{SEC:PUF_PoP}

\begin{figure*}[htbp]
	\centering
	\subfloat[Proof of Work (PoW) Consensus Algorithm]{\label{fig:PoW}\includegraphics[width=0.70\textwidth]{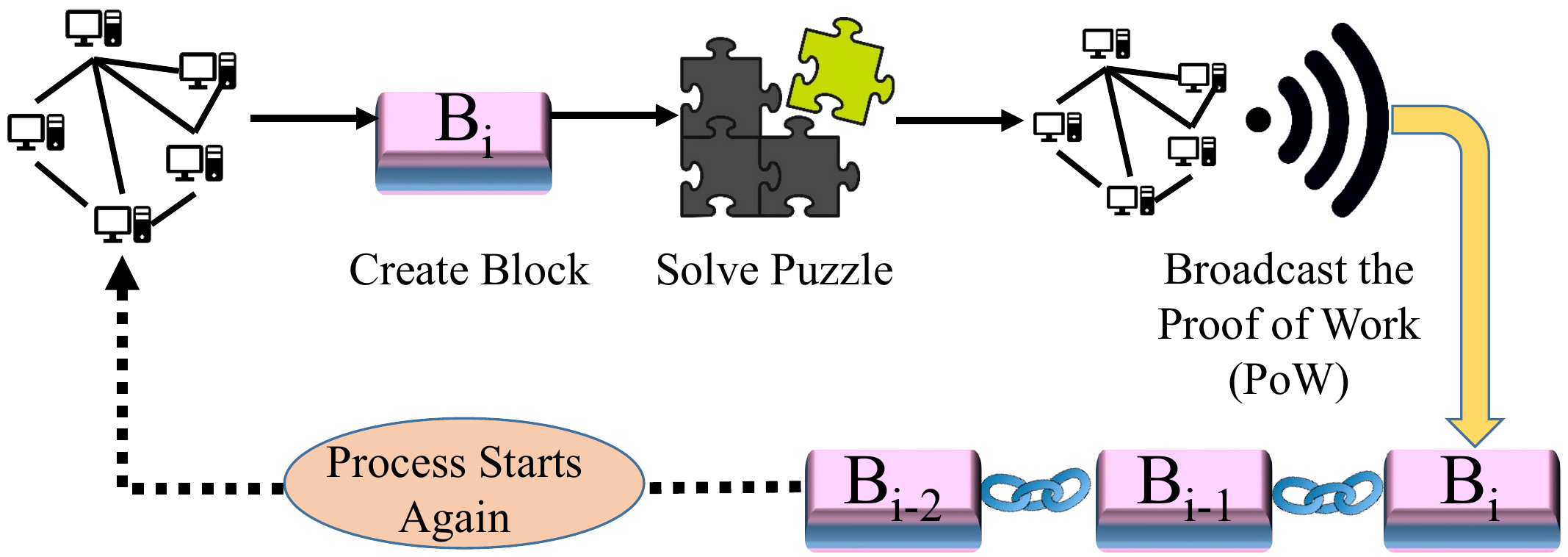}}
	\hspace{0.2cm}
	\subfloat[Proof of Stake (PoS) Consensus Algorithm]{\label{fig:PoS}\includegraphics[width=0.90\textwidth]{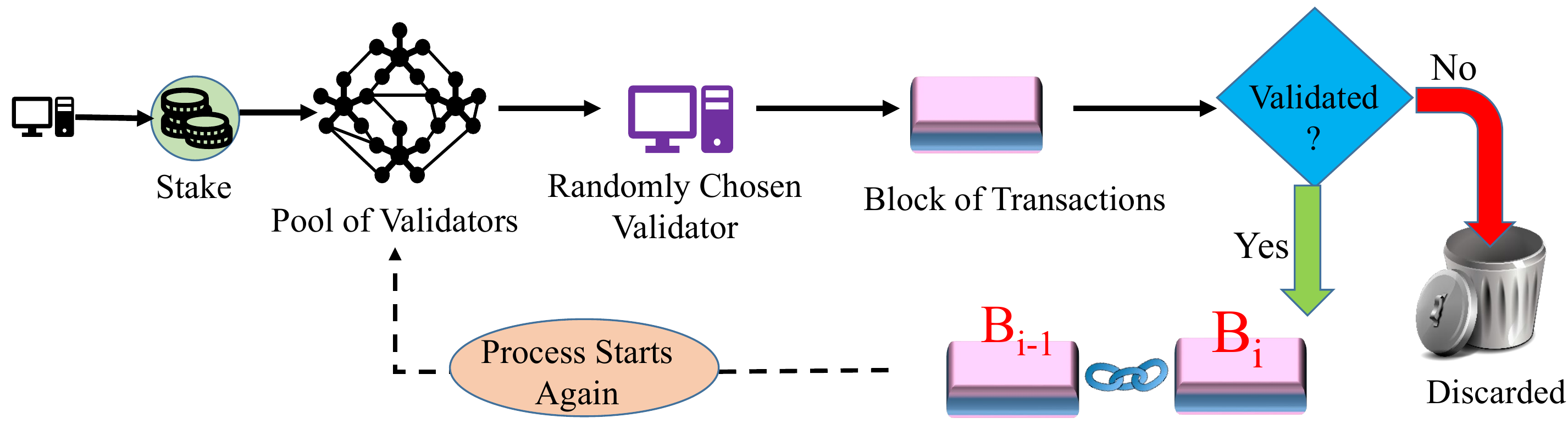}}
	\hspace{0.2cm}	
	\subfloat[Proof of Authentication (PoAh) Consensus Algorithm]{\label{fig:PoAh}\includegraphics[width=0.80\textwidth]{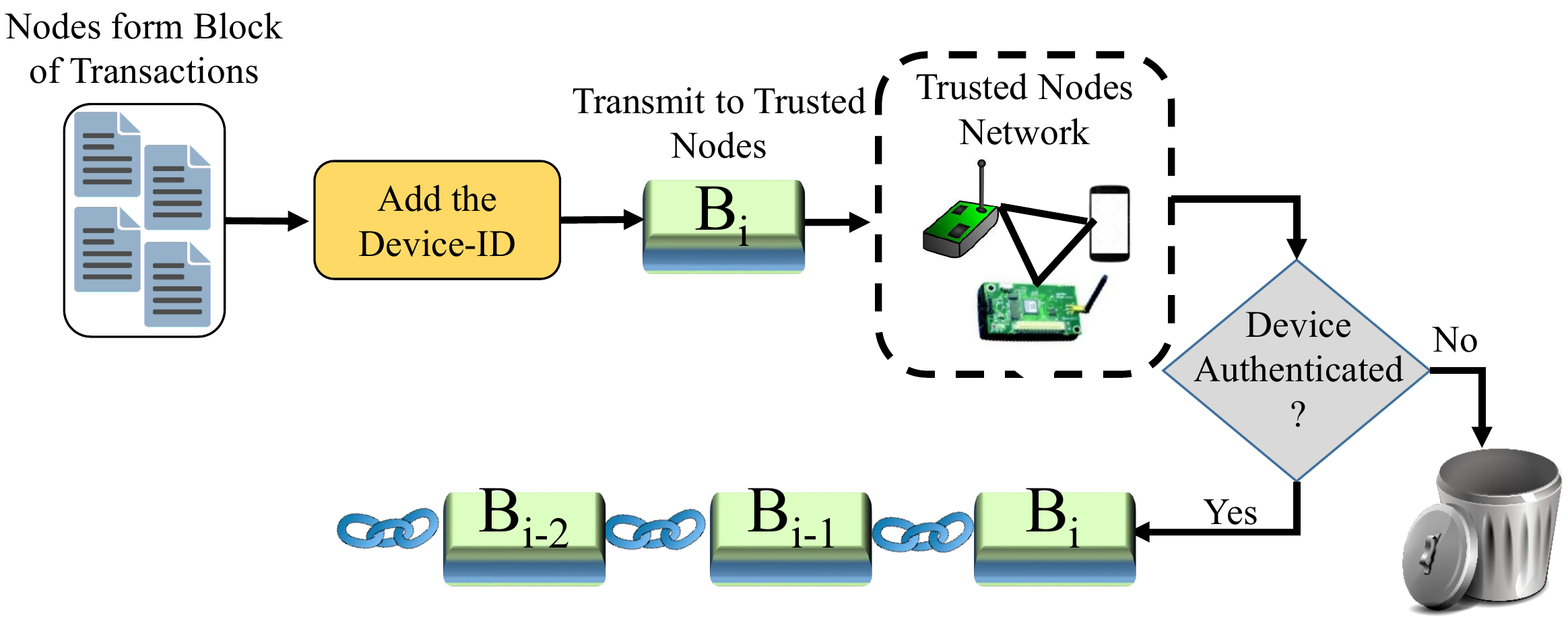}}
	\hspace{0.2cm}
	\subfloat[Proof PUF-Enabled Authentication (PoP) Consensus Algorithm]{\label{fig:PUFChain}\includegraphics[width=0.70\textwidth]{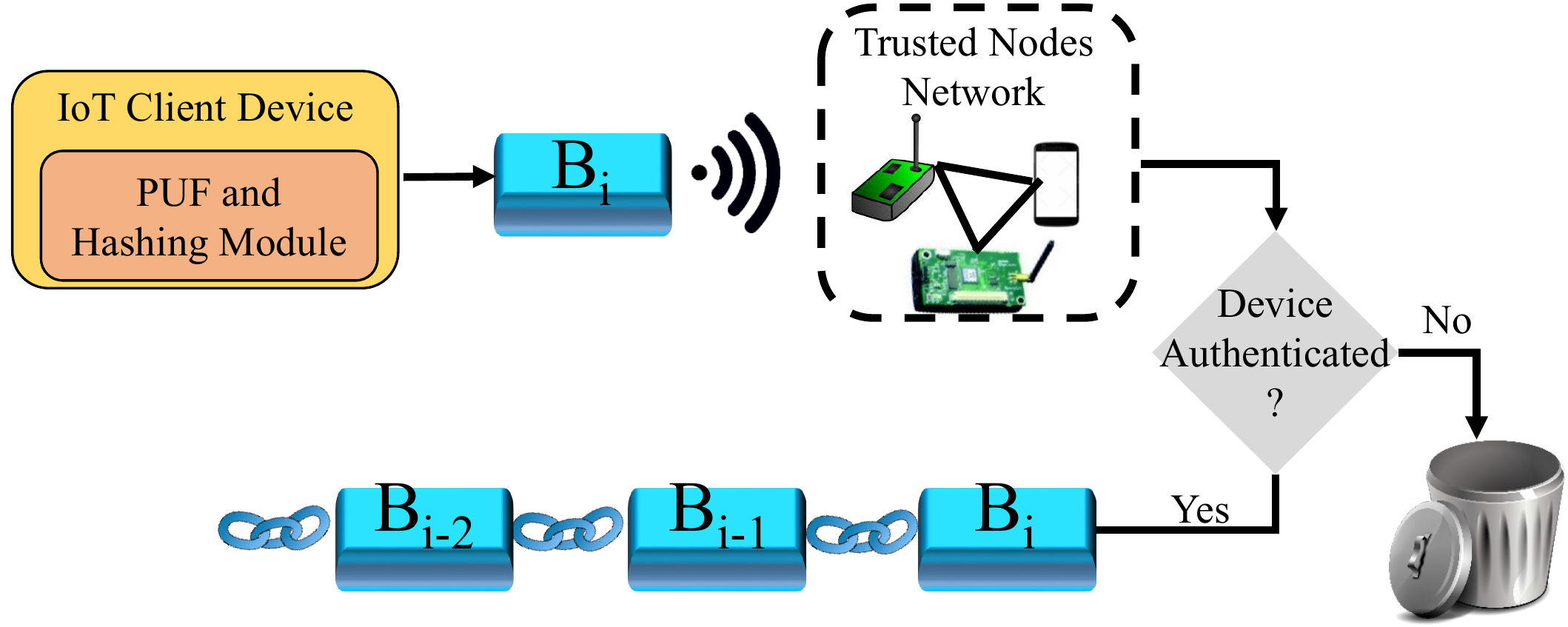}}
	\caption{PoW, PoS, PoAh, and PoP Consensus Algorithms.}
	\label{FIG:Four_Consensus_Algorithms}
\end{figure*}

\begin{figure*}[t]
	\centering
\subfloat[Device Enrollment Steps]{\label{fig:PoP_Enrollment_Flow}\includegraphics[width=0.60\textwidth]{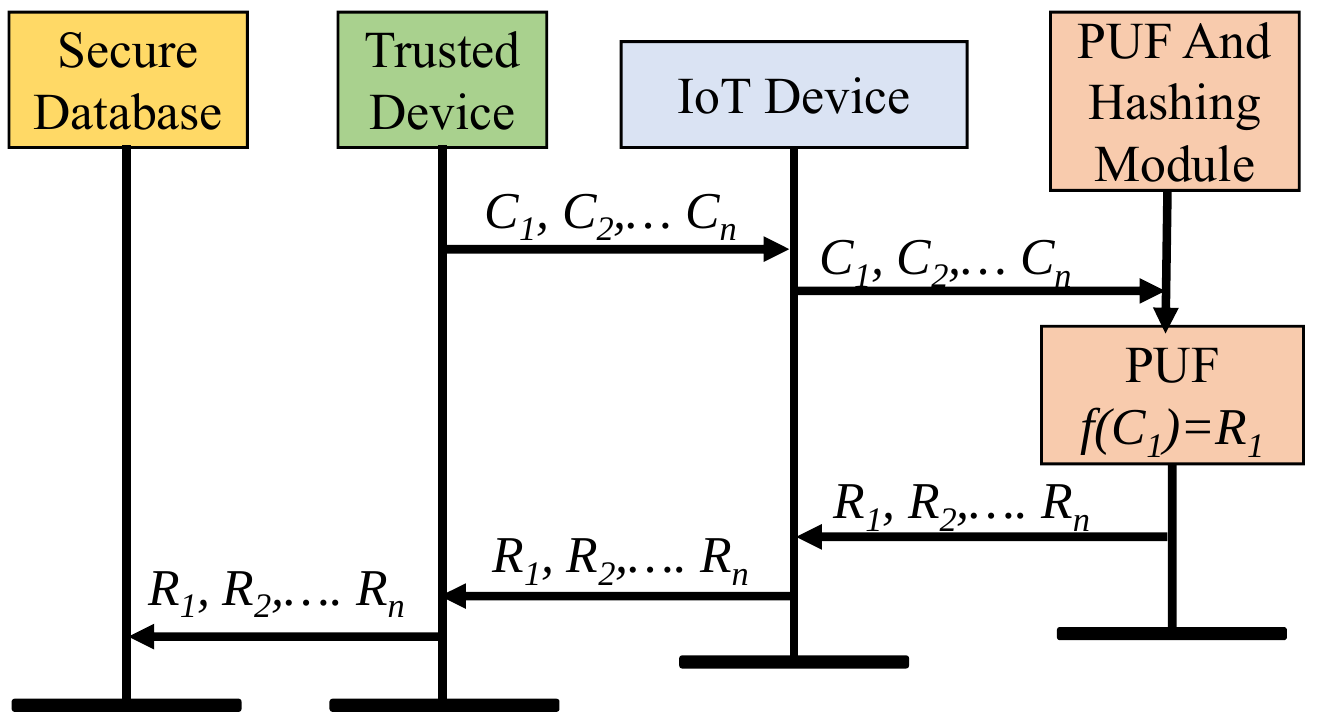}}
	\hspace{0.2cm}	
\subfloat[Transactions Initiation Steps]{\label{fig:PoP_Transaction_Flow}\includegraphics[width=0.45\textwidth]{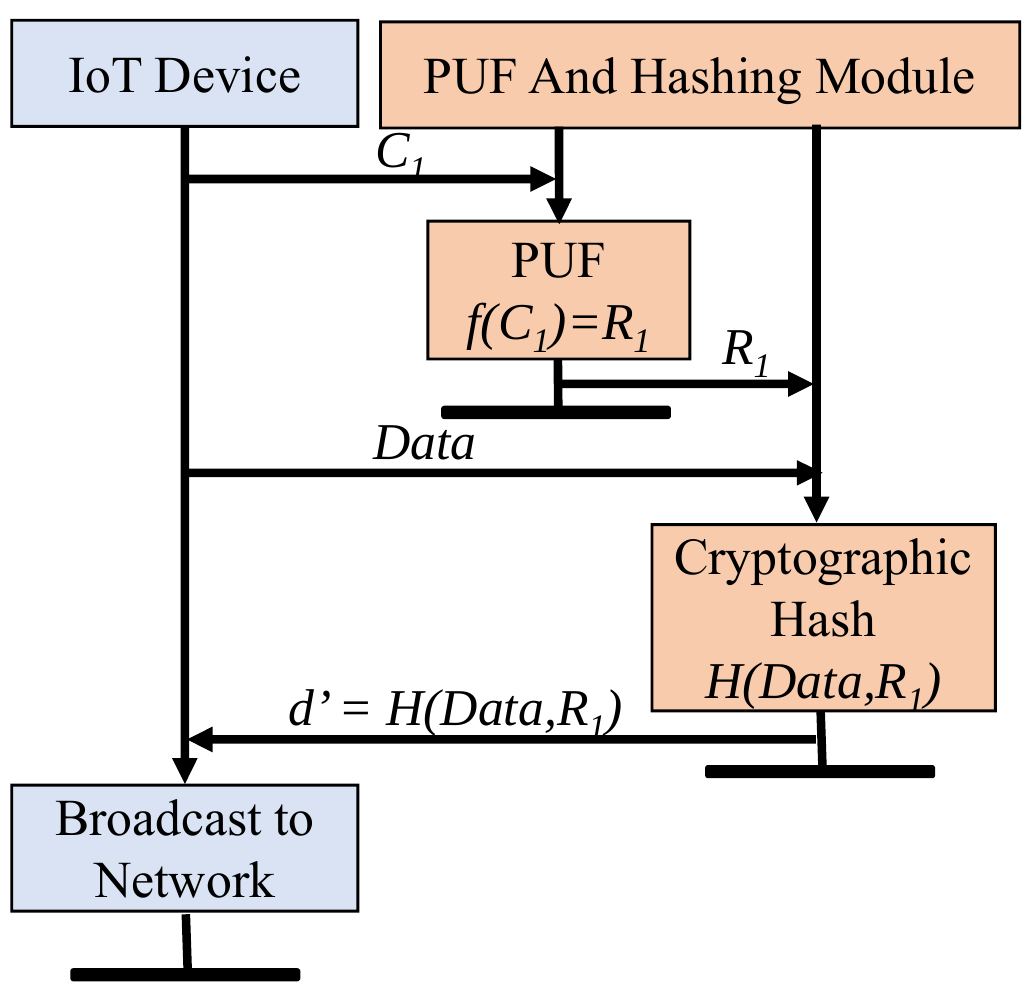}}
	\hspace{0.2cm}
\subfloat[Device Authentication Steps]{\label{fig:PoP_Authentication_Flow}\includegraphics[width=0.45\textwidth]{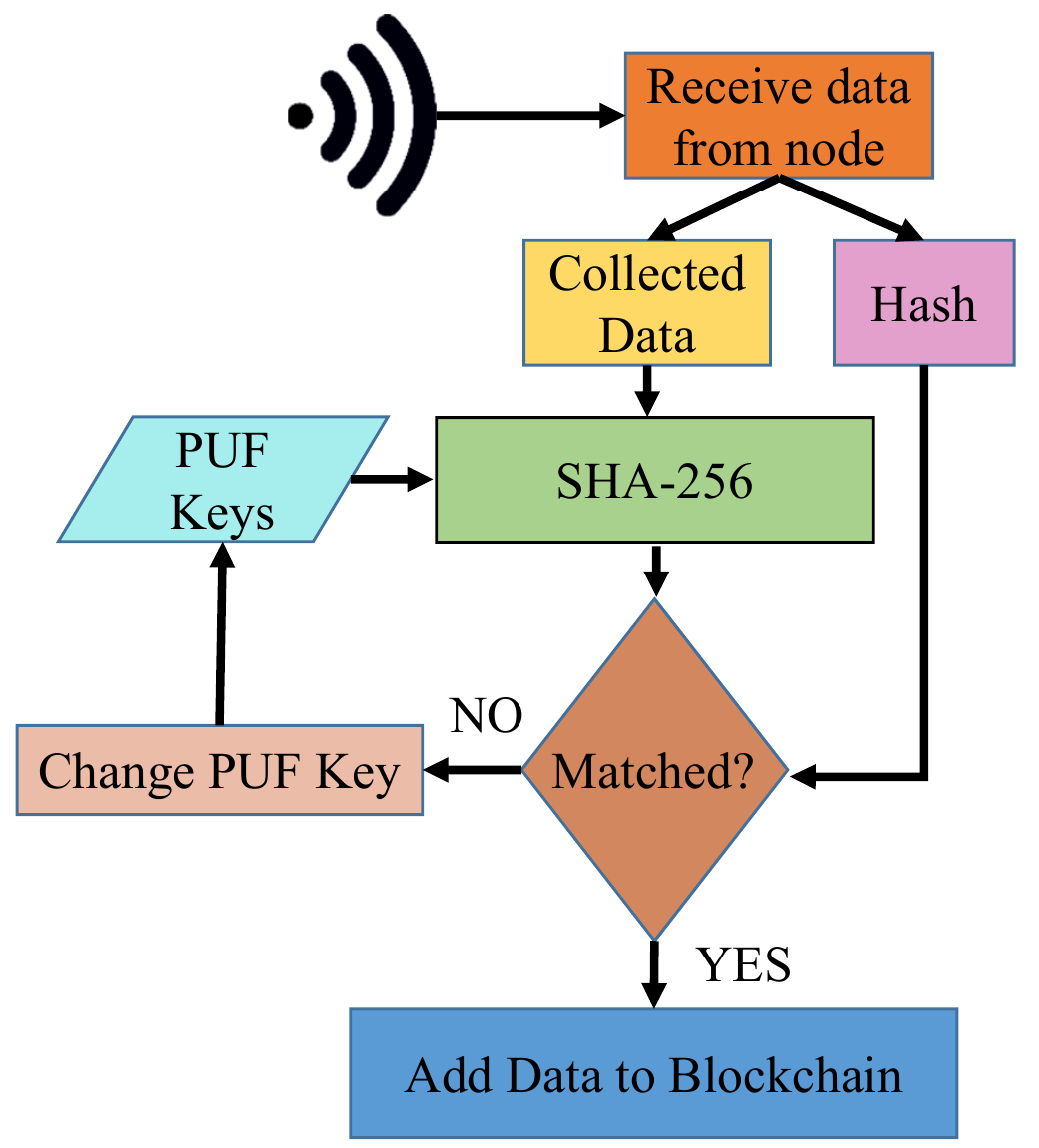}}
	\caption{Enrollment and Authentication Steps in Proposed PoP Consensus Algorithm.}
	\label{FIG:PoP_Consensus_Algorithm_Flow}
\end{figure*}

This section presents the novel Proof of PUF-Enabled Authentication (PoP), a PUF based blockchain consensus algorithm. The consensus algorithm is proposed to be implemented on an IoT network which has energy and processing power constraints.  In the case of PoP, the PUF module is responsible for generating the device's unique identification. The hash that is in the block is the cryptographic hash of all the previously considered data and also the PUF key that is uniquely generated at the PUF module of the device. The same key cannot be generated at any other device. The properties and working of PUF module are discussed in detail in Section \ref{SEC:PUF}. The proposed PoP Consensus Algorithm is presented in a comparative perspective with PoW, PoS, and PoAh in Fig. \ref{FIG:Four_Consensus_Algorithms}. The different steps involved in various phases of the proposed PoP consensus algorithm are presented in Fig. \ref{FIG:PoP_Consensus_Algorithm_Flow}.

\subsection{Device Enrollment Phase}
\label{sec:EnrollmentPhase}

The network of devices using the PUF chain consensus algorithm is close to any new devices that have not gone through the enrollment phase. This can be achieved with the help of the PUF module. Every device in the network contains a PUF module which can generate unique identification numbers that are later used in the authentication phase. Every trusted node in the network is connected to a secure database. During the enrollment phase, a series of challenge inputs will be selected for the PUF module. The inputs to the PUF module cannot be selected at random. They undergo a selection procedure before considering them for the PUF chain consensus algorithm. The challenges should satisfy a set of requirements to be considered as inputs to PUF. These set of requirements are discussed in Section \ref{SEC:PUF}. The selected challenge inputs are given to the PUF module in the IoT device and the responses are collected. These challenge response pairs are stored in the secure database which can be accessed by the trusted nodes of the network. Algorithm \ref{ALG:Enrollment_Phase} presents the algorithm of the enrollment process.

\begin{algorithm}[t]
	
	\caption{Device Enrollment Phase}
	\label{ALG:Enrollment_Phase}
	
	\SetKwData{Left}{left}\SetKwData{This}{this}\SetKwData{Up}{up}
	\SetKwFunction{Union}{Union}\SetKwFunction{FindCompress}{FindCompress}
	\SetKwInOut{Input}{Input}\SetKwInOut{Output}{Output}
		\Input{Challenges to PUF in IoT Device (PUF-N).} 
		\Output{Responses from PUF module $R_x$}
		\BlankLine
		Select the challenges for PUF $C_x$;\\
		\For{each challenge $C_x$}{
			\If{PUF requirements met}{$C_i \leftarrow C_x$; \\
			\tcp{Select the challenge($C_x$) as input for PUF ($C_i$)}}{
			\Else{Drop the challenge;}
			}
		}
		\For{each challenge input $C_i$}{
			PUF-N $\leftarrow$ $C_i$;\\
			Secure Database $\leftarrow$ $R_x$, $C_i$;\\
			\tcp{$R_x$ = PUF-N($C_i$)}
			\tcp{Store the challenge response pair in secure database.}
		}
	
\end{algorithm}

\subsection{Initiating a Transaction}

Once the device is successfully enrolled into the network, it becomes eligible to initiate successful transactions which a trusted node authenticates. The IoT device collects the data from the sensors and generates a transaction. The transaction data in this case contains the collected sensor data and the MAC address of the device. The MAC address serves as an identification for the rest of the nodes in the network and to the trusted node. Once the transaction is generated, it is sent to the hardware accelerator. It contains the PUF and is capable of computing the cryptographic hash. 

A challenge input ($C_i$) is selected for the PUF. This challenge input is one of the challenges that are stored in the secure database accessible to the trusted nodes. The challenge is given to the PUF module and the response is collected. The hardware accelerator then computes the hash of the data concatenated with the response generated. This hash is sent back to the IoT device which is broadcast to the network of devices. Algorithm \ref{ALG:Tranasaction_Initiation} presents the process of broadcasting a block to the network.

\begin{algorithm}[htbp]
	
	\caption{Transaction Initiation Phase}
	\label{ALG:Tranasaction_Initiation}
	
	\SetKwData{Left}{left}\SetKwData{This}{this}\SetKwData{Up}{up}
	\SetKwFunction{Union}{Union}\SetKwFunction{FindCompress}{FindCompress}
	\SetKwInOut{Input}{Input}\SetKwInOut{Output}{Output}
	\Input{Data from sensors and challenge $C_i$.}
	\Output{Signed Blocks}
	\BlankLine
	Collect the data from the sensors \emph{$D_n$};\\
	Select a challenge input $C_i$;
	\tcp{$C_i$ is one of the challenges that is present in secure database.}
	PUF-N $\leftarrow$ $C_i$;\\
	$R_x$ = PUF-N($C_i$)
	\tcp{Response $R_x$ is generated by PUF.}
	Generate the hash;\\	
	$H_n$ = H($D_n$,$R_x$)
	\tcp{SHA-256 cryptographic hasing algorithm is computed.}
	Broadcast ($D_n, H_n$)
	
\end{algorithm}

\subsection{Device Authentication Phase}

The block that is broadcast to the network has to be authorized by the trusted node before it gets added to the blockchain. Once the block is received by the trusted node, the data ($D_n$) and the hash ($H_n$) are retrieved. These two parameters help in authorizing the device. Algorithm \ref{ALG:Authentication_Phase} presents the process of authentication. $D_n$ contains the environmental data collected by the node and its MAC address. This is the ID that is used for identifying the device in the blockchain. The MAC address is used in addition to the unique identification generated by the PUF. 

With the help of the MAC address, the trusted node gets the PUF keys from the secure database. The PUF key and the data are sent to the hardware accelerator for performing the cryptographic hash. The hashing function used is similar across all the devices in the network. The data and the PUF key are hashed and the resulting hash is compared to the $H_n$ received from the node. If both the hashes match, the device is authenticated. If they do not match, the process continues for all the PUF keys that are stored during the enrollment phase of the device. 

When one of the hashes generated matches $H_n$, the device gets authenticated. If the device is not authenticated, the block will be dropped and not broadcast. Once the device is authenticated, the block gets added to the blockchain of the trusted node and then broadcast to the network. When other devices receive the block from the trusted node, they add it to their local blockchain.

\begin{algorithm}[htbp]
	
	\caption{Device Authentication Phase}
	\label{ALG:Authentication_Phase}
	
	\SetKwData{Left}{left}\SetKwData{This}{this}\SetKwData{Up}{up}
	\SetKwFunction{Union}{Union}\SetKwFunction{FindCompress}{FindCompress}
	\SetKwInOut{Input}{Input}\SetKwInOut{Output}{Output}
	\Input{Blocks from the node.}
	\Output{Authenticated Blocks}
	\BlankLine
	Receive the block data from the network;\\
	Block $\rightarrow$ \emph{$D_n$}, \emph{$H_n$};\\
	\tcp{$D_n$ is the data and $H_n$ is the hash retrieved from the block.}
	Retrieve Device-ID from \emph{$D_n$};\\
	\tcp{Device-ID is the MAC address of the IoT device which is different from PUF unique identifier.}
	Request set of $R_x$ from \emph{Secure Database} for corresponding device;\\
	\For{each response $R_x$}{
		$X'$ = H(D,$R_x$)\\
		\If{X' == $H_n$}{
			Device Authenticated;\\
			Add the block to the blockchain;\\
			Broadcast to the network;
		}
		\Else{Drop the block}
	}
	
\end{algorithm}


\section{Physical Unclonable Functions (PUF) as Hardware Security Primitive}
\label{SEC:PUF}

PUF is the representation of the manufacturing variations from the devices on the Integrated Circuit (IC) \cite{Yanambaka_TCE_2019_PMsec,Joshi_MPOT_2017-Nov}. The fabrication process of an IC introduces some variations into the design. The variations are unpredictable, unavoidable, uncontrollable and naturally occurring. Because of these nanoelectronic manufacturing variations, no two devices on a wafer look the same. The geometric variations of the devices will  cause the differences in performance of the device or the application itself. These nanoelectronic manufacturing variations are taken as an advantage in designing the PUF. The input to a PUF is called ``challenge'' and the output from the PUF is called the ``response''. For evaluation of PUFs, three figures of merit (FoM) are considered: Uniqueness, Reliability and Randomness \cite{Yanambaka_TCE_2019_PMsec,Joshi_MPOT_2017-Nov}.

\subsection{PUF Working Principle}

\begin{figure}[htbp]
	\centering
\subfloat[Process Variations Make PUF out of Intergrated Circuits]{\label{FIG:PUF_Working_Principle_PV}\includegraphics[width=0.995\textwidth]{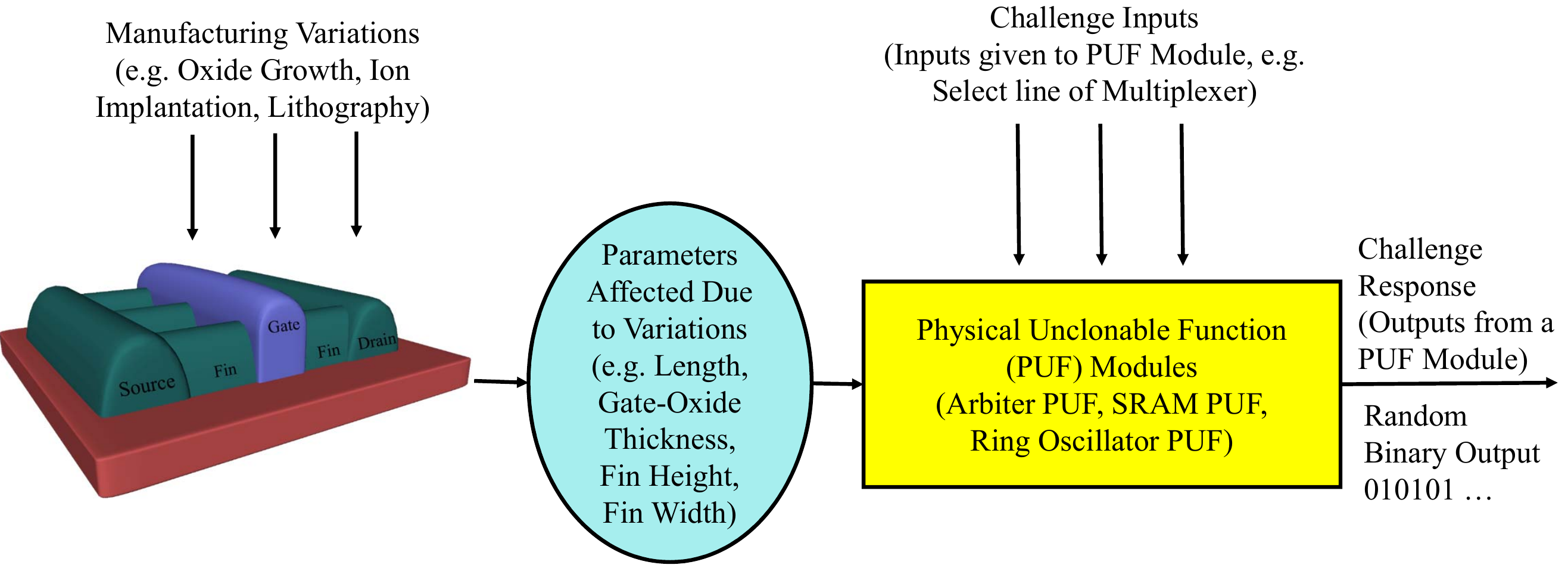}}\\
\hspace{0.2cm}
\\
\subfloat[Same Input Generates Different Outputs for Different PUFs]{\label{FIG:PUF_Working_Principle_CRP-Same-Input}\includegraphics[width=0.75\textwidth]{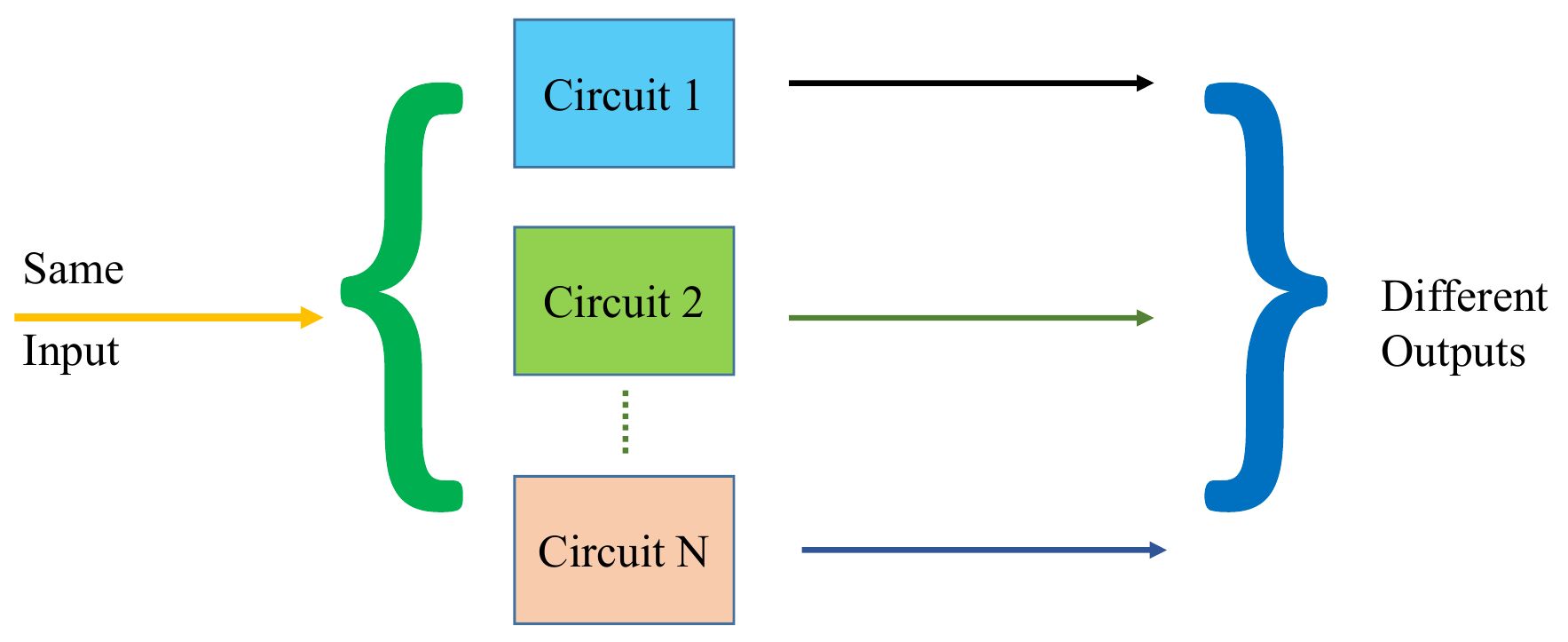}}
\hspace{0.2cm}
\\
\subfloat[Same Input Generates Different Outputs for Different PUFs]{\label{FIG:PUF_Working_Principle_CRP-Different-Inputs}\includegraphics[width=0.75\textwidth]{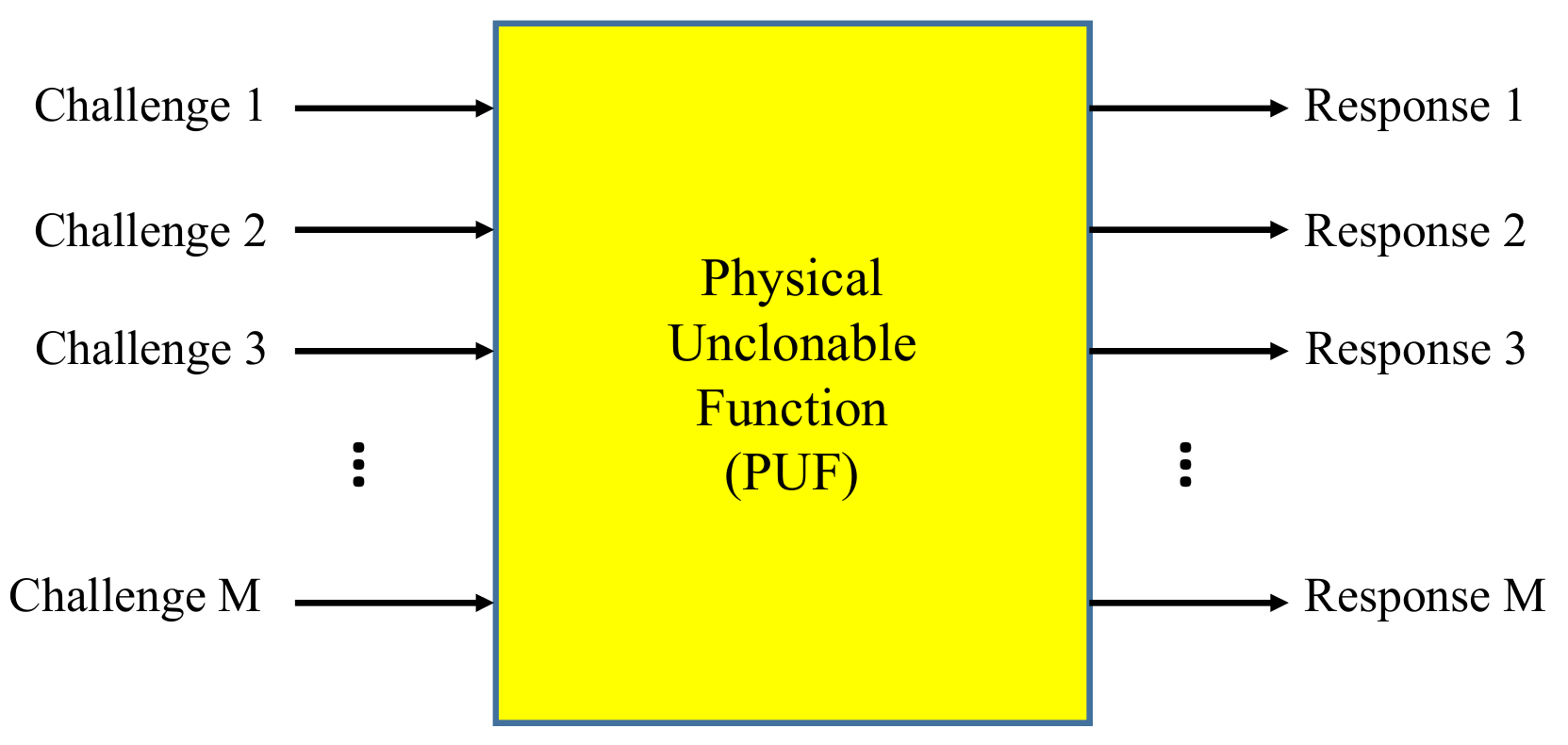}}
\caption{Working Principle of the Physical Unclonable Function \cite{Yanambaka_IEEE-TSM_2018-May,Yanambaka_ALOG_2017-Dec}.}
\label{FIG:PUF_Working_Principle}
\end{figure}

A PUF is a circuit in which process variations while manufacturing provide
unique characteristics to the input to output mapping, as presented in Fig. \ref{FIG:PUF_Working_Principle_PV} \cite{Yanambaka_IEEE-TSM_2018-May,Yanambaka_ALOG_2017-Dec,Wallrabenstein_FiCloud_2016_Elliptic-PUF}. The manufacturing variations are inevitable in any fabrication process and these physical variations in the integrated circuit will be unique for each  device. Therefore, the behavior of each device will be unique for the same given input. A single challenge and its corresponding response are called the ``\textit{challenge response pair}'' (CRP).
When various PUF modules are fabricated on
the same IC or the same wafer, for an identical input, no two PUF modules will give the same
output because of the naturally occurring variations, as illustrated in Fig. \ref{FIG:PUF_Working_Principle_CRP-Same-Input}. Moreover, as illustrated in Fig. \ref{FIG:PUF_Working_Principle_CRP-Different-Inputs}, if
the same challenge input is given to the same PUF module,
there will be no change in the output. If the challenge input is
changed for the same module, there will be a different output.
Overall, a key generated with a specific challenge input on a
module cannot be replicated on any other PUF module.

The variations introduced during the fabrication process are not the same across the wafer. Hence the outputs of PUF are also unpredictable. When a response is obtained from the PUF for a challenge input, it is naturally random, unlike the pseudorandomness present in algorithmically generated random numbers. A PUF is capable of multiple \textit{CRPs} but not all of them can be used for cryptographic applications. A PUF takes in a ``\textit{challenge}'' as an input and gives out the ``response'' as presented in Fig. \ref{FIG:PUF_Working_Principle} \cite{Yanambaka_IEEE-TSM_2018-May,Yanambaka_ALOG_2017-Dec}. In most cases, the CRP will be digital similar to the design used for PUFchain, ``\textit{Hybrid Oscillator Arbiter PUF}''. However, other PUF topologies can be considered as well. The CRPs of the PUF play a major role in deciding the classification of PUF and its level of security.

\subsection{Figures-of-Merit (FoMs) of PUF}

\begin{figure}[htbp]
	\centering
	\includegraphics[width=0.59\textwidth]{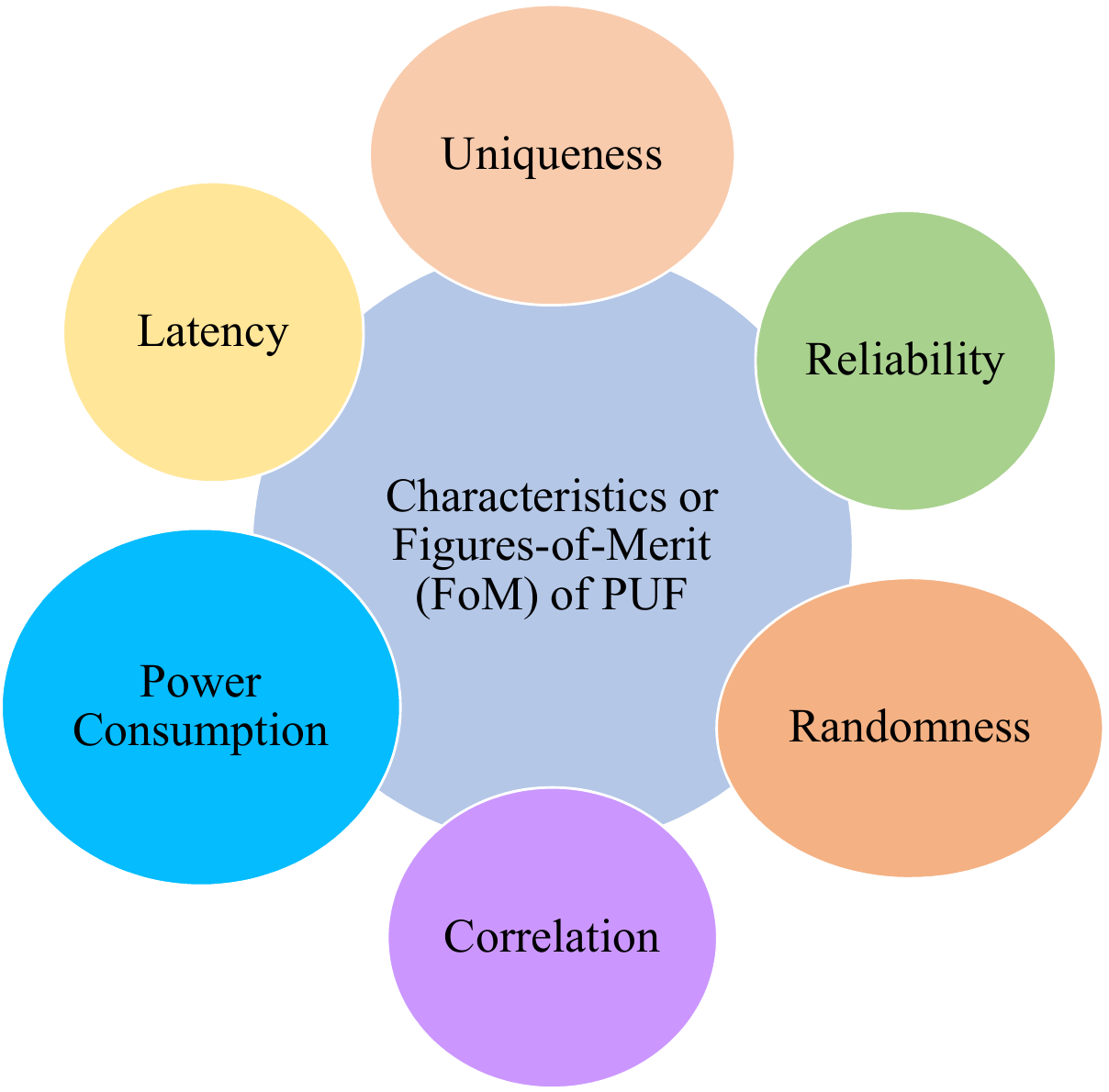}
	\caption{Design of Hybrid Ring-Oscillator Arbiter PUF \cite{Yanambaka_IEEE-TSM_2018-May,Yanambaka_ALOG_2017-Dec,Joshi_MPOT_2017-Nov}.}
	\label{FIG:PUF_FoMs}
\end{figure}

A set of properties has to be satisfied by the CRPs before they can be used for applications. We present various different FoMs of PUFs in Fig. \ref{FIG:PUF_FoMs} \cite{Yanambaka_IEEE-TSM_2018-May,Yanambaka_ALOG_2017-Dec,Joshi_MPOT_2017-Nov}.
 
\subsubsection{Uniqueness}

One of the major properties of a PUF is to generate a key that is unique to the device. For example, if $C_1$ is the challenge that produces the response $R_1$, by a PUF module $PUF_1$ this can be expressed as:
\begin{equation}
	PUF_1(C_1) = R_1
\end{equation}
For the CRP ${C_1,R_1}$ to be used for a cryptographic application, this should be unique to the module $PUF_1$. Given the same PUF module, any other challenge should not be able to generate the same response:
\begin{equation}
	PUF_1(C_x) \neq R_1 \text{sShould hold true } \forall \text{ }  x\neq 1
\end{equation}
Given a different PUF module, $PUF_x$ where $x \neq 1$, the response $R_1$ should not be generated for any challenge input. This is the uniqueness property of a PUF module. 

\subsubsection{Reliability}

The CRPs generated by a PUF are expected to be reliable, more importantly when the applications they are used for are cryptographic in nature. This is because, when data is encrypted with a key generated by a PUF, the module is expected to give the key whenever necessary for decryption. This is one of the properties of the PUF, where it should generate the same set of CRPs even when the module is subjected to various environmental variations such as temperature, power supply and aging. These three factors are the most affecting in case of a PUF. The uniqueness and reliability can be estimated using the Hamming distance. The ideal Hamming distance for the keys to be unique is 50\% and reliability is 0\%.

The uniqueness and reliability can be estimated using the Hamming distance. The output responses for a set of challenges are generated and the Hamming distance is calculated between the keys. This gives the uniqueness of the CRPs generated. The ideal Hamming distance for the keys to be unique is 50\%. The same challenge is given to the PUF under various conditions like temperature variations and aging effects and the responses are collected. The Hamming distance measured between these keys should be close to 0\% which means the keys are identical for the same challenge under various adverse conditions. This estimates the reliability of the PUF. 

\subsubsection{Randomness}

Randomness estimates the security aspect of the PUF keys. The Hybrid Oscillator Arbiter PUF takes the challenges and gives the responses in binary. These keys are also susceptible to various attacks. For the responses to be more robust and resistant to attacks, the number of 0's and 1's should be identical. For example, in a output key of $128-bits$, there should be 64 0-bits and 64 1-bits. This makes the responses from the PUF more secure and resistant to various attacks.

\subsubsection{Correlation}

Correlation is another important characteristic for a PUF to be useful in practical applications.
There should be no correlation between the modules when the keys are generated by these PUF modules. For example, when the modules are fabricated on the same wafer, or when they are fabricated in the same fabrication facility, there should be no correlation between the integrated circuits thereby giving a similarity between the output responses generated by
the PUF. Of course, various types of process variations including fab to fab, lot to lot, wafer to wafer, and die to die, are expected to take care of this \cite{Mohanty_Book_2015_Mixed-Signal}.

\subsubsection{Power Consumption}

Power consumption can be critical for PUFs in applications where the system or device is battery constrained. For example, Implantable Medical Devices (IMD), Wearable Medical Devices (WMD), or collectively Implantable and Wearable Medical Devices (IWMD) \cite{Yanambaka_TCE_2019_PMsec,Mohanty_IEEE-MCE_2019-Nov_Editorial}. Specifically, security should not create overhead on battery which has life expectancy of 8-10 years for IMDs like pacemakers. In such applications, increasing security by introducing a module which consumes more power is not an ideal solution. This can give rise to various issues and the  product may not compatible with the required specifications of the environment. So the PUF module which is being introduced into the IoT device should consume as little power as possible.

\subsubsection{Latency}

In the IoT, there is a huge number of devices connected to a network. They constantly
communicate with each other or a server and this is the key requirements for connected and sustainable smart cities in which IoT is the back bone \cite{Mohanty_MCE_2016-July_Smart-Cities}. The router or the network switch which connects them together will have to perform operations at a high rate. During this process, when performing various cryptographic operations, the keys generated by the PUF module should introduce minimal latency. The keys should be produced as fast as possible so that the router can send and receive messages with a minimal latency thereby reducing the network congestion. The speed with which the PUF module generates the output keys also plays a vital role in many applications other than IoT. For example, in smart cars and Unmanned Aerial Vehicles (UAVs) \cite{Mohanty_IEEE-MCE_2019-Nov_Editorial,Mohanty_ZINC_2018_Keynote}. Latency overhead can be a problem in smart cars in which action time should be in the range of milliseconds.

\subsection{PUF Types}

\begin{figure}[t]
	\centering
	\includegraphics[width=0.80\textwidth]{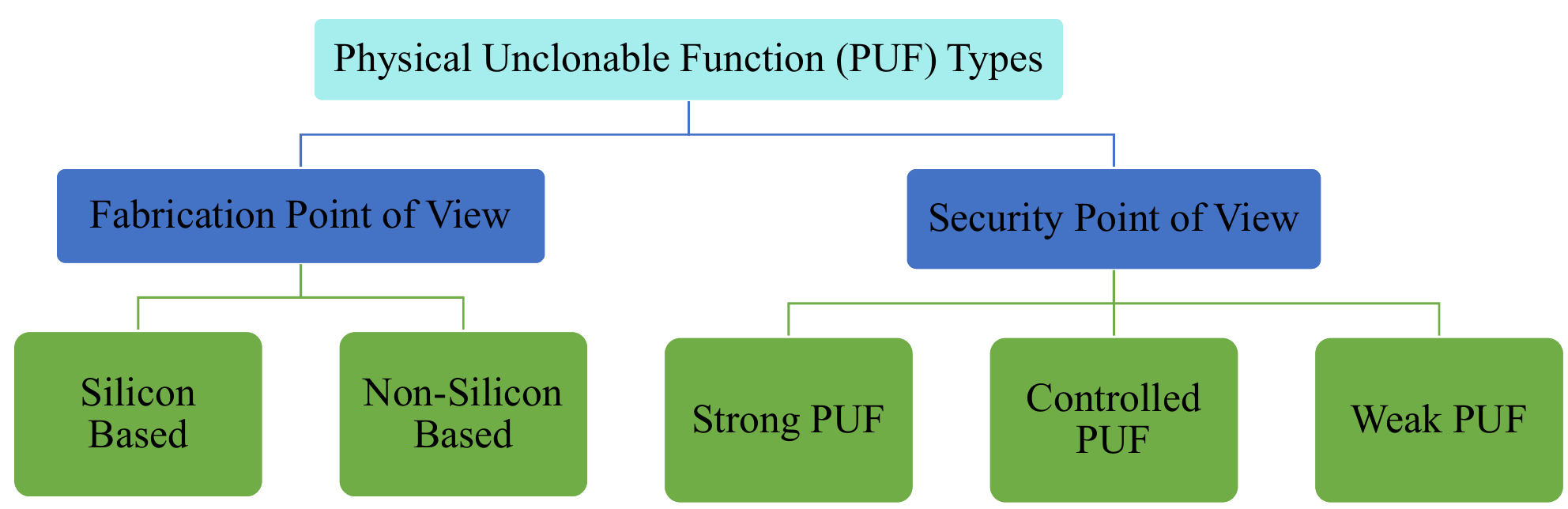}
	\caption{Different Types of PUF \cite{Joshi_MPOT_2017-Nov}.}
	\label{FIG:PUF_Types}
\end{figure}

Research has been ongoing on efficient PUF designs targeting various applications. A summary of various types of PUFs is presented in  Fig. \ref{FIG:PUF_Types} \cite{Joshi_MPOT_2017-Nov,Lao_TCAD_2014-May,Rose_NANOARCH_2013,Suh_DAC_2007}. In addition, a variety of PUF topologies including SRAM and ring oscillator exist in the literature. 

\begin{figure}[htbp]
	\centering
	\includegraphics[width=0.65\textwidth]{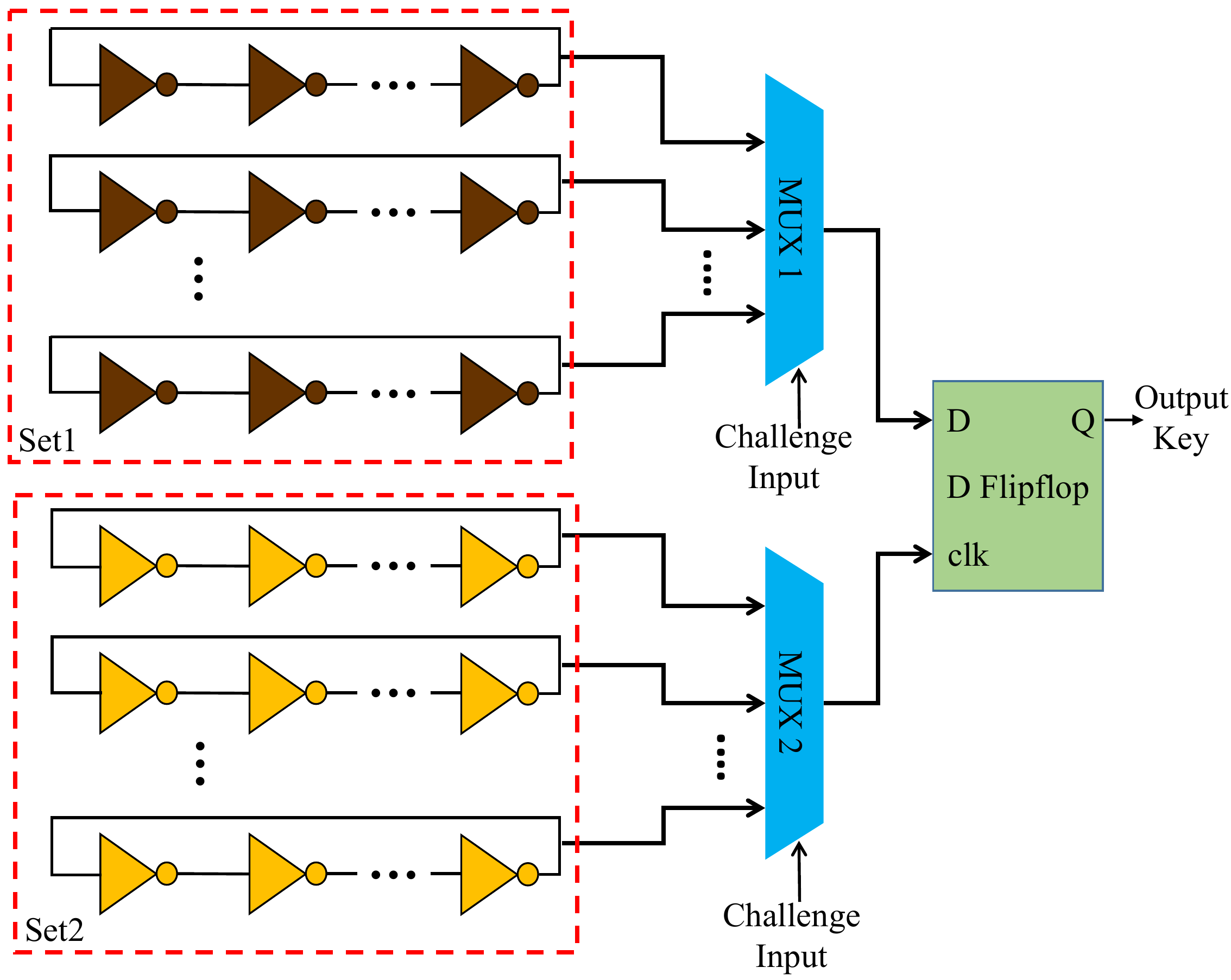}
	\caption{Design of Hybrid Ring-Oscillator Arbiter PUF \cite{Yanambaka_TCE_2019_PMsec}.}
	\label{FIG:PUF-Design_Hybrid-Ring-Oscillator-Arbiter}
\end{figure}

As a specific example, Fig. \ref{FIG:PUF-Design_Hybrid-Ring-Oscillator-Arbiter} shows the design of a Hybrid Oscillator Arbiter Physical Unclonable Function \cite{Yanambaka_ALOG_2017-Dec}. The ring oscillator is the main component in the design of the Hybrid Oscillator Arbiter PUF. All ring oscillators are identical to each other during the design phase but during the fabrication process variations are introduced into the design and hence no two oscillators are similar to each other.

As shown in the figure, the oscillators are split into two sets, \textit{SET1} and \textit{SET2}. All the oscillators in SET1 are connected to MUX1 and those in SET2 are connected to MUX2. The output from the multiplexers drives the D-Flipflop. The oscillators are responsible for producing the oscillations. The multiplexers will select the oscillations and feed them to the flipflop based on the challenge input given to the PUF. Because of the nanoelectronic manufacturing variations, the oscillation frequencies of the ring oscillators are going to be different and hence the signal at a given time \textit{t} at the D-input and Clock of the flipflop. Based on the signals at the inputs of flipflop, the output bit is generated. When the selection of ring oscillators is changed, the output key changes which helps in reconfiguring the device when necessary. 


\section{A Specific Case Study of PUFchain using Software and Hardware Platforms}
\label{SEC:Experimental_Results}

The experimental evaluation of the proposed PUFchain is presented in this section. For the evaluation of PUFchain, an IoT device and a hardware accelerator which contains the PUF and has the capability of cryptographic operations are necessary. As IoT devices, Raspberry Pi modules were used, where a Raspberry Pi 3 Model B+ is used as a trusted node. The hardware accelerator was developed on an FPGA.

\subsection{PUFchain Security Verification}
\begin{figure}[htbp]
	\centering
\subfloat[Experiment setup for PuFChain security verfication]{\label{fig:PUFChain_SecModel}\includegraphics[width=0.55\textwidth]{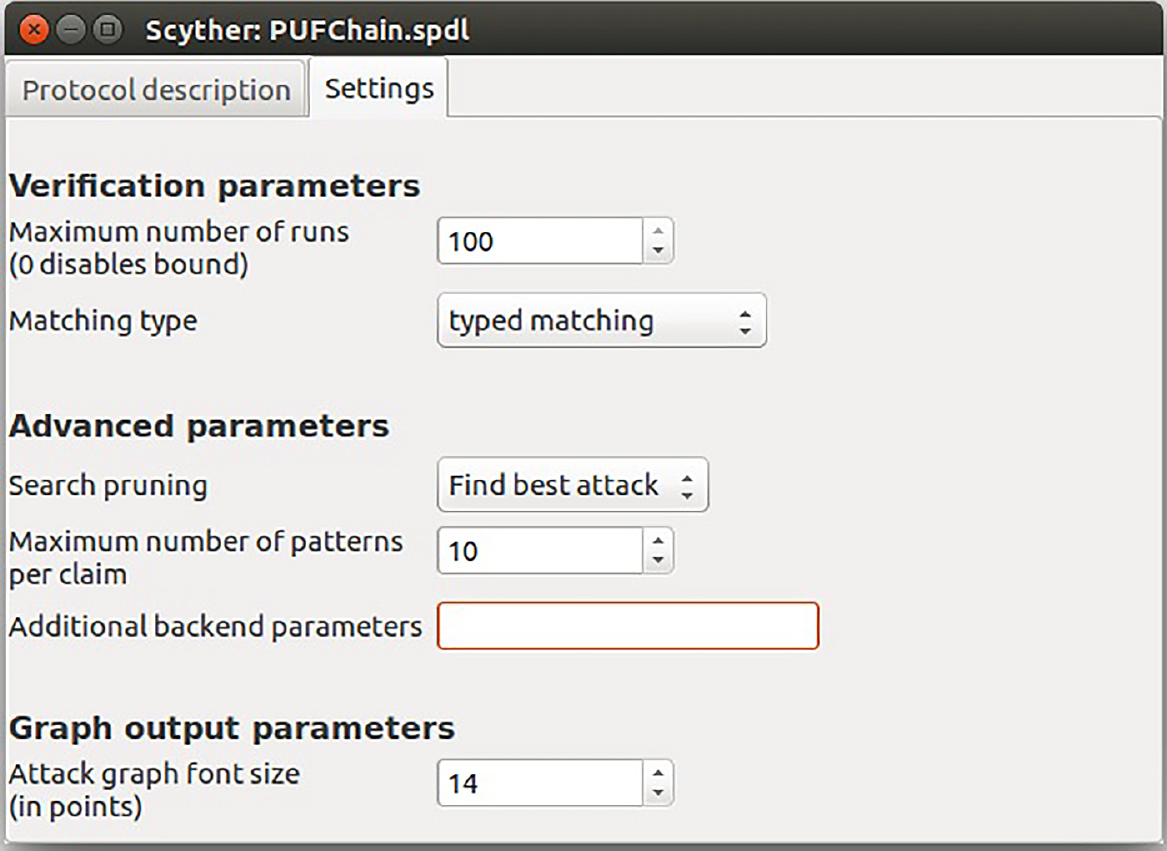}}\\
\hspace{0.2cm}
\subfloat[Scecurity verfication results]{\label{fig:PUFChain_SecRes}\includegraphics[width=0.55\textwidth]{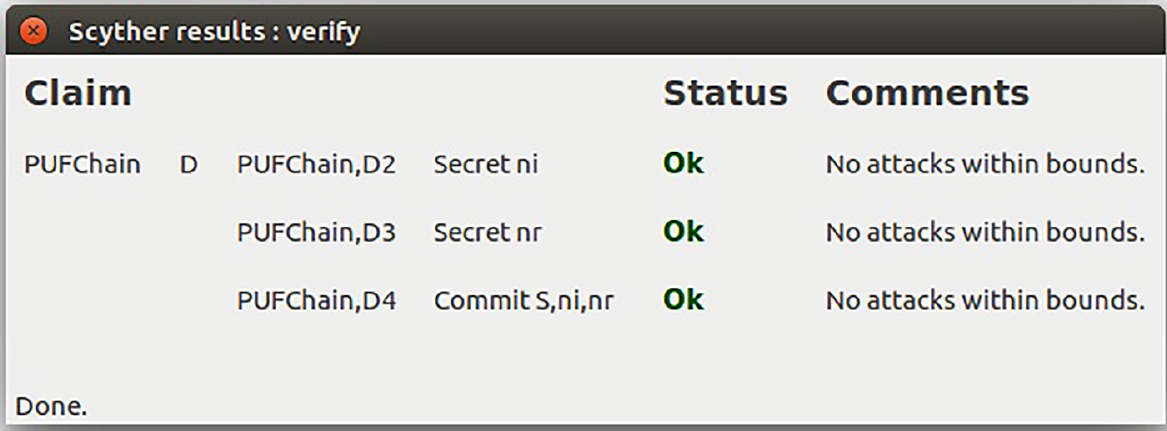}}
\caption{PUFChain security verification in Scyther}
	\label{FIG:PUFChain_Scyther}
\end{figure}

The PUFChain methodology is written in the Scyther simulation environment using the Security Protocol Description Language (.spdl) \cite{Scyther}. The simulation is conducted in Scyther v1.1.3 in the Ubuntu 18.04.3 OS. According to the features of Scyther, we define the role of D and S, where S is the source of the block and D is the miner or authenticator node in the networks. In the simulation, we have initialized all the features, as described in the model description. PUF random numbers are chosen randomly in this simulation. In our simulation, we have evaluated the scenario at the authenticated node (D) to find whether it authenticated the blocks thoroughly without any compromise. We experimented with 100 numbers of runs for each claim to found out the number of attacks at D as shown in Figure \ref{FIG:PUFChain_Scyther}. Apart from these, we follow the default properties of Scyther. 

We set the model to randomly generate the possible network attacks to verify our PUFChain against all the possible attacks scenario. In practice, attacks may be more sophisticated and efficient than brute force attacks. 

We did our simulation using variable numbers of data blocks in each run. Our experiment ranges from 100 to 1000 instances with 100 intervals as shown in Figure \ref{fig:PUFChain_SecModel}. We check authentication for each block and inside transactions, where the miners in the network check the blocks. There are two parameters we are validating in the results page, i.e. $ni$ for secret key and $nr$ for certificate. For individual blocks, miners in the network verify the blocks against all the randomly generated possible attacks. The security verification results of PUFChain are experimented against all the possible attacks and the result is shown in Figure \ref{fig:PUFChain_SecRes} which indicates that PUFChain is secure against potential network threats. 

\subsection{Real-Life Testbed of PUFchain using Off-The-Self Components}
\label{sec:Experimental_Setup}

\begin{figure*}[htbp]
	\centering
	\includegraphics[width=0.995\textwidth]{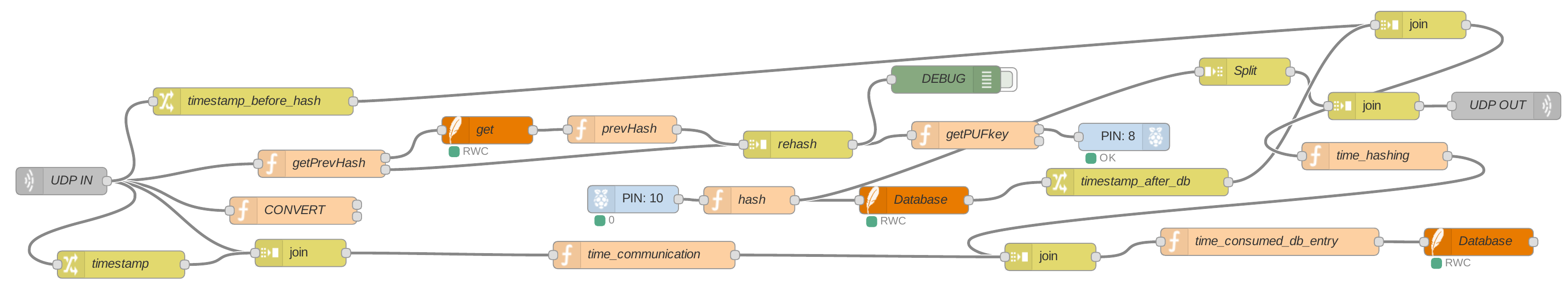}
	\caption{Node-RED Interface for PUFchain validation in real-life testbed.}
	\label{FIG:Node_Red_PUF}
\end{figure*}

The Node-RED development tool was used for developing the PUFchain blockchain and the PoP consensus algorithm. Node-RED is a flow-based programming tool that can be used on various platforms including single board computers like the Raspberry Pi. The interface is accessible through a web browser remotely using the IP address of the device. Fig. \ref{FIG:Node_Red_PUF} shows the Node-RED implementation of PUFchain on a Raspberry Pi single board computer. As shown in the figure, the data is transmitted and received using UDP over the network of devices. Node-RED allows the user to generate timestamps wherever necessary. Using this function, various timestamps are generated at different points through the implementation to measure the time taken by the data to traverse through the different processes. Fig. \ref{FIG:Node_Red_PUF} shows the implementation of the PUFchain in a trusted node. Once the device is validated using the PUF and hashing module, if the device is authenticated, the block gets added to the database at the local storage and then gets broadcast to the other nodes in the network. A similar process is followed at the client nodes to validate if the trusted node sent the block and is added to their respective local blockchains.

The experimental setup consists of the Raspberry Pi single board computers and an Altera\textsuperscript{\textregistered} DE2 FPGA module on which the PUF and the Hashing module were developed. The characterization of the PUFchain is shown in Table \ref{TABLE:PUFChain_Characterization}. To estimate the performance of the proposed PoP consensus algorithm and the PUFchain blockchain in an IoT environment, different single board computers were considered. 6 Raspberry Pi single board computers were considered for the evaluation of the proposed algorithm. Five of the Raspberry Pis are client nodes and the other is the trusted node. A Raspberry Pi 3 Model B+, which is equipped with a 1.4GHz 64-bit Quad Code ARMv8 processor, is used as the trusted node. Two Raspberry Pi 3 Model B boards and three Raspberry Pi 1 single board computers are used as clients. The Raspberry Pi 3 Model B is equipped with a Quad Core 1.2GHz processor and 1GB RAM and the Raspberry Pi 1 is equipped with a single core Cortex processor and 512MB RAM. For the evaluation of the consensus algorithm and the performance of the PUFchain blockchain architecture in an IoT environment, a total of 300 transactions were considered from the initiation, authentication and adding to the blockchain.

\begin{figure}[htbp]
	\centering
	\includegraphics[width=0.55\textwidth]{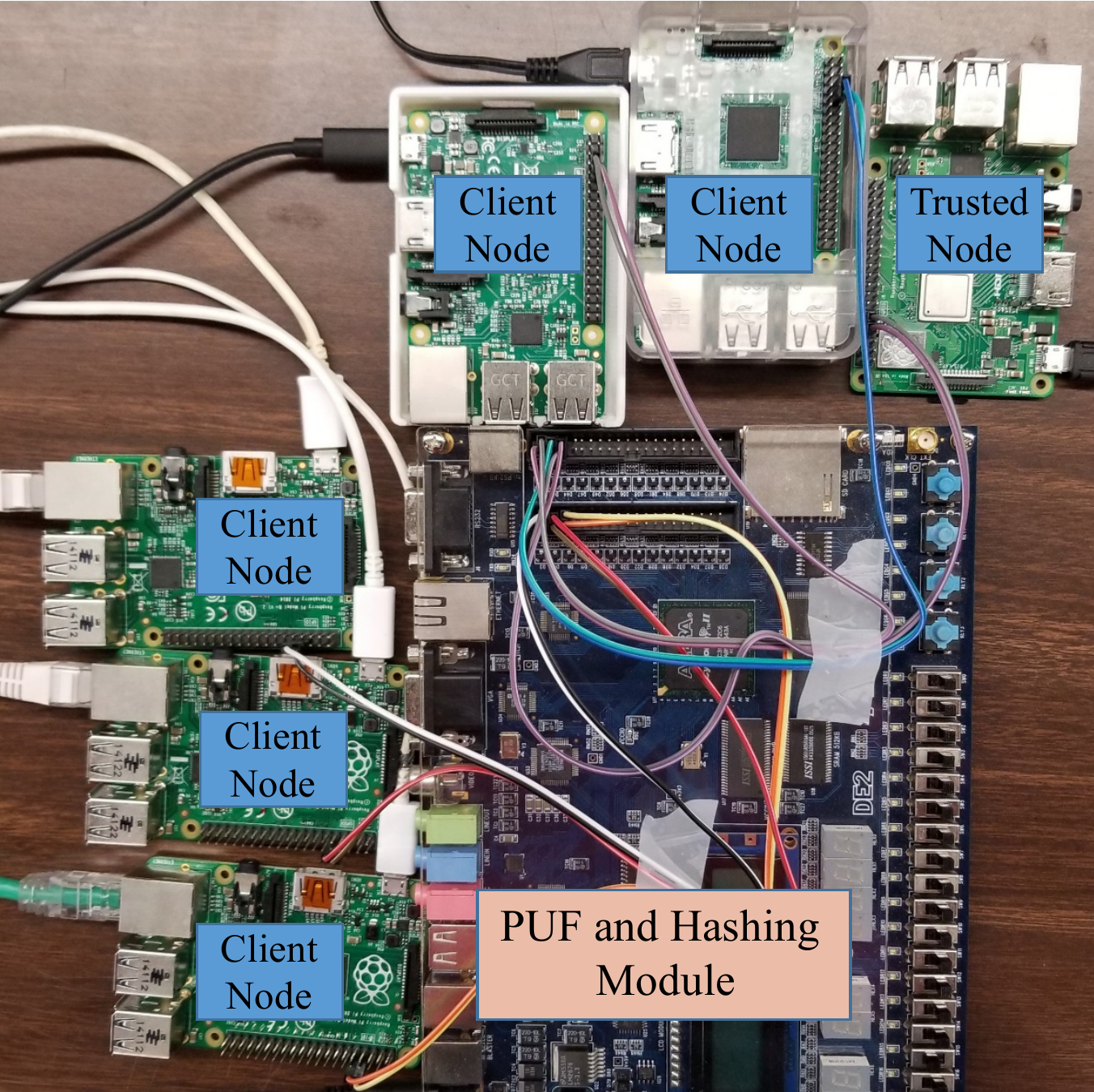}
	\caption{Experimental Setup of PUFchain}
	\label{FIG:Experimental_Setup}
\end{figure}

\begin{table}[htbp]
	\caption{Characterization of PUFchain Experimental Setup}
	\label{TABLE:PUFChain_Characterization}
	\centering
	\begin{tabular}{|p{5.9cm}|c|c|}
		
		\hline
		PUFchain Parameters & \multicolumn{2}{c|}{Specific Values}\\
		\hline \hline
		IoT Devices & Trusted Node & Client Node\\
		\cline{2-3}
		&Raspberry Pi 3 & Raspberry Pi 3 Model B\\
		&Model B+&Raspberry Pi 1\\ 
		\hline		
		Operating System & \multicolumn{2}{c|}{Raspbian 4.14}\\
		\hline
		Communication & Wireless & Wired and Wireless\\
		\hline
		PUF and Hashing Module & \multicolumn{2}{c|}{Altera DE-2}\\
		\hline
		Time taken to complete a transaction & \multicolumn{2}{c|}{192.3 ms}\\
		\hline
	\end{tabular}
	
\end{table}

\subsection{Transaction Time Analysis of PUFchain}

\begin{figure*}[t]
	\centering
	\subfloat[Raspberry Pi 1]{\label{fig:Raspberry Pi 1}\includegraphics[width=0.50\textwidth]{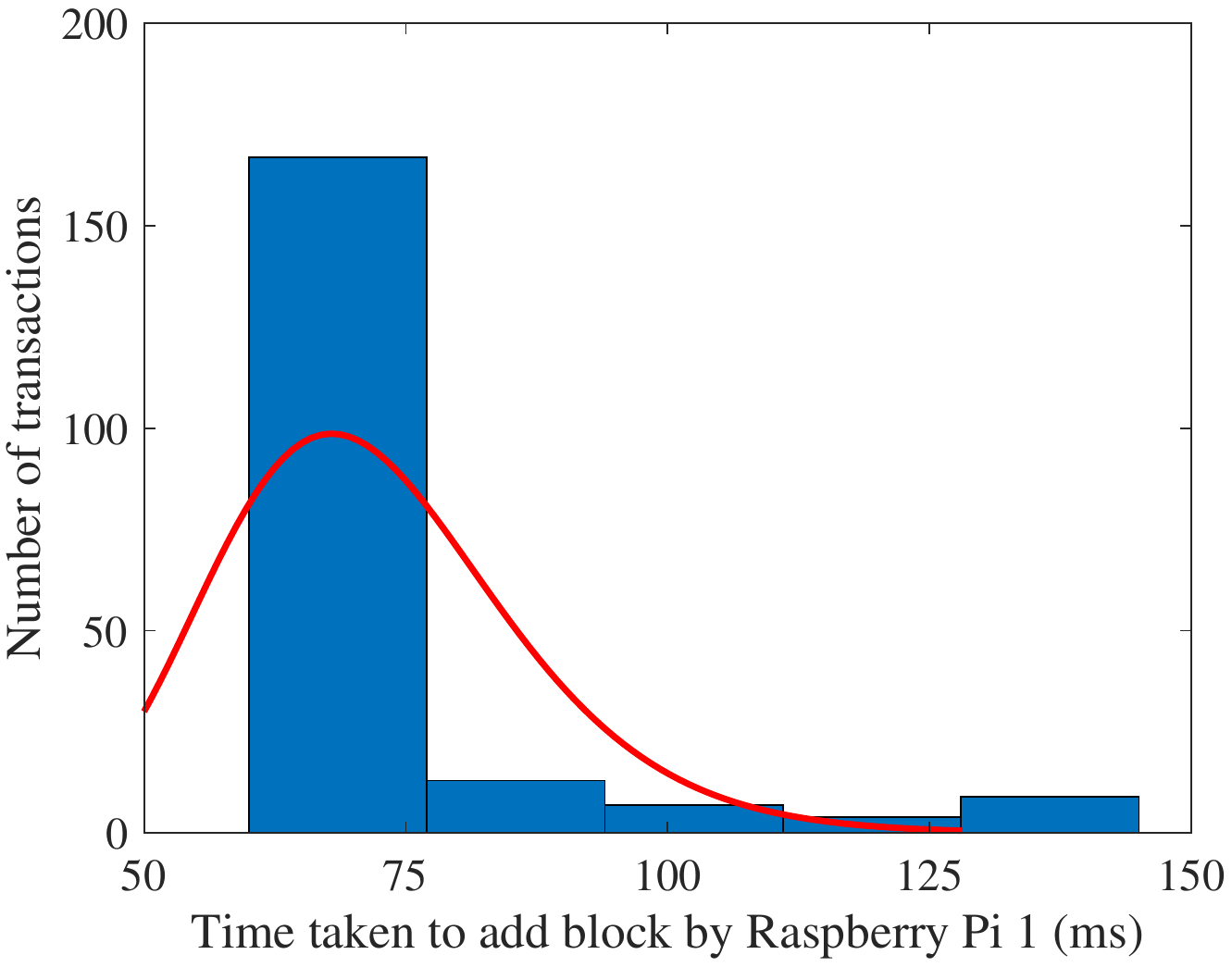}}
	\subfloat[Raspberry Pi 2]{\label{fig:Raspberry Pi 2}\includegraphics[width=0.50\textwidth]{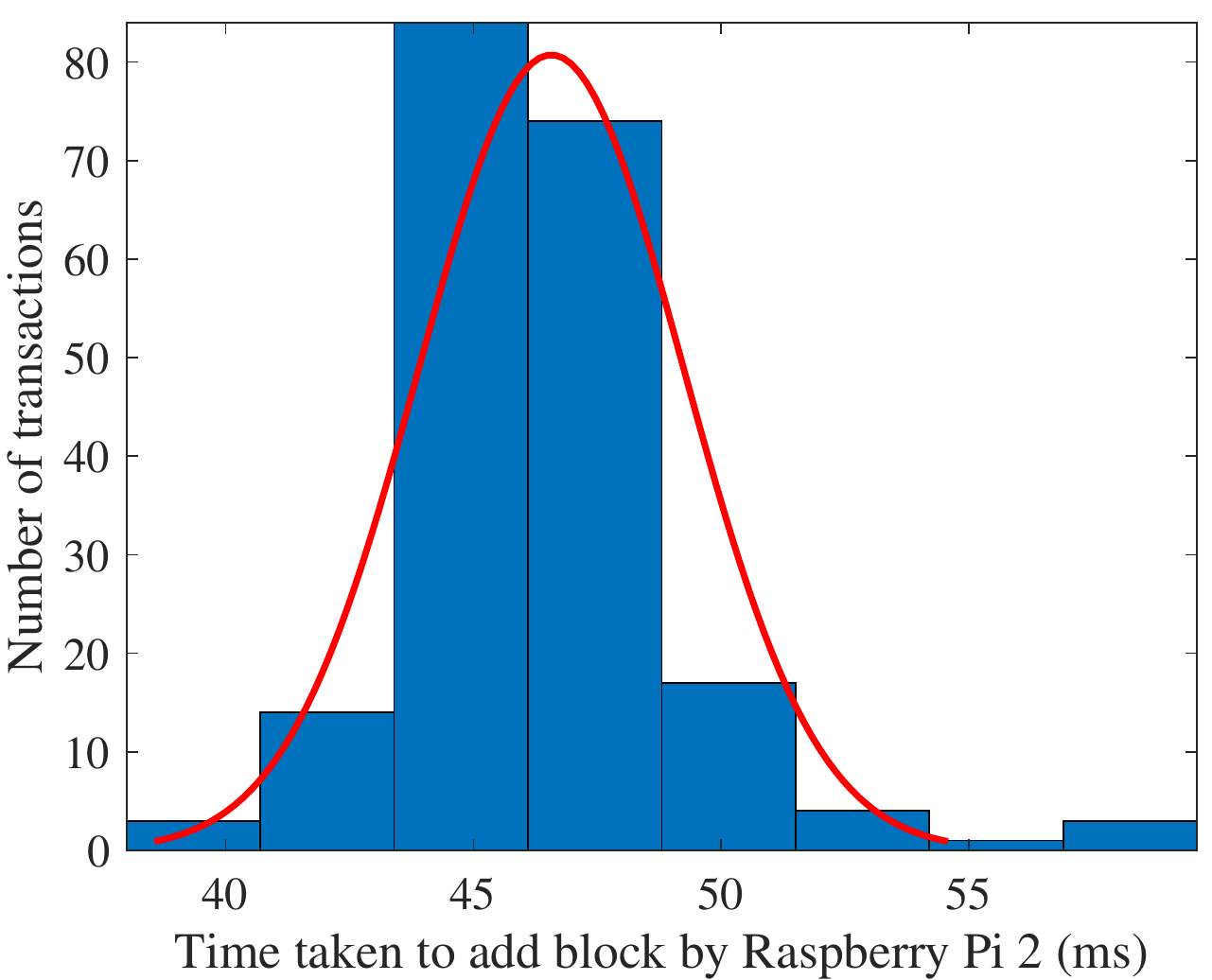}}\\
	\subfloat[Raspberry Pi 3]{\label{fig:Raspberry Pi 3}\includegraphics[width=0.50\textwidth]{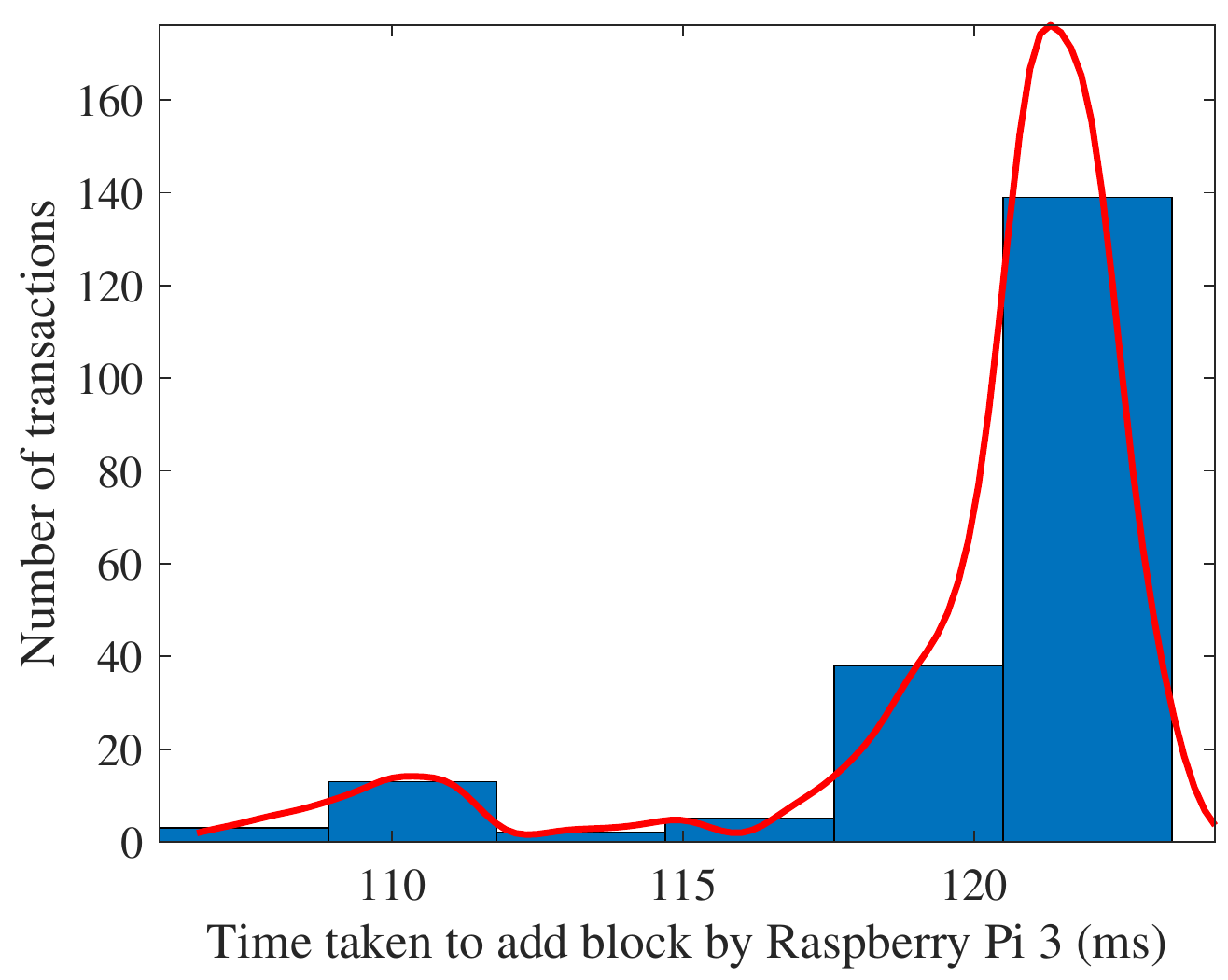}}
	\caption{Time taken by different single board computers to add the block to blockchain.}
	\label{FIG:Time_Taken}
\end{figure*}

As discussed in Section \ref{SEC:PUF_Chain}, the nodes in the network will create the blocks and broadcast them to the network. The device is not authenticated at this stage and the block has not originated from the trusted node. Hence, besides the trusted node, every other node in the network will drop the block that is broadcast. The trusted node listens to the network for the blocks and start the process of authentication. The data and the hash are extracted from the received block and the PUF keys are requested from the secure database. Then a hash is generated by the PUF and hashing module at the trusted node using the PUF key from the secure database. If the hash sent by the node and the hash computed at the trusted node match, the device is authenticated. The time taken to add the node to the database is shown in Fig. \ref{FIG:Time_Taken}. 

Once the device is authenticated, the trusted node will broadcast the block to the network. The other nodes in the network will receive the block and check the identification of the sender. If the sender is the trusted node, the other nodes will accept the block and add it to their local PUFchain. Fig. \ref{fig:Raspberry Pi 1} shows the time taken by a node which has a Raspberry Pi 1 as the IoT device, to receive the block from the trusted node and add it to the local blockchain. Once the block reaches the node, the IoT device, Raspberry Pi 1 in this case, will access the secure database to retrieve the trusted node PUF key using the device identification. The hash of the PUF key combined with the data received from the block is computed at every node. Once the hashes match, the block is sent by the trusted node and the block can be added to the PUF chain. If the hash does not match, another PUF key which belongs to the trusted node is requested from the secure database and this process is repeated until all the PUF keys are checked or a valid identification is attained, whichever comes first. If none of the keys can produce the desired result, the block is considered to be tampered with and will be discarded. 

\begin{figure}[!h]
	\centering
	\includegraphics[width=0.55\textwidth]{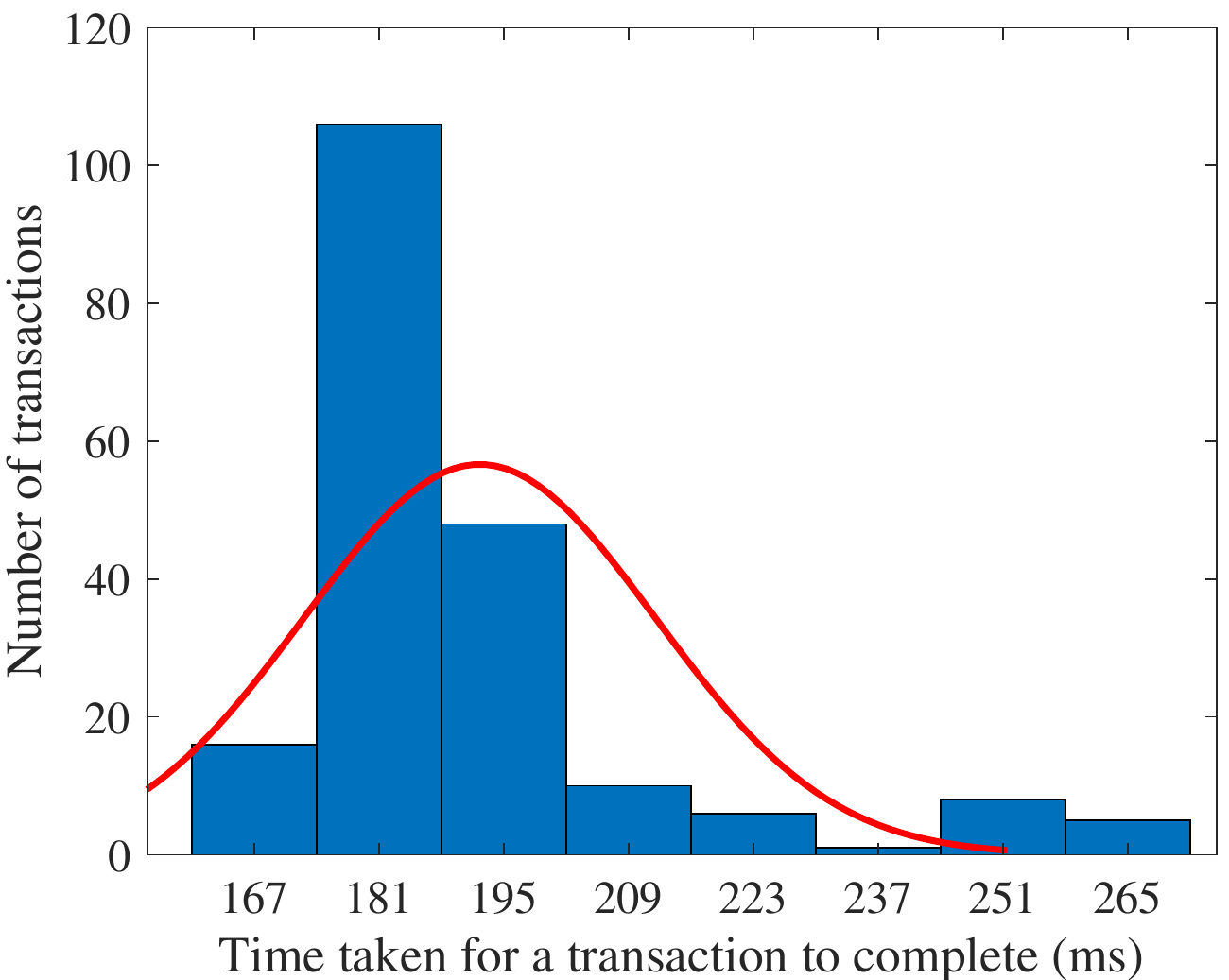}
	\caption{Time taken for a transaction to be authenticated and added to blockchain.}
	\label{FIG:Transaction_Authentication_Time}
\end{figure}

In this experimental setup, Raspberry Pi 1 and Raspberry Pi 2 are the IoT devices which are constantly collecting the data and storing the authenticated blocks coming from the trusted node. Fig. \ref{fig:Raspberry Pi 1} and Fig. \ref{fig:Raspberry Pi 2} show the time taken by the two devices to complete the above process from the time the block arrives at the node till the time it gets added to the PUFchain. Table \ref{Table:AddTime} shows the time taken by the nodes to add the block to their local blockchain. The time consumed by the IoT devices is presented for 300 transactions from the initiation, authentication to the adding of block to the local blockchains at the client nodes. Fig. \ref{FIG:Transaction_Authentication_Time} shows the time taken by 300 transactions from initiation to the final addition to blockchain. The average time taken for a transaction to be complete is 198ms. 

\begin{table}[htbp]
	\caption{Time taken to add a block to the Blockchain once it is received}
	\label{Table:AddTime}
	\centering
	\begin{tabular}{|c|c|c|}
		\hline
		Node & Mean (ms) & Standard Deviation (ms) \\
		\hline
		\hline
		Raspberry Pi 1 & 72.27 & 18.07\\
		\hline
		Raspberry Pi 2 & 46.5 & 2.66 \\
		\hline
		Raspberry Pi 3 & 120.03 & 3.44\\
		(Trusted Node)&&\\
		\hline
		
	\end{tabular}
\end{table}

As shown in Fig. \ref{FIG:Node_Red_PUF}, timestamps were generated at different stages of the process in Node-RED to calculate the time taken for a transaction. When the message is received at the server, the timestamp $t_{sr}$ is generated. The data goes through the validation process and traverses through various functions. Once the validation is complete and the block added to the blockchain, the timestamp $t_{sv}$ is generated. The time taken by the server node for validating and adding the block is:

\begin{equation}
	\delta t_{sa} = t_{sv} - t_{sr}
\end{equation}

Similarly, once the message is received at the client node from the server, the timestamp $t_{cr}$ is generated. Once the client node validates that the server sent the block and adds it to the blockchain, the timestamp $t_{cv}$ is generated. The time taken by the client nodes to check if the server sent the data and add to the blockchain is the following:

\begin{equation}
	\delta t_{ca} = t_{cv} - t_{cr}
\end{equation}

Every block of data contains the initial timestamp when the data was collected from the sensors, $t_i$. The total time taken for the transaction is given as:

\begin{equation}
	\delta t_{tx} = t_{cv} - t_i
\end{equation}

\subsection{Power Consumption Analysis of PUFchain}
\label{sec:Power_Consumption}


Power consumption is one of the major aspects of a blockchain. Various consensus algorithms were proposed for different applications across multiple domains. But power consumption has been a major issue that restricts the blockchain to certain areas of applications. Requiring high processing power and dedicated hardware for mining the blocks requires more power in the case of blockchains that were proposed before. 

\begin{figure}[htbp]
	\centering
	\includegraphics[width=0.78\textwidth]{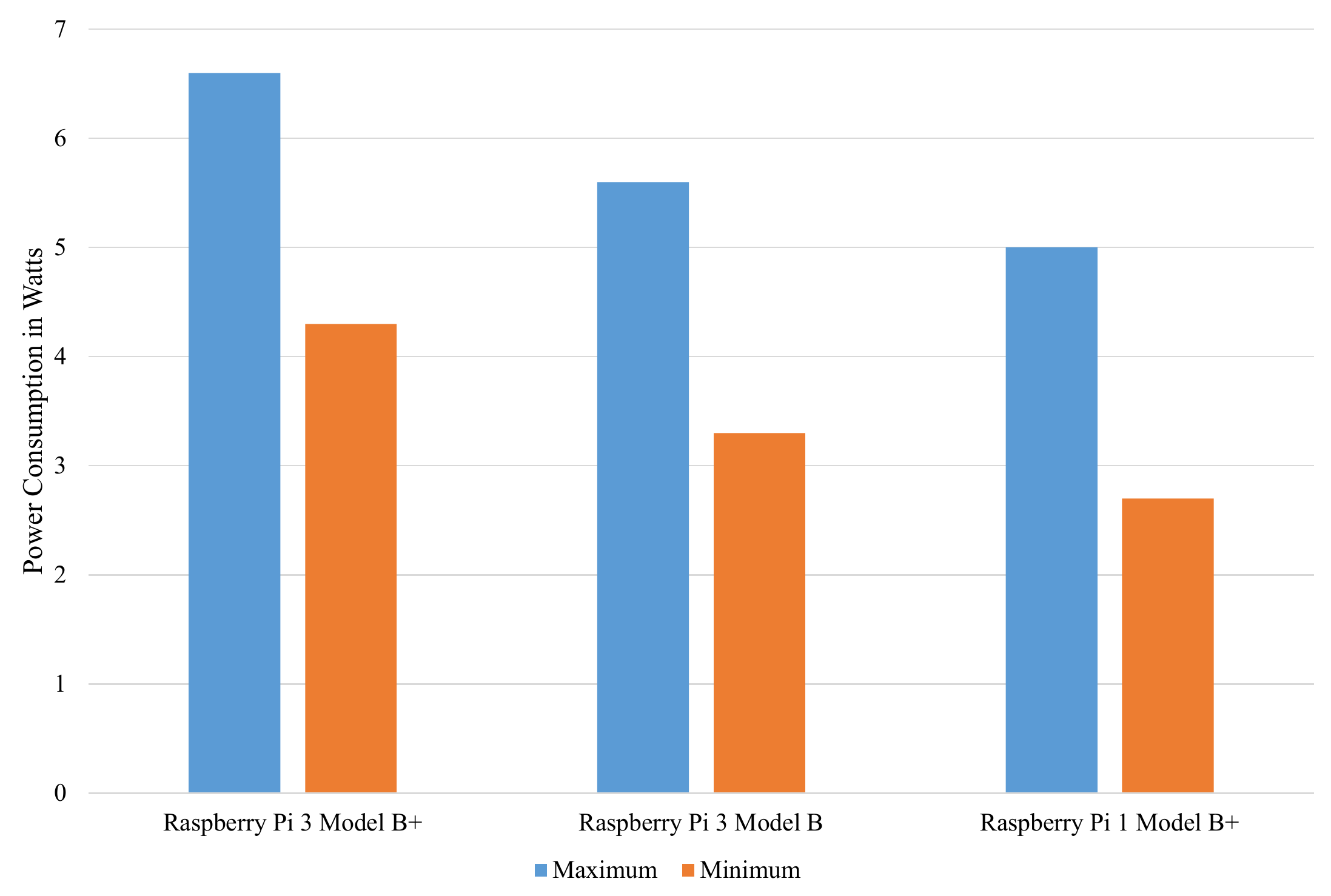}
	\caption{Power Consumption of PUFchain}
	\label{FIG:Power_Consumption}
\end{figure}

Power consumption requirements of IoT applications are low which makes integration of blockchain technology difficult. PoP uses light cryptographic algorithms and low power consumption hardware. The PUF used for the implementation of PUFchain is the Power Optimized Hybrid Oscillator Arbiter PUF which consumes low power. The single board computers used for PUFchain also are low power consuming which makes them ideal for developing a blockchain for IoT environment. Fig. \ref{FIG:Power_Consumption} shows the power consumption of various single board computers in the PUFchain environment. As discussed in the previous subsections, three models of Raspberry Pi are used in the experimental setup. The maximum power consumption is when the block processing is performed and the minimum is when the system is idle. 

The PUFs are implemented on an FPGA development board, Altera \textsuperscript{\textregistered} DE2. There are many components on the board besides the FPGA itself. To get more accurate value of power consumption, an initial value of power consumption is recorded when the PUF is not implemented on the FPGA and a value after the implementation and running. This gives a more accurate value of power consumption of the FPGA alone. Let us consider the power consumption of the FPGA when the PUF was not implemented on the FPGA as $P_{FPGA}$ and the power consumption of the FPGA after the PUF is implemented as $P_{FPGAPUF}$. The power consumed by the PUF module on the FPGA development board is given as:

\begin{equation}
	P_{PUF} = P_{FPGAPUF} - P_{FPGA}
\end{equation}

Let us consider that the Raspberry Pi consumes power $P_{rpi}$. The power consumed by the Raspberry Pi and the FPGA is given as:

\begin{equation}
	P_{PUFchain} = P_{PUF} + P_{Rpi}
\end{equation}

The FPGA consumes a maximum power of 3.5W when the processing is performed  and a minimum power of 1.2W when it is idle. The overall power consumption of the Raspberry Pi 3 Model B+ and the FPGA combined is 6.6W and the minimum is 4.3W. Raspberry Pi 1 Model B consumes less power of them all where maximum power consumption is 5W and the minimum is 2.7W.

\subsection{Evaluation of Proposed PUF}
\label{sec:PUF_Evaluation}

For the evaluation of the Hybrid Oscillator Arbiter PUF, various key properties that need to be satisfied by the developed module and the output keys generated were discussed in Section \ref{SEC:PUF}. The Hybrid Oscillator Arbiter PUF was developed on the Altera \textsuperscript{\textregistered} DE2 board which is equipped with a Cyclone-2 FPGA. The PUF module and the hashing algorithms were developed on the FPGA and were used in the PUFchain implementation. 

Table \ref{TABLE:PUF_Characterization} shows the characterization table for the PUF module. The PUF modules that were developed on the FPGA consist of 512 oscillators. From the developed oscillators, the number of keys that were obtained from each PUF are 500. There is also the possibility of giving more challenge inputs and obtaining responses for them but, for the current experiment, initially 500 keys were obtained from the PUF module and were checked for the properties. As shown in Table \ref{TABLE:PUF_Characterization}, 128 keys that were obtained from the PUF satisfy the properties better than the others. 

\begin{table}[htbp]
	\caption{Characterization of PUF}
	\label{TABLE:PUF_Characterization}
	\centering
	\begin{tabular}{|c|c|}
		\hline
		PUF	Parameters & Calculated Values\\
		\hline
		\hline
		PUF Architecture& Hybrid Oscillator Arbiter PUF\\
		\hline
		No. of Oscillators & 512\\
		\hline
		No. of Keys Generated & 500\\
		\hline
		No. of keys satisfied properties & 128 \\
		\hline
		Uniqueness & 47.02 \% \\
		\hline
		Reliability & 1.25 \% \\
		\hline
		Randomness & 47 \%\\
		\hline
		Mean Oscillation Frequency & 250 MHz \\
		
		\hline
	\end{tabular}
\end{table}

\begin{figure*}[t]
	\centering
	\subfloat[Uniqueness]{\label{fig:Uniqueness_TrustedNode}\includegraphics[width=0.50\textwidth]{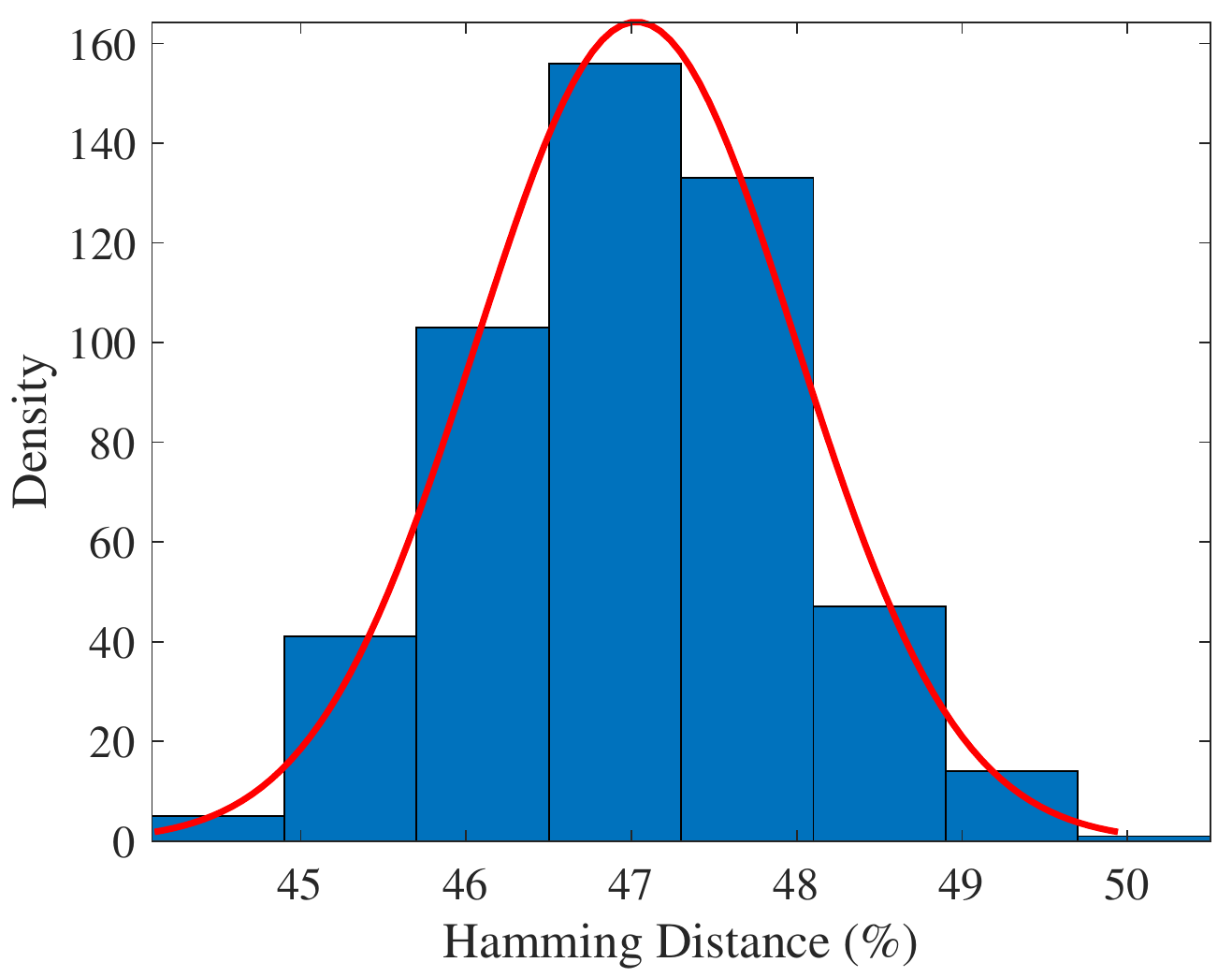}}
	\subfloat[Reliability]{\label{fig:Reliability_TrustedNode}\includegraphics[width=0.50\textwidth]{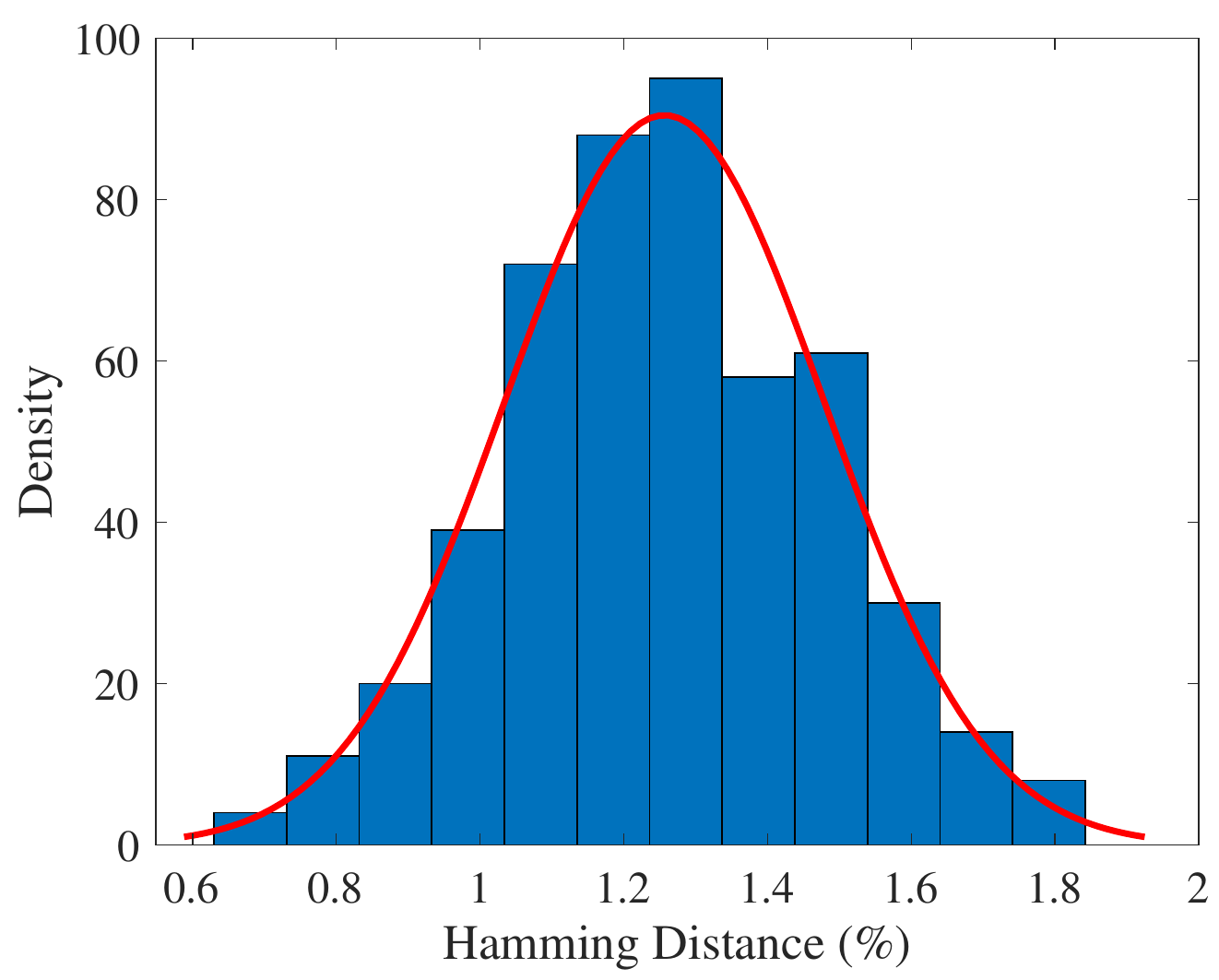}}\\
	\subfloat[Randomness]{\label{fig:Randomness_TrustedNode}\includegraphics[width=0.50\textwidth]{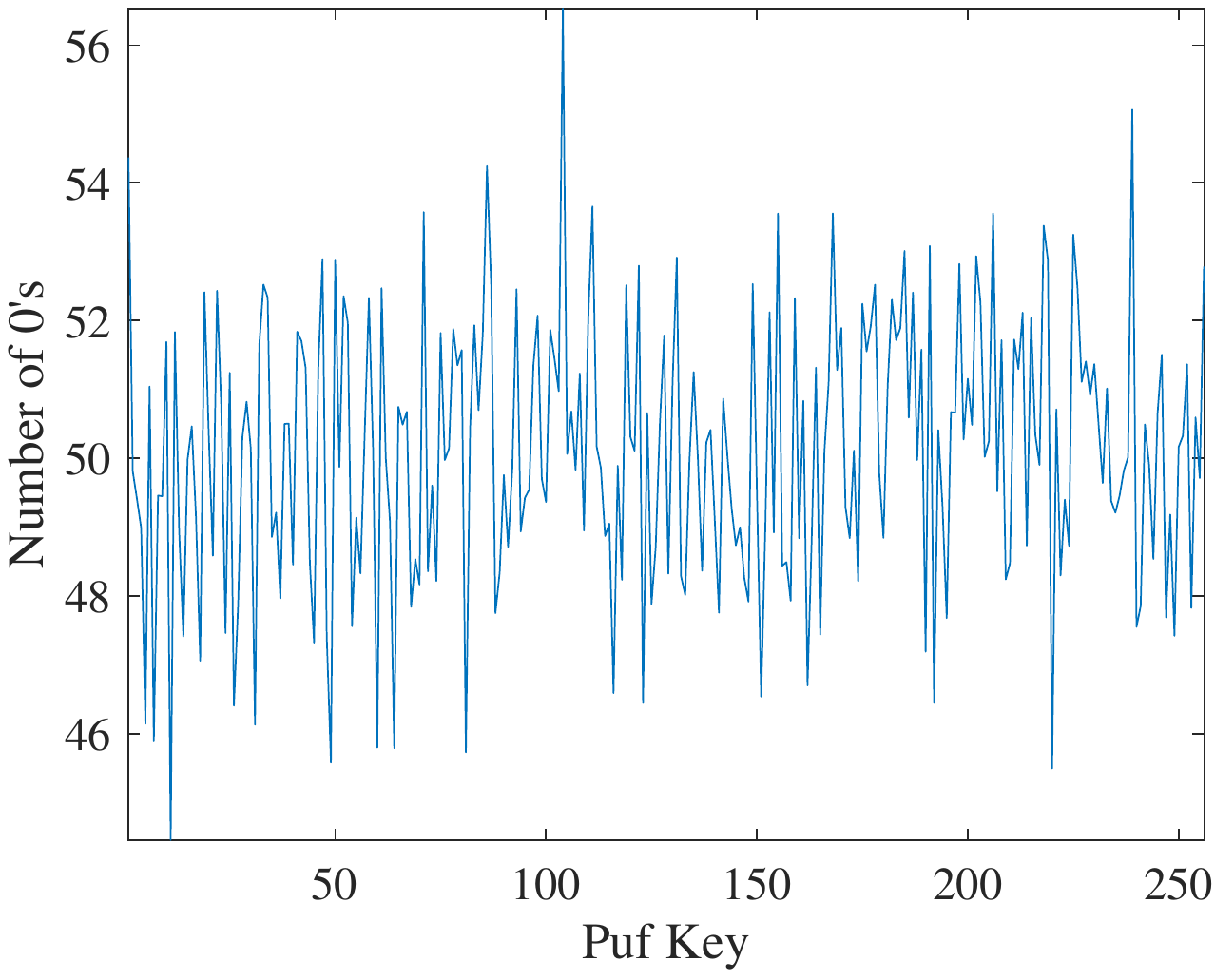}}
	\caption{Properties of PUF in the Trusted Node.}
	\label{FIG:PUF_Properties}
\end{figure*}

Fig. \ref{FIG:PUF_Properties} shows the properties of PUF, the uniqueness, the reliability and the randomness that were discussed in Section \ref{SEC:PUF}. The plots are presented for the PUF module that was used for the trusted node. Satisfactory results were obtained from the other PUF modules which are presented in Table \ref{TABLE:PUF_Properties}. The Hamming distance for the uniqueness property of the PUF module in the trusted node is 47.02\% and the reliability is 1.25\%. PUF 1 presented in Table \ref{TABLE:PUF_Properties} is the module that is present in the trusted node. The rest of the PUFs, PUF 2 through PUF 6 are the client nodes. 

\begin{table}[htbp]
	\caption{Hybrid Oscillator Arbiter PUF Uniqueness}
	\label{TABLE:PUF_Properties}
	\centering
	\begin{tabular}{|c|c|c|}
		\hline
		PUF Module & Uniqueness& Reliability\\
		\hline
		\hline
		PUF 1 (Trusted Node) & 47.02 & 1.25 \\
		\hline
		PUF 2 & 48.53 & 2.26\\
		\hline
		PUF 3 & 47.90 & 3.49\\
		\hline
		PUF 4 & 43.15 & 2.77\\
		\hline
		PUF 5 & 49.85 & 1.72\\
		\hline
		PUF 6 & 50.07 & 2.07\\
		\hline
	\end{tabular}
\end{table}

\subsection{Comparative Perspectives of PUFchain with other Blockchains}

Table \ref{TABLE:PUFChain_PoAh_Comparison} presents a comparison of PUFchain with PoAh. The blockchain consensus algorithms used for both of the implementations are permissioned and the mining is based on the authentication of the devices. One of the major difference between the two is in the security primitive used for the algorithm, hashing. The hash of PoAh is the cryptographic hash of the block and the device ID but in the case of the PUFchain, the PUF key is added to the hash. This makes the entire system resilient to attacks. The added PUF and hashing hardware reduces the computing burden on the IoT device. With an ultra low power design of PUF, and a low power IoT device, the overall power consumption can be decreased. The time taken for the completion of the transaction also sees a major difference. There is a 79.15\% decrease in transaction completion time.

\begin{table}[htbp]
	\caption{Comparison of PUFchain and PoAh}
	\label{TABLE:PUFChain_PoAh_Comparison}
	\centering
	\begin{tabular}{|p{4.1cm}|p{3.5cm}|p{3.2cm}|}
		\hline
		Parameter& PoAh based Blockchain \cite{Puthal_ICCE_2019} & PUFchain (\textbf{The Current Paper})\\
		\hline
		\hline
		Blockchain Type & Permission Based & Permission Based\\
		\hline
		Mining & Authentication Based & Authentication Based\\
		\hline
		Security primitive & Hashing & Hashing and added PUF Key\\
		\hline
		Overhead & Device ID & Device ID \\
		\hline
		Hardware needed & IoT Device capable of performing hashing & IoT Device \\
		\hline
		
		\hline
		\multicolumn{3}{|c|}{Time taken to add the received block}\\
		\hline
		\hline
		BlackPi & 843ms & 120.03ms\\
		\hline
		ClearPi (Raspberry Pi 3) & 85ms & 46.5ms\\
		\hline
		ClearPi (Raspberry Pi 1) & 162.4ms & 120.03ms\\
		\hline
		Time taken for a complete transaction & 950ms & 198ms\\
		\hline	
		\multicolumn{3}{|c|}{Power Consumption Range}\\
		\hline
		BlackPi & 3.1W|3.6W & 4.3W|6.6W\\
		\hline
		ClearPi (Raspberry Pi 3) & 2.1W|2.5W & 3.3W|5.6W\\
		\hline
		ClearPi (Raspberry Pi 1) & 1.5W|1.8W& 2.7W|5W\\
		\hline
	\end{tabular}
	
\end{table}

The well-established PoW while running in high-performance computing resources has a latency in the order of 10 mins. PoAh while running in limited computer resources has a latency in the order of 3 secs.  Thus,  PoAh is at least 200$\times$ faster than PoW which is used in traditional blockchain \cite{Puthal_IEEEP_2019,Puthal_ICCE_2019}. The proposed PoP algorithm is 5$\times$ faster than PoAh. Thus, PoP is approximately 1,000$\times$ faster than the well-established PoW. Overall, it can be concluded that PoP is highly scalable for large datasets which will run significantly faster,  uses minimal resources, and has a minimal energy consumption footprint.



\section{Conclusions and Future Directions of Research}
\label{SEC:Conclusion}

The blockchain has become on of the major technologies that have the potential to assist various application environments such as the IoT and solve various issues. Integration of blockchain into technologies such as IoT requires fine tuning of the consensus algorithms and blockchain architectures. Issues such as scalability, transaction time, fake block creation, security and privacy are still becoming hindrances in integration of blockchain and IoT. As a solution to some of the issues such as scalability, processing time and security, a novel blockchain consensus algorithm, Proof of PUF-Enabled Authentication (PoP) or Hardware Assisted Proof-of-Authentication (HA-PoAh) and a novel blockchain architecture, PUFchain are presented in the paper. The time taken to complete a transaction from initiation phase to adding the block to the blockchain takes 198ms. PUFchain integrates Physical Unclonable Functions (PUFs) to the blockchain architecture which adds an extra layer of security and makes the entire environment resistant to various attacks. The PUF key generated is not transmitted over the communication channel in the PoP consensus algorithm which makes the system resilient to communication attacks. 

Exploring PUF integration in other consensus algorithms to study their speed up as compared to without PUF is a possible future research direction. Future research is also be possible in examining the impact of PUF integration in public, private, and permissioned blockchains. As a future extension, an ultra low power design of PUF could be integrated into the designs which can reduce the overall power consumption of the system. Scalability is another aspect of blockchain that has been an issue in integrating the technology into IoT architectures. PUFchain solves the issue of scalability to some extent but the architecture could be improved due to the increasing of IoT devices number by the day.


\section*{Acknowledgments}


This article is an extended version of our article \cite{Mohanty_IEEE-MCE_2020-Mar_PUFchain} and conference presentation \cite{Yanambaka_iSES_2019_PUFchain}.

\bibliographystyle{IEEEtran}


\section*{About the Authors}

\begin{minipage}[htbp]{\columnwidth}
	\begin{wrapfigure}{l}{1.40in}
		\vspace{-0.5cm}
		\includegraphics[width=1.40in,keepaspectratio]{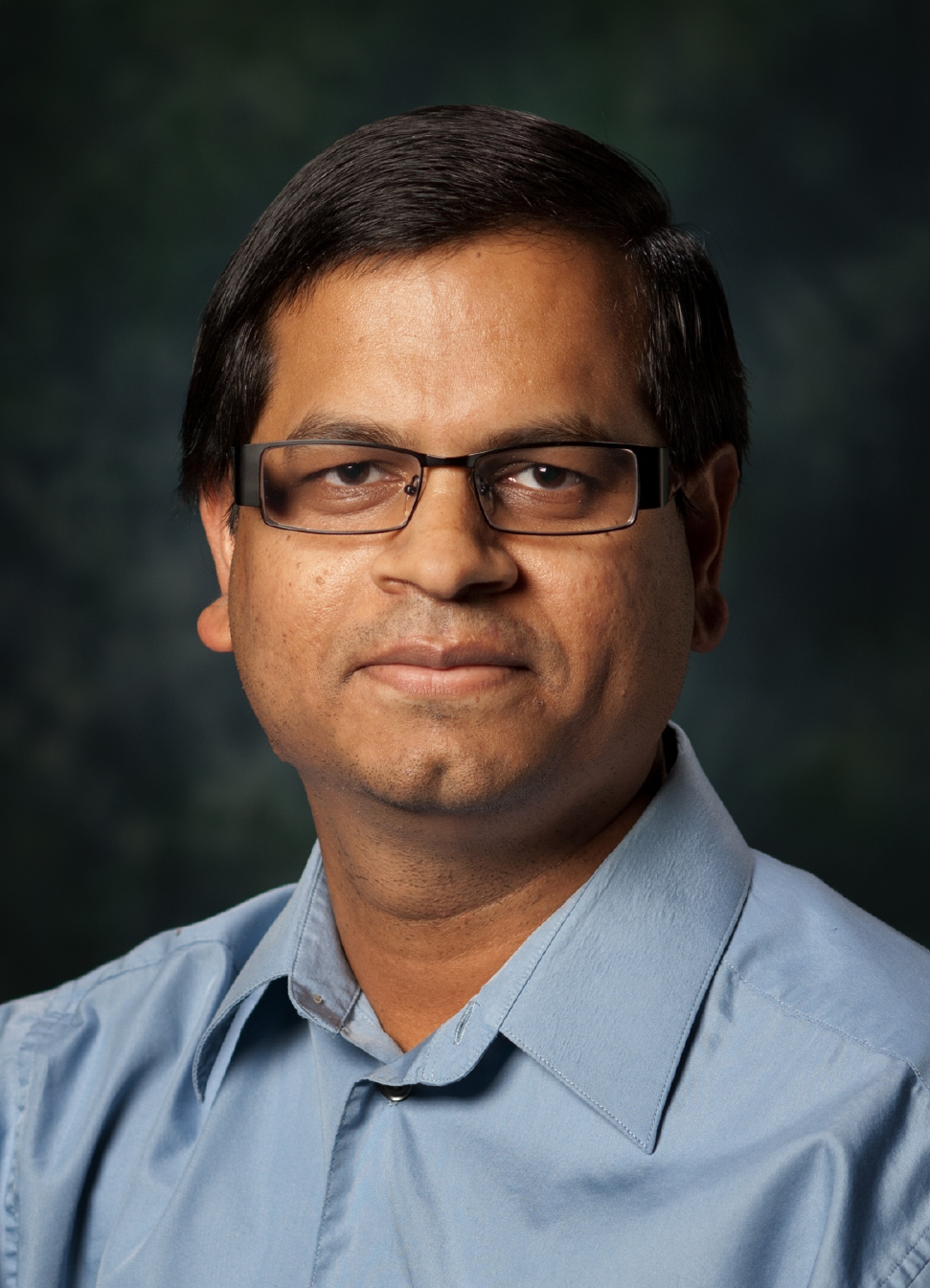}
		\vspace{-0.9cm}
	\end{wrapfigure}
	\noindent
\textbf{Saraju P. Mohanty} (SM'08) received the bachelor's degree (Honors) in electrical engineering from the Orissa University of Agriculture and Technology, Bhubaneswar, in 1995, the master's degree in Systems Science and Automation from the Indian Institute of Science, Bengaluru, in 1999, and the Ph.D. degree in Computer Science and Engineering from the University of South Florida, Tampa, in 2003.
He is a Professor with the University of North Texas. His research is in ``Smart Electronic Systems'' which has been funded by National Science Foundations (NSF), Semiconductor Research Corporation (SRC), U.S. Air Force, IUSSTF, and Mission Innovation Global Alliance. He has authored 300 research articles, 4 books, and invented 4 U.S. patents. His has Google Scholar citations with an H-index of 32 and i10-index of 110. He was a recipient of nine best paper awards, the IEEE-CS-TCVLSI Distinguished Leadership Award in 2018 for services to the IEEE and to the VLSI research community, and the 2016 PROSE Award for Best Textbook in Physical Sciences and Mathematics category from the Association of American Publishers for his Mixed-Signal System Design book published by McGraw-Hill. 
He has delivered 8 keynotes and served on 5 panels at various International Conferences. He has been serving on the editorial board of several peer-reviewed international journals, including IEEE Transactions on Consumer Electronics (TCE), and IEEE Transactions on Big Data (TBD). 
He is currently the Editor-in-Chief (EiC) of the IEEE Consumer Electronics Magazine (MCE). 
He has been serving on the Board of Governors (BoG) of the IEEE Consumer Electronics Society, and has served as the Chair of Technical Committee on Very Large Scale Integration (TCVLSI), IEEE Computer Society (IEEE-CS) during 2014-2018. He is the founding steering committee chair for the IEEE International Symposium on Smart Electronic Systems (iSES), steering committee vice-chair of the IEEE-CS Symposium on VLSI (ISVLSI), and steering committee vice-chair of the OITS International Conference on Information Technology (ICIT). He has mentored 2 post-doctoral researchers, and supervised 10 Ph.D. dissertations, 26 M.S. theses, and 10 undergraduate projects. 
\end{minipage}


	
\vspace{1.5cm}
\begin{minipage}[htbp]{\columnwidth}
	\begin{wrapfigure}{l}{1.40in}
		\vspace{-0.5cm}
		\includegraphics[width=1.40in,keepaspectratio]{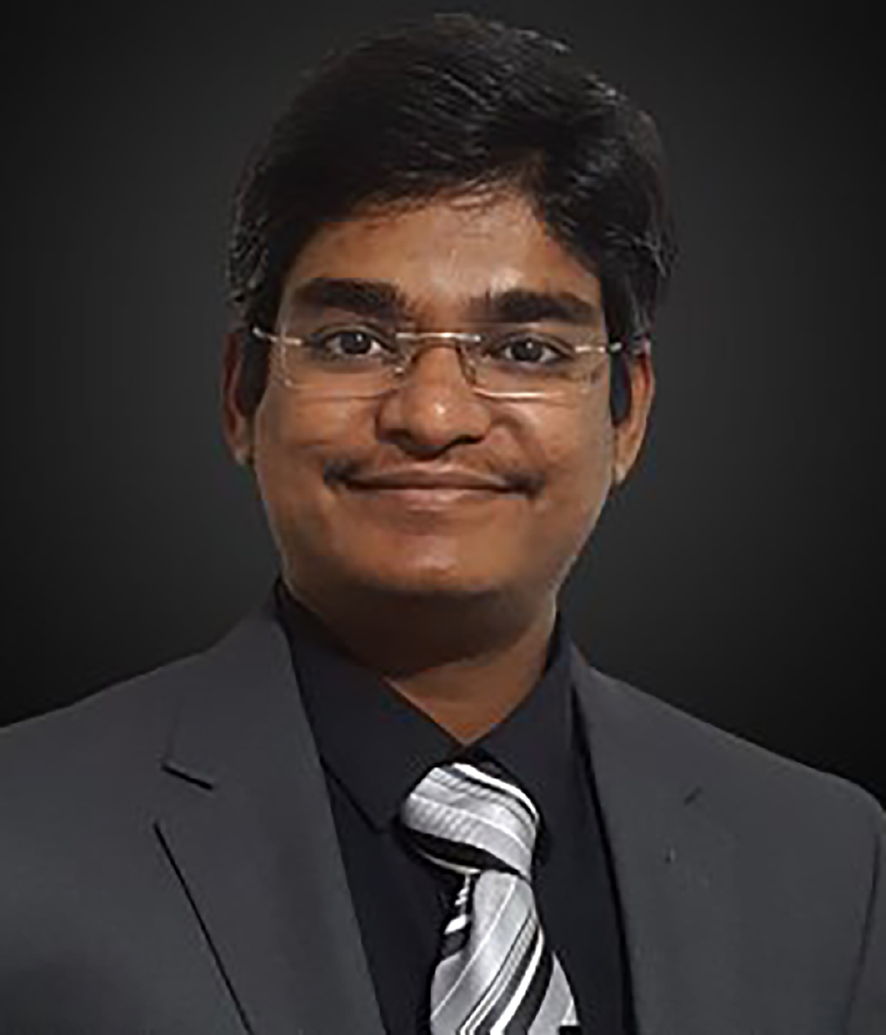}
		\vspace{-0.9cm}
	\end{wrapfigure}
\noindent 
\textbf{Venkata P. Yanambaka} (M'19) received the Bachelor of Technology degree in electronics and communications from the JNTU, India, in 2014.  He obtained his Ph.D. at the System Electronic Systems Laboratory (SESL) at the Department of Computer Science and Engineering, University of North Texas. He is currently an Assistant Professor in the School of Engineering and Technology, Central Michigan University. His research interests are in Security in Internet of Things (IoT), Energy-Efficient Circuits and Systems, and Application-Specific Systems Design. He has authored of a 12 research articles which include multiple journals/transactions articles. 
He has a regular reviewer of various peer-reviewed journals and conferences.
\end{minipage}

	
\begin{minipage}[htbp]{\columnwidth}
	\vspace{1.0cm}	
	\begin{wrapfigure}{l}{1.4in}
		\vspace{-0.5cm}
		\includegraphics[width=1.4in,keepaspectratio]{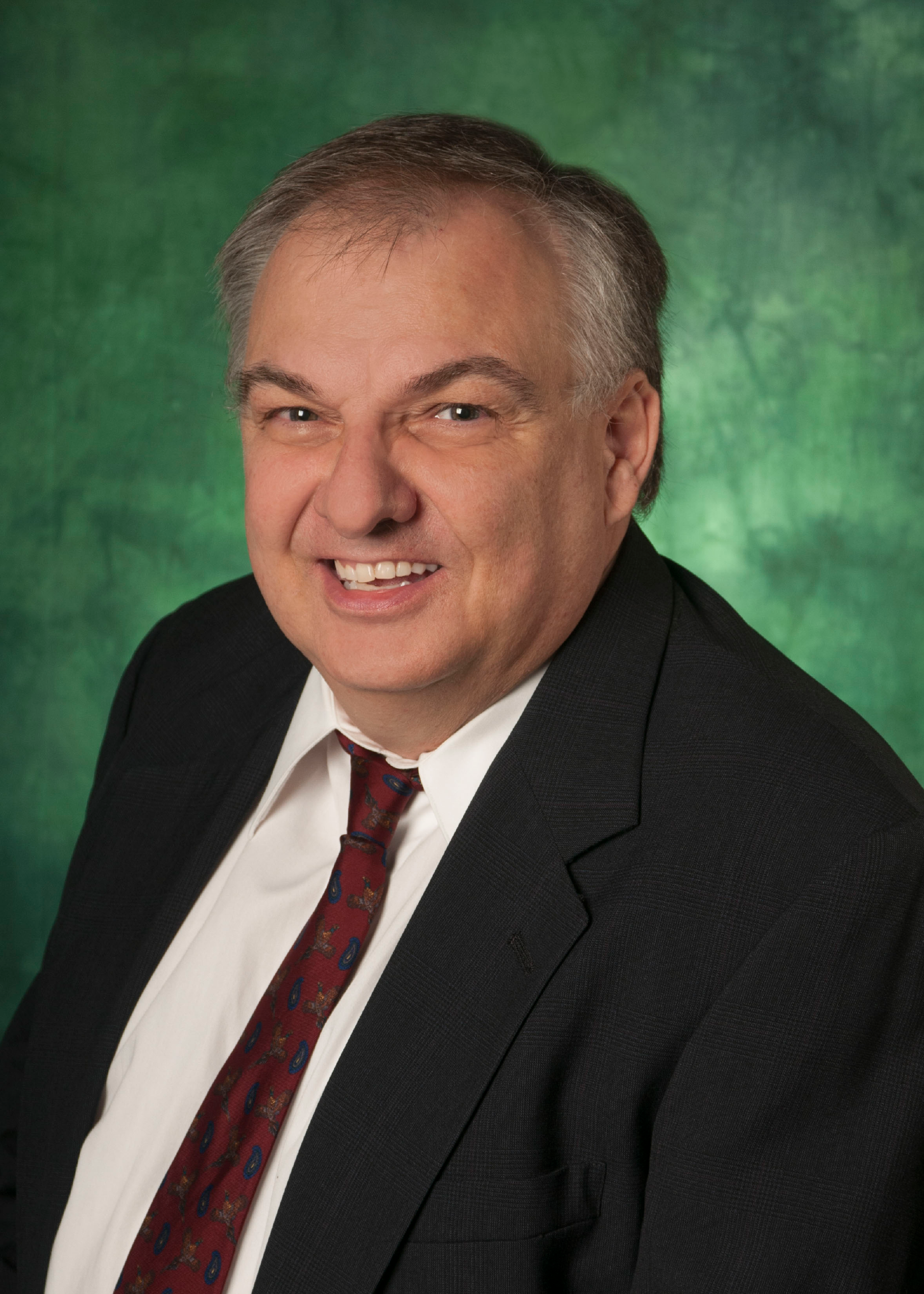}
		\vspace{-0.9cm}
	\end{wrapfigure}
	\noindent
\textbf{Elias Kougianos} (SM'07) received a BSEE from the University of Patras, Greece in 1985 and an MSEE in 1987, an MS in Physics in 1988 and a Ph.D. in EE in 1997, all from Louisiana State University. 
From 1988 through 1997 he was with Texas Instruments, Inc., in Houston and Dallas, TX.
Initially he concentrated on process integration of flash memories and later as a researcher in the areas of Technology CAD and VLSI CAD development. 
In 1997 he joined Avant! Corp. (now Synopsys) in Phoenix, AZ as a Senior Applications engineer and in 2001 he joined Cadence Design Systems, Inc., in Dallas, TX as a Senior Architect in Analog/Mixed-Signal Custom IC design. He has been at UNT since 2004. He is a Professor in the Department of Electrical Engineering, at the University of North Texas (UNT), Denton, TX. His research interests are in the area of Analog/Mixed-Signal/RF IC design and simulation and in the development of VLSI architectures for multimedia applications. 
He is an author of over 120 peer-reviewed journal and conference publications.
\end{minipage}


\begin{minipage}[htbp]{\columnwidth}
	\vspace{1.0cm}	
	\begin{wrapfigure}{l}{1.4in}
		\vspace{-0.5cm}
		\includegraphics[width=1.4in,keepaspectratio]{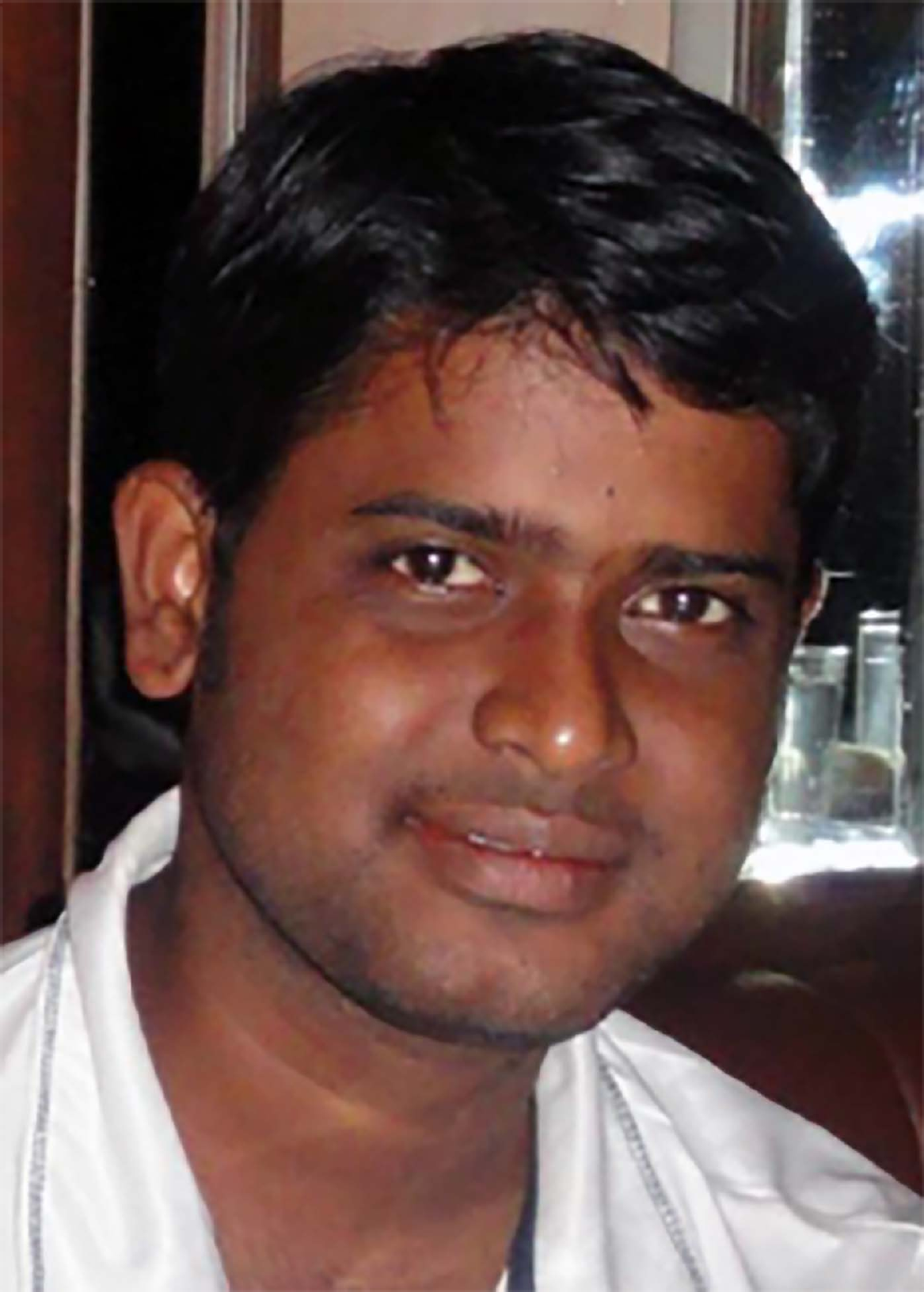}
		\vspace{-0.9cm}
	\end{wrapfigure}
	\noindent
\textbf{Deepak Puthal} (M'16) received the Ph.D. degree in computer science from the University of Technology Sydney (UTS), Australia. He is currently a Lecturer at School of Computing, Newcastle University, Newcastle upon Tyne, UK. He is an author/co-author of more than 100 peer-reviewed publications in international conferences and journals, including ACM and IEEE transactions. His research interests include cyber security, Internet of Things, distributed computing, and edge/fog computing. He has been a Program Chair and a Program Committee member in several IEEE and ACM sponsored conferences. He was a recipient of the 2017 IEEE Distinguished Doctoral Dissertation Award from the IEEE Computer Society and STC on Smart Computing. He served as a Co-Guest Editor of several reputed journals, including Concurrency and Computation: Practice and Experience, Wireless Communications and Mobile Computing, and Information Systems Frontier. He is an Associate Editor of the IEEE Transactions on Big Data, and IEEE Consumer Electronics Magazine.

\end{minipage}



\end{document}